\documentclass[english,prd,superscriptaddress,nofootinbib,preprintnumbers]{revtex4}
\usepackage[latin1]{inputenc}
\usepackage{bm}
\usepackage{amsmath}
\usepackage{amssymb}
\usepackage{graphicx}
\usepackage{esint}

\makeatletter

\providecommand{\tabularnewline}{\\}

\@ifundefined{textcolor}{}
{%
 \definecolor{BLACK}{gray}{0}
 \definecolor{WHITE}{gray}{1}
 \definecolor{RED}{rgb}{1,0,0}
 \definecolor{GREEN}{rgb}{0,1,0}
 \definecolor{BLUE}{rgb}{0,0,1}
 \definecolor{CYAN}{cmyk}{1,0,0,0}
 \definecolor{MAGENTA}{cmyk}{0,1,0,0}
 \definecolor{YELLOW}{cmyk}{0,0,1,0}
}

\usepackage{bm}\usepackage{color}

\usepackage{amsfonts}\usepackage{dcolumn}

\def\be{\begin{equation}}
\def\ee{\end{equation}}
\def\ba{\begin{eqnarray}}
\def\ea{\end{eqnarray}}
\def\bs{\begin{subequations}}
\def\es{\end{subequations}}

\def\Mpl{M_{\rm pl}}

\usepackage{color}

\usepackage{babel}

\makeatother

\begin{document}

\title{Shapes of primordial non-Gaussianities in the Horndeski's \\
most general scalar-tensor
theories}

\author{Antonio De Felice}

\affiliation{ThEP's CRL, NEP, The Institute for Fundamental Study, Naresuan University,
Phitsanulok 65000, Thailand}

\affiliation{Thailand Center of Excellence in Physics, Ministry of Education,
Bangkok 10400, Thailand}

\author{Shinji Tsujikawa}

\affiliation{Department of Physics, Faculty of Science, Tokyo University of Science,
1-3, Kagurazaka, Shinjuku-ku, Tokyo 162-8601, Japan}

\begin{abstract}
In the Horndeski's most general scalar-tensor theories, we derive
the three-point correlation function of scalar non-Gaussianities generated
during single-field inflation in the presence of slow-variation corrections
to the leading-order term. Unlike previous works, the resulting bispectrum
is valid for any shape of non-Gaussianities. In the squeezed limit,
for example, this gives rise to the same consistency relation as that
derived by Maldacena in standard single-field slow-roll inflation.
We estimate the shape close to the squeezed one at which the effect
of the term inversely proportional to the scalar propagation speed
squared begins to contribute to the bispectrum. 
We also show that the leading-order
bispectrum can be expressed by the linear combination of two convenient
bases whose shapes are highly correlated with equilateral and orthogonal
types respectively. We present concrete models in which the orthogonal
and enfolded shapes can dominate over the equilateral one.
\end{abstract}

\date{\today}

\maketitle

\section{Introduction}

\label{inrosec} 

 The potential presence of primordial non-Gaussianities in the CMB
temperature anisotropies can be a powerful probe for the physics in
the early Universe--especially for inflation \cite{inflation}. The
inflationary paradigm generally predicts nearly scale-invariant density
perturbations \cite{oldper} with a suppressed tensor-to-scalar ratio,
whose prediction is consistent with the CMB power spectrum measured
by COBE \cite{COBE} and WMAP \cite{WMAP1,WMAP9}. The detection of
scalar non-Gaussianities not only breaks the degeneracy among many
inflationary models, but it also offers the possibility to discriminate
between the inflationary paradigm and other alternative scenarios
(such as curvaton \cite{curvaton}) \cite{Salopek}-\cite{Ribeiro}.

There are several different shapes of non-Gaussianities depending
on the wave numbers ${\bm{k}}_{1}$, ${\bm{k}}_{2}$, and ${\bm{k}}_{3}$
satisfying the condition ${\bm{k}}_{1}+{\bm{k}}_{2}+{\bm{k}}_{3}=0$
\cite{Wandelt}-\cite{CremiGa}. The simplest one is the so-called
local shape, which has a peak in the squeezed limit (i.e., the limit
where the modulus of the momenta approaches $k_{3}\to0$ and $k_{1}\simeq k_{2}$).
The second shape corresponds to the equilateral configuration with
a peak at $k_{1}=k_{2}=k_{3}$. A factorizable shape whose scalar
product with the equilateral template vanishes is called the orthogonal
one. There is another shape dubbed the enfolded one, which is a linear
combination of the equilateral and orthogonal templates.

{}From the bispectrum ${\cal A}_{{\cal R}}$ of the three-point correlation
function of curvature perturbations ${\cal R}$, the non-linear parameter
characterizing the strength of non-Gaussianities is defined by 
$f_{{\rm NL}}=(10/3)\,{\cal A}_{{\cal R}}/\sum_{i=1}^{3}k_{i}^{3}$.
For purely adiabatic Gaussian perturbations we have that $f_{{\rm NL}}=0$,
but the presence of non-Gaussian perturbations leads to the deviation
from $f_{{\rm NL}}=0$. The WMAP 9 year data provide the following
bounds on the non-linear parameters of local, equilateral, and orthogonal
non-Gaussianities, respectively \cite{WMAP9fnl}: 
\begin{eqnarray}
 &  & f_{{\rm NL}}^{{\rm local}}=37.2\pm19.9~~~~~~~\,(68\,\%\,{\rm CL})\,,
 \qquad f_{{\rm NL}}^{{\rm local}}=37\pm40~~~~~~~~~~~\,\,(95\,\%\,{\rm CL})\,,\label{localcon}\\
 &  & f_{{\rm NL}}^{{\rm equil}}=51\pm136~~~~~~~~~~\,(68\,\%\,{\rm CL})\,,\qquad f_{{\rm NL}}^{{\rm equil}}=51\pm272~~~~~~~~~~\,(95\,\%\,{\rm CL})\,,\label{equilcon}\\
 &  & f_{{\rm NL}}^{{\rm ortho}}=-245\pm100~~~~~~(68\,\%\,{\rm CL})\,,
 \qquad f_{{\rm NL}}^{{\rm ortho}}=-245\pm200~~~~~~(95\,\%\,{\rm CL})\,.\label{orthocon}
\end{eqnarray}
Since the non-linear parameter of the enfolded shape is given by 
$f_{{\rm NL}}^{{\rm enfold}}=(f_{{\rm NL}}^{{\rm equil}}-f_{{\rm NL}}^{{\rm ortho}})/2$
\cite{Meerburg,CremiGa}, we obtain the following bounds from Eqs.~(\ref{equilcon})
and (\ref{orthocon}): 
\begin{equation}
f_{{\rm NL}}^{{\rm enfold}}=148\pm118~~~~~~~~(68\,\%\,{\rm CL})\,,
\qquad f_{{\rm NL}}^{{\rm enfold}}=148\pm236~~~~~~~(95\,\%\,{\rm CL})\,.
\label{enfoldcon}
\end{equation}
For the local, orthogonal, and enfolded shapes the model with purely
Gaussian perturbations ($f_{{\rm NL}}=0$) is outside the 68 \% observational
contour, but, apart from the orthogonal case, it is still consistent
with the WMAP constraints at 95 \% CL.

In standard single-field inflation based on a canonical scalar field,
Maldacena \cite{Maldacena} showed that the non-linear parameter in
the squeezed limit is given by $f_{{\rm NL}}^{{\rm local}}=(5/12)(1-n_{{\cal R}})$,
where $n_{{\cal R}}$ is the scalar spectral index. Creminelli and
Zaldarriaga \cite{Cremi} pointed out that the same non-Gaussianity
consistency relation holds for any single-field model under the condition
that only one mode of curvature perturbations survives after the Hubble
radius crossing while the other one decays\footnote{If the decaying mode 
is non-negligible relative to the growing mode,
the Maldacena's consistency relation can be violated \cite{Arroja,Namjoo}.
} (see Refs.~\cite{Cheung,Ganc,Petelsqueezed,Watanabe} for related works).

In the context of single-field k-inflation \cite{Mukhanov}, the bispectrum
of curvature perturbations was first derived by Seery and Lidsey in
2005 \cite{Seery}. Since the scalar propagation speed squared $c_{s}^{2}$
can be much smaller than 1 \cite{Garriga}, it is possible to realize
the large equilateral non-linear parameter $|f_{{\rm NL}}^{{\rm equil}}|\sim1/c_{s}^{2}\gg1$.
If we naively take the squeezed limit for the leading-order bispectrum
derived in Refs.~\cite{Seery,Chen}, the term proportional to $1/c_{s}^{2}$ does
not disappear. This comes from the fact that the slow-variation corrections
to the bispectrum need to be taken into account to estimate the local-type
non-Gaussianity correctly. In fact, Chen \textit{et al.} \cite{Chen}
showed that the Maldacena's consistency relation is recovered in the
squeezed limit by carefully computing all the possible slow-variation corrections
to the leading-order bispectrum. 
Thus the slow-variation single-field
k-inflation models with $c_{s}^{2}\ll1$ lead to small local non-Gaussianities, even
though the equilateral non-linear parameter can be large.

In the Horndeski's most general scalar-tensor theories with second-order
equations of motion \cite{Horndeski,DGSZ,Char,Koba11}, the leading-order
three-point correlation function of curvature perturbations was derived
on the quasi de Sitter background \cite{Gao,ATnongau} (see Refs.~\cite{Mizuno,ATnongau0,KobaGa}
for the scalar non-Gaussianities in related Galileon models and Refs.~\cite{Gaotensor}
for the bispectrum of tensor perturbations in the Horndeski's theories).
Although the result is valid for the estimation of the equilateral
non-linear parameter, the bispectrum is not general enough to be used
for any shape of non-Gaussianities. In this paper we take into account
all the possible slow-variation corrections to the leading-order bispectrum
in the Horndeski's 
theories\footnote{In the effective field theory of inflation (which allows the equations 
of motion higher than second order), a similar approach was 
taken by Cheung {\it et al.} \cite{Cheung} to show that the Maldacena's 
consistency relation holds in the squeezed limit.
While the authors in this paper mainly focused on the local shape, 
we derive the full bispectrum in the Horndeski's theory
which can be used for any shape of non-Gaussianities.}. 
Not only we reproduce the Maldacena's
consistency relation in the squeezed limit, but we identify the
shape close to the squeezed one at which the term $1/c_{s}^{2}$ begins
to contribute to the bispectrum.

Given our general expression of the bispectrum in the most general
single-field scalar-tensor theories, we can evaluate the non-linear
parameters of several different shapes to confront each inflationary
model with observations. In particular the result 
$|f_{{\rm NL}}^{{\rm local}}| \ll 1$
is robust for any slow-variation single-field model, so the detection of
non-Gaussianities in the squeezed limit will allow us to falsify the
slow-variation single-field scenario. Note that in realistic observations the shape
is not completely squeezed, in which case the bispectrum can be affected
by the appearance of the term $1/c_{s}^{2}$ mentioned above. Our
results are useful to distinguish such difference accurately.

If $c_{s}^{2}\ll1$, then the non-linear parameters $|f_{{\rm NL}}^{{\rm equil}}|$,
$|f_{{\rm NL}}^{{\rm ortho}}|$, and $|f_{{\rm NL}}^{{\rm enfold}}|$
can be much larger than the order of 1. Which shape dominates over
the other ones depends on the models of inflation. In Ref.~\cite{ATnongau}
it was shown that the correlation between the equilateral template
and the shapes arising from the Horndeski's theories is quite high,
but linear combinations of equilateral operators can give rise to
a significantly different shape for a wide range of 
coefficients \cite{Meerburg,CremiGa}.
In this regard we anticipate that there may be some models in which
the shape orthogonal to the equilateral template provides an important
contribution to the bispectrum.

In this paper we show that the leading-order three-point correlation
function in the Horndeski's theories can be expressed by a linear
combination of two bases whose shapes are highly correlated with equilateral
and orthogonal shapes respectively. This decomposition is useful because
the contributions from the equilateral and orthogonal shapes can be
easily estimated for concrete models of inflation. We show that in
k-inflation with the covariant Galileon terms there are cases in which
the correlations with the orthogonal and enfolded templates are 
larger than that with the equilateral one. Thus
the shapes of non-Gaussianities allow us to discriminate such models
from observations.

This paper is organized as follows. In Sec.~\ref{linear} we review
the background and linear perturbation equations in the Horndeski's
theories. In Sec.~\ref{threepoint} we derive the three-point correlation
function of curvature perturbations in the presence of slow-variation corrections
to the leading-order bispectrum. In Sec.~\ref{leesec} the non-linear
parameter $f_{{\rm NL}}$ is evaluated in the squeezed, equilateral,
and enfolded limits, respectively. In Sec.~\ref{shapesec} we express
the leading-order bispectrum in terms of equilateral and orthogonal
bases. In Sec.~\ref{concretesec} we show concrete models of inflation
in which the orthogonal and enfolded shapes can dominate over the equilateral one.
Sec.~\ref{concludesec} is devoted to conclusions. In Appendix we
show the details of the slow-variation corrections to the bispectrum.

\section{Equations of motion for the background and linear perturbations}
\label{linear} 

The action corresponding to the most general scalar-tensor theories
is given by \cite{Horndeski,DGSZ} %
\begin{equation}
{\cal S}=\int d^{4}x\sqrt{-g}\biggl[\frac{\Mpl^{2}}{2}\, R+P(\phi,X)-G_{3}(\phi,X)\,\Box\phi+\mathcal{L}_{4}+\mathcal{L}_{5}\biggr]\,,\label{action}
\end{equation}
where $g$ is the determinant of the metric $g_{\mu\nu}$, $M_{{\rm pl}}$
is the reduced Planck mass, $R$ is a Ricci scalar, and 
\begin{eqnarray}
\mathcal{L}_{4} & = & G_{4}(\phi,X)\, R+G_{4,X}\,[(\Box\phi)^{2}-(\nabla_{\mu}\nabla_{\nu}\phi)\,(\nabla^{\mu}\nabla^{\nu}\phi)]\,,\\
\mathcal{L}_{5} & = & G_{5}(\phi,X)\, G_{\mu\nu}\,(\nabla^{\mu}\nabla^{\nu}\phi)-\frac{1}{6}G_{5,X}[(\Box\phi)^{3}-3(\Box\phi)\,(\nabla_{\mu}\nabla_{\nu}\phi)\,(\nabla^{\mu}\nabla^{\nu}\phi)+2(\nabla^{\mu}\nabla_{\alpha}\phi)\,(\nabla^{\alpha}\nabla_{\beta}\phi)\,(\nabla^{\beta}\nabla_{\mu}\phi)]\,.
\end{eqnarray}
Here $P$ and $G_{i}$'s ($i=3,4,5$) are functions in terms of $\phi$
and $X=-\partial^{\mu}\phi\partial_{\mu}\phi/2$ with the partial
derivatives $G_{i,X}\equiv\partial G_{i}/\partial X$, and $G_{\mu\nu}=R_{\mu\nu}-g_{\mu\nu}R/2$
is the Einstein tensor ($R_{\mu\nu}$ is the Ricci tensor). 

We consider the following ADM metric \cite{ADM} with scalar metric perturbations
$\alpha$, $\psi$, and ${\cal R}$ about the flat 
Friedmann-Lema\^{i}tre-Robertson-Walker (FLRW) background
\begin{equation}
ds^{2}=-[(1+\alpha)^{2}-a(t)^{-2}\, e^{-2{\cal R}}\,(\partial\psi)^{2}]\, dt^{2}
+2\partial_{i}\psi\, dt\, dx^{i}+a(t)^{2}e^{2{\cal R}}d\bm{x}^{2}\,,\label{eq:metric}
\end{equation}
where $a(t)$ is the scale factor with cosmic time $t$. We choose
the uniform field gauge $\delta\phi=0$, which fixes the time-component
of a gauge-transformation vector $\xi^{\mu}$. The spatial part of
$\xi^{\mu}$ is fixed by gauging away a perturbation $E$ that appears
as a form $E_{,ij}$ in the metric (\ref{eq:metric}).

The background equations of motion are given by 
\begin{eqnarray}
 &  & 3\Mpl^{2}H^{2}F+P+6HG_{{4,\phi}}\dot{\phi}+\left(G_{{3,\phi}}-12\,{H}^{2}G_{{4,X}}+9\,{H}^{2}G_{{5,\phi}}-P_{{,X}}\right){\dot{\phi}^{2}}\nonumber \\
 &  & +\left(6\, G_{{4,\phi X}}-3\, G_{{3,X}}-5\, G_{{5,X}}{H}^{2}\right)H{\dot{\phi}^{3}}+3\left(G_{{5,\phi X}}-2\, G_{{4,{\it XX}}}\right)H^{2}\dot{\phi}^{4}-{H}^{3}G_{{5,{\it XX}}}{\dot{\phi}^{5}}=0\,,\label{E1d}\\
 &  & (1-4\delta_{G4X}-2\delta_{G5X}+2\delta_{G5\phi})\epsilon=\delta_{PX}+3\delta_{G3X}-2\delta_{G3\phi}+6\,\delta_{G4X}-\delta_{G4\phi}-6\,\delta_{G5\phi}+3\,\delta_{G5X}+12\,\delta_{G4XX}+2\,\delta_{G5XX}\nonumber \\
 &  & ~~~~~~~~~~~~~~~~~~~~~~~~~~~~~~~~~~~~~~~~~~~~-10\,\delta_{G4\phi X}+2\,\delta_{G4\phi\phi}-8\,\delta_{G5\phi X}+2\,\delta_{G5\phi\phi}-\delta_{\phi}(\delta_{G3X}+4\,\delta_{G4X}-\delta_{G4\phi}\nonumber \\
 &  & ~~~~~~~~~~~~~~~~~~~~~~~~~~~~~~~~~~~~~~~~~~~~+8\,\delta_{G4XX}+3\,\delta_{G5X}-4\,\delta_{G5\phi}+2\,\delta_{G5XX}-2\delta_{G4\phi X}-4\,\delta_{G5\phi X})\,,\label{E1}
\end{eqnarray}
where $H=\dot{a}/a$ is the Hubble parameter (a dot represents a derivative with 
respect to $t$), $F=1+2G_{4}/\Mpl^{2}$, and 
\begin{eqnarray}
\hspace{-0.4cm} &  & \epsilon=-\frac{\dot{H}}{H^{2}}\,,\quad\delta_{\phi}=\frac{\ddot{\phi}}{H\dot{\phi}}\,,\quad\delta_{PX}=\frac{P_{,X}X}{\Mpl^{2}H^{2}F}\,,\quad\delta_{G3X}=\frac{G_{3,X}\dot{\phi}X}{\Mpl^{2}HF}\,,\quad\delta_{G3\phi}=\frac{G_{3,\phi}X}{\Mpl^{2}H^{2}F}\,,\quad\delta_{G4X}=\frac{G_{4,X}X}{\Mpl^{2}F}\,,\nonumber \\
\hspace{-0.4cm} &  & \delta_{G4\phi}=\frac{G_{4,\phi}\dot{\phi}}{\Mpl^{2}HF}\,,\quad\delta_{G4\phi X}=\frac{G_{4,\phi X}\dot{\phi}X}{\Mpl^{2}HF}\,,\quad\delta_{G4\phi\phi}=\frac{G_{4,\phi\phi}X}{\Mpl^{2}H^{2}F}\,,\quad\delta_{G4XX}=\frac{G_{4,XX}X^{2}}{\Mpl^{2}F}\,,\quad\delta_{G5\phi}=\frac{G_{5,\phi}X}{\Mpl^{2}F}\,,\nonumber \\
\hspace{-0.4cm} &  & \delta_{G5X}=\frac{G_{5,X}H\dot{\phi}X}{\Mpl^{2}F}\,,\quad\delta_{G5XX}=\frac{G_{5,XX}H\dot{\phi}X^{2}}{\Mpl^{2}F}\quad\delta_{G5\phi X}=\frac{G_{5,\phi X}X^{2}}{\Mpl^{2}F}\,,\quad\delta_{G5\phi\phi}=\frac{G_{5,\phi\phi}\dot{\phi}X}{\Mpl^{2}HF}\,,\label{slowva}
\end{eqnarray}
whose magnitudes are much smaller than 1 during inflation. 
The terms
$\delta_{G4\phi X}$, $\delta_{G4\phi\phi}$, $\delta_{G5\phi X}$,
$\delta G_{5\phi\phi}$ as well as $\delta_{G3\phi X}=G_{3,\phi X}X^{2}/(M_{{\rm pl}}^{2}H^{2}F)$
and $\delta_{G3\phi\phi}=G_{3,\phi\phi}\dot{\phi}X/(M_{{\rm pl}}^{2}H^{3}F)$
are second-order of $\epsilon$.
{}From Eq.~(\ref{E1}) it follows that 
\begin{equation}
\epsilon=\delta_{PX}+3\delta_{G3X}-2\delta_{G3\phi}+6\,\delta_{G4X}
-\delta_{G4\phi}-6\,\delta_{G5\phi}+3\,\delta_{G5X}+12\,\delta_{G4XX}+
2\,\delta_{G5XX}+{\cal O}(\epsilon^2)\,.\label{epap}
\end{equation}
For the quantity $\delta_{F}=\dot{F}/(HF)$ we have
\begin{equation}
\delta_{F}=2\delta_{G4\phi}+{\cal O} (\epsilon^2)\,.
\label{delF}
\end{equation}

Using the relations between ${\cal R}$, $\psi$, and $\alpha$ that
follows from Hamiltonian and momentum constraints, the second-order
action for perturbations reduces to \cite{Koba11,Gao,ATnongau,DeFe2012} 
\begin{equation}
{\cal S}_{2}=\int dtd^{3}x\, a^{3}Q\left[\dot{{\cal R}}^{2}-\frac{c_{s}^{2}}{a^{2}}\,(\partial{\cal R})^{2}\right]\,,\label{secondaction}
\end{equation}
where 
\begin{eqnarray}
Q & = & \frac{w_{1}(4w_{1}w_{3}+9w_{2}^{2})}{3w_{2}^{2}}\,,\label{Qdef}\\
c_{s}^{2} & = & \frac{3(2w_{1}^{2}w_{2}H-w_{2}^{2}w_{4}+4w_{1}\dot{w}_{1}w_{2}-2w_{1}^{2}\dot{w}_{2})}{w_{1}(4w_{1}w_{3}+9w_{2}^{2})}\,,
\end{eqnarray}
and 
\begin{eqnarray}
w_{1} & = & \Mpl^{2}F-4XG_{4,X}-2HX\dot{\phi}G_{5,X}+2XG_{5,\phi}\,,\\
w_{2} & = & 2\Mpl^{2}HF-2X\dot{\phi}G_{3,X}-16H(XG_{4,X}+X^{2}G_{4,XX})+2\dot{\phi}(G_{4,\phi}+2XG_{4,\phi X})\nonumber \\
 &  & {}-2H^{2}\dot{\phi}(5XG_{5,X}+2X^{2}G_{5,XX})+4HX(3G_{5,\phi}+2XG_{5,\phi X})\,,\\
w_{3} & = & -9M_{{\rm pl}}^{2}H^{2}F+3(XP_{,X}+2X^{2}P_{,XX})+18H\dot{\phi}(2XG_{3,X}+X^{2}G_{3,XX})-6X(G_{3,\phi}+XG_{3,\phi X})\nonumber \\
 &  & +18H^{2}(7XG_{4,X}+16X^{2}G_{4,XX}+4X^{3}G_{4,XXX})-18H\dot{\phi}(G_{4,\phi}+5XG_{4,\phi X}+2X^{2}G_{4,\phi XX})\nonumber \\
 &  & {}+6H^{3}\dot{\phi}(15XG_{5,X}+13X^{2}G_{5,XX}+2X^{3}G_{,5XXX})-18H^{2}X(6G_{5,\phi}+9XG_{5,\phi X}+2X^{2}G_{5,\phi XX})\,,\\
w_{4} & = & \Mpl^{2}F-2XG_{5,\phi}-2XG_{5,X}\ddot{\phi}\,.
\end{eqnarray}
For later convenience
we introduce the following parameter 
\begin{equation}
\epsilon_{s}\equiv\frac{Qc_{s}^{2}}{M_{{\rm pl}}^{2}F}\simeq\epsilon
+\delta_{G3X}+\delta_{G4\phi}+8\delta_{G4XX}+\delta_{G5X}+2\delta_{G5XX}+{\cal O}(\epsilon^{2})\,.\label{eps}
\end{equation}

At linear level the curvature perturbation obeys
the equation of motion
\begin{equation}
\frac{\delta {\cal L}_2}{\delta {\cal R}}\bigg|_1 \equiv
-2 \left[ \frac{d}{dt}(a^{3}Q\dot{{\cal R}})-aQc_{s}^{2}\partial^{2}{\cal R} \right]=0\,.
\end{equation}
We decompose ${\cal R}$ into the Fourier components, as
\begin{equation}
{\cal R}(\tau,{\bm{x}})=\frac{1}{(2\pi)^{3}}\int d^{3}{k}\,{\cal R}(\tau,{\bm{k}})e^{i{\bm{k}}\cdot{\bm{x}}}\,,\qquad{\cal R}(\tau,{\bm{k}})=u(\tau,{\bm{k}})a({\bm{k}})+u^{*}(\tau,{-\bm{k}})a^{\dagger}(-{\bm{k}})\,,\label{RFourier}
\end{equation}
where $\tau=\int a^{-1}\, dt$, ${\bm{k}}$ is the comoving wave number,
$a({\bm{k}})$ and $a^{\dagger}({\bm{k}})$ are the annihilation and
creation operators, respectively, satisfying the commutation relations
$\left[a({\bm{k}}_{1}),a^{\dagger}({\bm{k}}_{2})\right]=(2\pi)^{3}\delta^{(3)}({\bm{k}}_{1}-{\bm{k}}_{2})$
and $\left[a({\bm{k}}_{1}),a({\bm{k}}_{2})\right]=\left[a^{\dagger}({\bm{k}}_{1}),a^{\dagger}({\bm{k}}_{2})\right]=0$.

Introducing a rescaled field $v=zu$ with $z=a\sqrt{2Q}$, 
it follows that
\begin{equation}
v''+\left(c_{s}^{2}k^{2}-\frac{z''}{z}\right)v=0\,,\label{veq}
\end{equation}
where a prime represents a derivative with respect to $\tau$. Under
the slow-variation approximation the term $z''/z$ can be expressed as 
\begin{equation}
\frac{z''}{z}=2(aH)^{2}\left(1-\frac{1}{2}\epsilon+\frac{3}{4}\eta_{sF}-\frac{3}{2}s\right)+{\cal O}(\epsilon^{2})\,,\label{zdd}
\end{equation}
where 
\begin{equation}
\eta_{sF}\equiv\frac{(\epsilon_{s}F)^{\cdot}}{H(\epsilon_{s}F)}
=\eta_{s}+\delta_{F}\,,
\qquad\eta_{s}\equiv\frac{\dot{\epsilon}_{s}}{H\epsilon_{s}}\,,
\qquad
s \equiv \frac{\dot{c}_s}{Hc_s}\,.
\label{etasF}
\end{equation}
Taking the dominant contribution in Eq.~(\ref{zdd}) and using the
approximate relation $a\simeq-1/(H\tau)$, we have $z''/z\simeq2/\tau^{2}$.
The solution to Eq.~(\ref{veq}), which recovers the Bunch-Davies
vacuum state ($v=e^{-ic_{s}k\tau}/\sqrt{2c_{s}k}$) in the asymptotic
past ($k\tau\to-\infty$), is given by 
\begin{equation}
u(\tau,k)=\frac{i\, H\, e^{-ic_{s}k\tau}}{2(c_{s}k)^{3/2}\sqrt{Q}}\,(1+ic_{s}k\tau)\,.\label{usol}
\end{equation}
The slow-variation terms in Eq.~(\ref{zdd}) provide the corrections
to the mode function (\ref{usol}). Later we shall discuss the effect of
such corrections on the primordial non-Gaussianities.

The power spectrum ${\cal P}_{{\cal R}}(k_{1})$ of curvature
perturbations, some time after the Hubble radius crossing, is defined
by $\langle0|{\cal R}(0,{\bm{k}}_{1}){\cal R}(0,{\bm{k}}_{2})|0\rangle
=(2\pi^{2}/k_{1}^{3}){\cal P}_{{\cal R}}(k_{1})\,(2\pi)^{3}\delta^{(3)}({\bm{k}}_{1}+{\bm{k}}_{2})$.
{}From Eq.~(\ref{usol}) it follows that 
\begin{equation}
{\cal P}_{{\cal R}}=\frac{H^{2}}{8\pi^{2}\Mpl^{2}\epsilon_{s}Fc_{s}}\,,
\label{scalarpower}
\end{equation}
which should be evaluated at $c_{s}k=aH$. 
The spectral index $n_{{\cal R}}$ is given by 
\begin{equation}
n_{{\cal R}}-1\equiv\frac{d\ln{\cal P}_{{\cal R}}}{d\ln k}\bigg|_{c_{s}k=aH}
=-2\epsilon-\eta_{sF}-s\,.\label{nR}
\end{equation}
The corrections to the solution (\ref{usol}) only give rise to the
${\cal O}(\epsilon^2)$ terms in Eq.~(\ref{nR}).

Similarly the power spectrum ${\cal P}_{h}$ and the spectral index $n_t$ 
of gravitational waves are given, respectively, by \cite{Koba11,ATnongau}
\begin{equation}
{\cal P}_{h}=\frac{H^{2}}
{2\pi^{2}Q_t c_t^3} \simeq 
\frac{2H^2}{\pi^2 M_{\rm pl}^2 F}\,,\qquad
n_t =\frac{d\ln{\cal P}_h}{d\ln k}\bigg|_{c_{t}k=aH}
=-2\epsilon-\delta_{F}\,,
\end{equation}
where $Q_t=w_1/4=(M_{\rm pl}^2 F/4)
(1-4\delta_{G4X}-2\delta_{G5X}+2\delta_{G5\phi})$ and 
$c_t^2=w_4/w_1 \simeq 1+4\delta_{G4X}+2\delta_{G5X}-4\delta_{G5\phi}$.
When both ${\cal P}_{{\cal R}}$ and ${\cal P}_h$ remain constant, 
the tensor-to-scalar ratio can be evaluated as 
\begin{equation}
r=\frac{{\cal P}_h}{{\cal P}_{\cal R}} \simeq 16c_s \epsilon_s\,.
\label{tsratio}
\end{equation}

\section{Three-point correlation functions in the presence of correction terms}
\label{threepoint} 

In the Horndeski's theories the third-order action of 
perturbations was derived in Ref.~\cite{Gao,ATnongau}.
Here we do not repeat the details, but we summarize the main 
results. Under the approximation that all of the slow-variation 
terms in Eq.~(\ref{slowva}) are much smaller than 1, 
the third-order action reads
\begin{eqnarray}
{\cal S}_{3} & = & \int dt\, d^{3}x\biggl\{ a^{3}{\cal C}_{1}\Mpl^{2}{\cal R}
\dot{{\cal R}}^{2}+a\,{\cal C}_{2}\Mpl^{2}{\cal R}(\partial{\cal R})^{2}
+a^{3}{\cal C}_{3}\Mpl\dot{{\cal R}}^{3}+a^{3}{\cal C}_{4}\dot{{\cal R}}
(\partial_{i}{\cal R})(\partial_{i}{\cal X})+a^{3}({\cal C}_{5}/\Mpl^{2})\partial^{2}{\cal R}(\partial{\cal X})^{2}\nonumber \\
 &  & {}+a{\cal C}_{6}\dot{{\cal R}}^{2}\partial^{2}{\cal R}+{\cal C}_{7}\left[\partial^{2}{\cal R}(\partial{\cal R})^{2}-{\cal R}\partial_{i}\partial_{j}(\partial_{i}{\cal R})(\partial_{j}{\cal R})\right]/a+a({\cal C}_{8}/\Mpl)\left[\partial^{2}{\cal R}\partial_{i}{\cal R}\partial_{i}{\cal X}-{\cal R}\partial_{i}\partial_{j}(\partial_{i}{\cal R})(\partial_{j}{\cal X})\right]\nonumber \\
 &  & {}+{\cal F}_{1}\frac{\delta{\cal L}_{2}}{\delta{\cal R}}\biggr|_{1}\biggr\}\,,
 \label{L3}
\end{eqnarray}
where $\partial^{2}{\cal X}=Q\dot{{\cal R}}$.
The dimensionless coefficients ${\cal C}_{i}$ ($i=1,\cdots,8$) and the coefficient
${\cal F}_1$ are \cite{ATnongau} 
\begin{eqnarray}
{\cal C}_{1} & = & -\frac{3F\epsilon_{s}}{c_{s}^{2}}\left(\frac{1}{c_{s}^{2}}-1\right)+\frac{F\epsilon_{s}}{c_{s}^{4}}\left(\epsilon_{s}-\eta_{s}-4\delta_{G3X}-12\delta_{G4X}-32\delta_{G4XX}+12\delta_{G5\phi}-10\delta_{G5X}-8\delta_{G5XX}\right)+{\cal O}(\epsilon^{3}),\nonumber \\
\\
{\cal C}_{2} & = & F\epsilon_{s}\left(\frac{1}{c_{s}^{2}}-1\right)+\frac{F\epsilon_{s}}{c_{s}^{2}}\left(\epsilon_{s}+\eta_{s}-2s+4\delta_{G4X}+2\delta_{G5X}-4\delta_{G5\phi}\right)+{\cal O}(\epsilon^{3}),\\
{\cal C}_{3} & = & \frac{F\epsilon_{s}}{c_{s}^{2}}\frac{M_{{\rm pl}}}{H}\left(\frac{1}{c_{s}^{2}}-1\ -\frac{2\lambda}{\Sigma}\right)+\frac{F\epsilon_{s}}{c_{s}^{2}}\frac{M_{{\rm pl}}}{H}\biggl\{\frac{1}{c_{s}^{2}}(\delta_{G3X}+4\delta_{G4X}+3\delta_{G5X}-\delta_{G4\phi}-4\delta_{G5\phi}+8\delta_{G4XX}+2\delta_{G5XX})\nonumber \\
 &  & {}-\left(3+2\lambda_{3X}\right)\delta_{G3X}-8\left(5+2\lambda_{4X}\right)\delta_{G4XX}-4(4+\lambda_{5X})\delta_{G5XX}\nonumber \\
 &  & +\delta_{G4\phi}+8\delta_{G5\phi}-8\delta_{G4X}-9\delta_{G5X}-6\frac{c_{s}^{2}}{\epsilon_{s}}\left[(1+\lambda_{3X})\delta_{G3X}^{2}+\xi(\delta^{2})\right]\biggr\}+{\cal O}(\epsilon^{3}),\label{eq:fNL3}\\
{\cal C}_{4} & = & -\frac{2\epsilon_{s}}{c_{s}^{2}}+{\cal O}(\epsilon^{2})\,,\\
{\cal C}_{5} & = & \frac{1}{4F}\left(\epsilon_{s}-4\delta_{G3X}-8\delta_{G4XX}+8\delta_{G5X}+4\delta_{G5XX}\right)+{\cal O}(\epsilon^{2})\,,\\
{\cal C}_{6} & = & 2F\left(\frac{M_{{\rm pl}}}{H}\right)^{2}\left[(1+\lambda_{3X})\delta_{G3X}+4(3+2\lambda_{4X})\delta_{G4XX}+\delta_{G5X}+(5+2\lambda_{5X})\delta_{G5XX}\right]+{\cal O}(\epsilon^{2}),\\
{\cal C}_{7} & = & -\frac{2}{3}F\left(\frac{M_{{\rm pl}}}{H}\right)^{2}\left(\delta_{G3X}+6\delta_{G4XX}+\delta_{G5X}+\delta_{G5XX}\right)+{\cal O}(\epsilon^{2}),\\
{\cal C}_{8} & = & 2\frac{M_{{\rm pl}}}{H}\left(\delta_{G3X}+4\delta_{G4XX}\right)+{\cal O}(\epsilon^{2})\,,\\
{\cal F}_1 &=&  -\frac{L_{1}\mu_1+6X\dot{\phi}G_{5,X}}{6w_{1}^{2}}\{(\partial_{k}{\cal R})(\partial_{k}{\cal X})-\partial^{-2}\partial_{i}\partial_{j}[(\partial_{i}{\cal R})(\partial_{j}{\cal X})]\}-\frac{L_{1}}{c_{s}^{2}}{\cal R}\dot{{\cal R}}\nonumber \\
 &  & +\frac{L_{1}(L_{1}\mu_1+12X\dot{\phi}G_{5,X})}
 {12w_{1}a^{2}}\{(\partial{\cal R})^{2}-\partial^{-2}\partial_{i}\partial_{j}[(\partial_{i}{\cal R})(\partial_{j}{\cal R})]\}\,,
\end{eqnarray}
where $\lambda_{3X}=XG_{3,XX}/G_{3,X}$, $\lambda_{4X}=XG_{4,XXX}/G_{4,XX}$,
$\lambda_{5X}=XG_{5,XXX}/G_{5,XX}$,
$\mu_1=3M_{{\rm pl}}^{2}F-24(XG_{4,X}+X^{2}G_{4,XX})-6H\dot{\phi}(5XG_{5,X}+2X^{2}G_{5,XX})
+6X(3G_{5,\phi}+2XG_{5,\phi X})$, 
$L_1=2w_1/w_2$, and 
\begin{eqnarray}
\Sigma & = & \frac{w_{1}(4w_{1}w_{3}+9w_{2}^{2})}{12M_{{\rm pl}}^{4}}\,,\\
\lambda & = & \frac{F^{2}}{3}[3X^{2}P_{,XX}+2X^{3}P_{,XXX}+3H\dot{\phi}(XG_{3,X}+5X^{2}G_{3,XX}+2X^{3}G_{3,XXX})-2(2X^{2}G_{3,\phi X}+X^{3}G_{3,\phi XX})\nonumber \\
 &  & +6H^{2}(9X^{2}G_{4,XX}+16X^{3}G_{4,XXX}+4X^{4}G_{4,XXXX})-3H\dot{\phi}(3XG_{4\phi,X}+12X^{2}G_{4,\phi XX}+4X^{3}G_{4,\phi XXX})\nonumber \\
 &  & +H^{3}\dot{\phi}(3XG_{5,X}+27X^{2}G_{5,XX}+24X^{3}G_{5,XXX}+4X^{4}G_{5,XXXX})\nonumber \\
 &  & -6H^{2}(6X^{2}G_{5,\phi X}+9X^{3}G_{5,\phi XX}+2X^{4}G_{5,\phi XXX})]\,.
\end{eqnarray}
The explicit form of the second-order term $\xi (\delta^2)$ in Eq.~(\ref{eq:fNL3})
is given in Appendix of Ref.~\cite{ATnongau}.
The coefficient ${\cal F}_{1}$ involves the terms with the spatial and time 
derivatives of ${\cal R}$ and ${\cal X}$. These provide 
the corrections to the three-point correlation function higher than
first order in slow-variation parameters\footnote{Note that in Ref.~\cite{Maldacena}
the term ${\cal R}^2$ is present in the expression of ${\cal F}_1$, 
which gives rise to the first-order contribution $\eta_s$.
We absorb this term to other coefficients, so that the field definition 
in Ref.~\cite{Maldacena} is unnecessary.}. 
Since we are interested in the bispectrum up to first order, we neglect the contribution of the
term ${\cal F}_{1}(\delta{\cal L}_{2}/\delta{\cal R})|_{1}$ in the
following discussion. We also evaluated other boundary terms and found
that they only lead to the contribution higher than the order $\epsilon$.

The vacuum expectation value of ${\cal R}$ for the three-point operator
in the asymptotic future ($\tau \to 0$) is 
\begin{equation}
\langle{\cal R}({\bm{k}}_{1}){\cal R}({\bm{k}}_{2}){\cal R}({\bm{k}}_{3})
\rangle=-i\int_{-\infty}^{0}d\tau\, a\,\langle0|\,[{\cal R}(0,{\bm{k}}_{1})
{\cal R}(0,{\bm{k}}_{2}){\cal R}(0,{\bm{k}}_{3}),{\cal H}_{{\rm int}}(\tau)]\,|0\rangle\,.
\label{Rvacuum}
\end{equation}
The interacting Hamiltonian ${\cal H}_{{\rm int}}$ is related to 
the third-order Lagrangian ${\cal L}_3$ as
${\cal H}_{{\rm int}}=-{\cal L}_{3}$, where 
${\cal S}_{3}=\int dt\,{\cal L}_{3}$.
We write the three-point correlation function in the form 
\begin{equation}
\langle{\cal R}({\bm{k}}_{1}){\cal R}({\bm{k}}_{2}){\cal R}({\bm{k}}_{3})\rangle
=(2\pi)^{3}\delta^{(3)}({\bm{k}}_{1}+{\bm{k}}_{2}+{\bm{k}}_{3})({\cal P}_{{\cal R}})^{2}{\cal F}_{{\cal R}}(k_{1},k_{2},k_{3})\,,\label{bispe}
\end{equation}
where 
\begin{equation}
{\cal F}_{{\cal R}}(k_{1},k_{2},k_{3})=\frac{(2\pi)^{4}}{\prod_{i=1}^{3}k_{i}^{3}}
{\cal A}_{{\cal R}}(k_1,k_2,k_3)\,.\label{calF}
\end{equation}
If we use the leading-order solution (\ref{usol}) for the mode function
and neglect the variation of the terms ${\cal C}_{i}$'s for the integration
of Eq.~(\ref{Rvacuum}) with the approximation $a\simeq-1/(H\tau)$,
the resulting bispectrum is \cite{Gao,ATnongau} 
\begin{eqnarray}
{\cal A}_{{\cal R}} & \supset & \frac{c_{s}^{2}}{4\epsilon_{s}F}{\cal C}_{1}S_{1}
+\frac{1}{4\epsilon_{s}F}{\cal C}_{2}S_{2}+\frac{3c_{s}^{2}}{2\epsilon_{s}F}\frac{H}{M_{{\rm pl}}}{\cal C}_{3}S_{3}+\frac{1}{8}{\cal C}_{4}S_{4}+\frac{\epsilon_{s}F}{4c_{s}^{2}}{\cal C}_{5}S_{5}+\frac{3}{\epsilon_{s}F}\left(\frac{H}{M_{{\rm pl}}}\right)^{2}{\cal C}_{6}S_{6}\nonumber \\
 &  & +\frac{1}{2\epsilon_{s}Fc_{s}^{2}}\left(\frac{H}{M_{{\rm pl}}}\right)^{2}{\cal C}_{7}S_{7}+\frac{1}{8c_{s}^{2}}\frac{H}{M_{{\rm pl}}}{\cal C}_{8}S_{8}\,,\label{ARl}
\end{eqnarray}
where 
\begin{eqnarray}
 &  & S_{1}=\frac{2}{K}\sum_{i>j}k_{i}^{2}k_{j}^{2}-\frac{1}{K^{2}}\sum_{i\neq j}k_{i}^{2}k_{j}^{3},\qquad S_{2}=\frac{1}{2}\sum_{i}k_{i}^{3}+\frac{2}{K}\sum_{i>j}k_{i}^{2}k_{j}^{2}-\frac{1}{K^{2}}\sum_{i\neq j}k_{i}^{2}k_{j}^{3},\qquad S_{3}=\frac{(k_{1}k_{2}k_{3})^{2}}{K^{3}},\nonumber \\
 &  & S_{4}=\sum_{i}k_{i}^{3}-\frac{1}{2}\sum_{i\neq j}k_{i}k_{j}^{2}-\frac{2}{K^{2}}\sum_{i\neq j}k_{i}^{2}k_{j}^{3}\,,\qquad S_{5}=\frac{1}{K^{2}}\left[\sum_{i}k_{i}^{5}+\frac{1}{2}\sum_{i\neq j}k_{i}k_{j}^{4}-\frac{3}{2}\sum_{i\neq j}k_{i}^{2}k_{j}^{3}-k_{1}k_{2}k_{3}\sum_{i>j}k_{i}k_{j}\right],\nonumber \\
 &  & S_{6}=S_{3},\qquad S_{7}=\frac{1}{K}\left(1+\frac{1}{K^{2}}\,\sum_{i>j}k_{i}k_{j}+\frac{3k_{1}k_{2}k_{3}}{K^{3}}\right)\left[\frac{3}{4}\,\sum_{i}k_{i}^{4}-\frac{3}{2}\sum_{i>j}k_{i}^{2}k_{j}^{2}\right],\nonumber \\
 &  & S_{8}=\frac{1}{K^{2}}\left[\frac{3}{2}\, k_{1}k_{2}k_{3}\sum_{i}k_{i}^{2}-\frac{5}{2}\, k_{1}k_{2}k_{3}K^{2}-6\sum_{i\neq j}k_{i}^{2}k_{j}^{3}-\sum_{i}k_{i}^{5}+\frac{7}{2}\, K\sum_{i}k_{i}^{4}\right]\,,
\end{eqnarray}
and $K=k_{1}+k_{2}+k_{3}$. 
The five shape functions $S_{i}$'s ($i=1,\cdots,5$)
are present in the context of k-inflation \cite{Seery,Chen}. 
In the Horndeski's theories the additional
functions $S_{7}$ and $S_{8}$ appear, but they can be expressed
by using other shape functions as \cite{Petel}
\begin{equation}
S_{7}=-\frac{3}{2}(3S_{1}-S_{2})+18S_{3}\,,\qquad S_{8}=3S_{1}-S_{2}+3S_{4}\,.\label{S78}
\end{equation}
Since the three functions $S_{6}$, $S_{7}$, and $S_{8}$ vanish
in the limit $k_{3}\to0$, the last three terms in Eq.~(\ref{ARl})
do not contribute to the local non-Gaussianities.

The bispectrum ${\cal A}_{\cal R}$ coming from 
the contributions of ${\cal C}_{1}$, ${\cal C}_{2}$,
${\cal C}_{3}$, ${\cal C}_{6}$, ${\cal C}_{7}$ 
are 0-th order of $\epsilon$, while the bispectrum from 
${\cal C}_{4}$ and ${\cal C}_{8}$ are first order. Since
the term ${\cal C}_{5}$ leads to the bispectrum at the order
of $\epsilon^2$, we can neglect its contribution. In the case where the
leading-order terms of ${\cal A}_{{\cal R}}$ vanish (which occurs
for local non-Gaussianities), we need to take into account next-order
corrections to the bispectrum coming from the integrals that involve
the terms ${\cal C}_{1}$, ${\cal C}_{2}$, ${\cal C}_{3}$, ${\cal C}_{6}$,
${\cal C}_{7}$ in Eq.~(\ref{L3}). Using the linear equation of
motion $\delta{\cal L}_{2}/\delta{\cal R}|_{1}=0$, the ${\cal C}_{6}$,
${\cal C}_{7}$, and ${\cal C}_{8}$ dependent terms can be absorbed
into the first five terms in Eq.~(\ref{L3}) \cite{Petel}. Then
the third-order action (\ref{L3}) reads 
\begin{equation}
{\cal S}_{3}=\int dt\, d^{3}x\biggl\{ a^{3}\tilde{{\cal C}}_{1}\Mpl^{2}{\cal R}\dot{{\cal R}}^{2}+a\,\tilde{{\cal C}}_{2}\Mpl^{2}{\cal R}(\partial{\cal R})^{2}+a^{3}\tilde{{\cal C}}_{3}\Mpl\dot{{\cal R}}^{3}+a^{3}\tilde{{\cal C}}_{4}\dot{{\cal R}}(\partial_{i}{\cal R})(\partial_{i}{\cal X})+a^{3}(\tilde{{\cal C}}_{5}/\Mpl^{2})\partial^{2}{\cal R}(\partial{\cal X})^{2}\biggr\}\,.\label{L3d}
\end{equation}

The coefficients $\tilde{{\cal C}}_{i}$ ($i=1,\cdots,5$), which
give rise to the corrections up to the order of $\epsilon$ in ${\cal A}_{{\cal R}}$,
are \cite{Petel} 
\begin{eqnarray}
\tilde{{\cal C}}_{1} & = & {\cal C}_{1}-\frac{3H^{2}}{2c_{s}^{4}M_{{\rm pl}}^{2}}(6+2\epsilon+7\eta_{sF}-5\eta_{7}){\cal C}_{7}+\frac{3H\epsilon_{s}F}{2c_{s}^{4}M_{{\rm pl}}}{\cal C}_{8}\,,\label{C1trans}\\
\tilde{{\cal C}}_{2} & = & {\cal C}_{2}+\frac{3H^{2}}{2c_{s}^{2}M_{{\rm pl}}^{2}}(2-2\epsilon+\eta_{sF}+\eta_{7}-4s){\cal C}_{7}-\frac{H\epsilon_{s}F}{2c_{s}^{2}M_{{\rm pl}}}{\cal C}_{8}\,,\\
\tilde{{\cal C}}_{3} & = & {\cal C}_{3}+\frac{H}{3c_{s}^{2}M_{{\rm pl}}}(6+3\eta_{sF}-4s-\eta_{6}){\cal C}_{6}+\frac{H}{c_{s}^{4}M_{{\rm pl}}}(6+3\eta_{sF}-s-2\eta_{7}){\cal C}_{7}\,,\\
\tilde{{\cal C}}_{4} & = & {\cal C}_{4}+\frac{3H}{c_{s}^{2}M_{{\rm pl}}}{\cal C}_{8}\,,\label{C4trans}\\
\tilde{{\cal C}}_{5} & = & {\cal C}_{5}\,,
\end{eqnarray}
where $\eta_{i}\equiv\dot{{\cal C}}_{i}/(H{\cal C}_{i})$, with $i=6,7,8$.
The three-point correlation function analogous to (\ref{ARl}) is given by 
\begin{eqnarray}
{\cal A}_{{\cal R}} & \supset & \frac{c_{s}^{2}}{4\epsilon_{s}F}\tilde{{\cal C}}_{1}S_{1}+\frac{1}{4\epsilon_{s}F}\tilde{{\cal C}}_{2}S_{2}+\frac{3c_{s}^{2}}{2\epsilon_{s}F}\frac{H}{M_{{\rm pl}}}\tilde{{\cal C}}_{3}S_{3}+\frac{1}{8}\tilde{{\cal C}}_{4}S_{4}\,,\label{ARl2}
\end{eqnarray}
where we dropped the $\tilde{{\cal C}}_{5}$-dependent term. 
The difference between Eqs.~(\ref{ARl})
and (\ref{ARl2}) is that the bispectrum (\ref{ARl2}) includes the
corrections coming from the time-variations of ${\cal C}_{6}$ 
and ${\cal C}_{7}$. However Eqs.~(\ref{ARl})
and (\ref{ARl2}) are equivalent at leading order.
The terms $\tilde{{\cal C}}_{i}$ ($i=1,\cdots,4$) can be expressed as
\begin{eqnarray}
\tilde{{\cal C}}_{1} & = & \tilde{{\cal C}}_{1}^{{\rm lead}}+\frac{F}{c_{s}^{4}}\left[\epsilon_{s}\delta{\cal C}_{1}+(2\epsilon+7\eta_{sF}-5\eta_{7})\delta{\cal C}_{7}+3\epsilon_{s}\delta{\cal C}_{8}\right]\,,\\
\tilde{{\cal C}}_{2} & = & \tilde{{\cal C}}_{2}^{{\rm lead}}+\frac{F}{c_{s}^{2}}\left[\epsilon_{s}\delta{\cal C}_{2}+(2\epsilon-\eta_{sF}-\eta_{7}+4s)\delta{\cal C}_{7}-\epsilon_{s}\delta{\cal C}_{8}\right]\,,\\
\tilde{{\cal C}}_{3} & = & \tilde{{\cal C}}_{3}^{{\rm lead}}+\frac{FM_{{\rm pl}}}{c_{s}^{2}H}\left[\epsilon_{s}\delta{\cal C}_{3}+\frac{2}{3}\left(3\eta_{sF}-4s-\eta_{6}\right)\delta{\cal C}_{6}-\frac{2}{3c_{s}^{2}}(3\eta_{sF}-s-2\eta_{7})\delta{\cal C}_{7}\right]\,,\\
\tilde{{\cal C}}_{4} & = & -\frac{2\epsilon_{s}}{c_{s}^{2}}+\frac{6}{c_{s}^{2}}\delta{\cal C}_{8}\,,
\end{eqnarray}
where the leading-order terms are 
\begin{eqnarray}
\tilde{{\cal C}}_{1}^{{\rm lead}} & = & -\frac{3F}{c_{s}^{2}}\left(\frac{1}{c_{s}^{2}}-1\right)\epsilon_{s}+\frac{6F}{c_{s}^{4}}\delta{\cal C}_{7}\,,
\label{C1lead}\\
\tilde{{\cal C}}_{2}^{{\rm lead}} & = & F\left(\frac{1}{c_{s}^{2}}-1\right)\epsilon_{s}-\frac{2F}{c_{s}^{2}}\delta{\cal C}_{7}=-\frac{c_{s}^{2}}{3}\tilde{{\cal C}}_{1}^{{\rm lead}}\,,\label{C2lead}\\
\tilde{{\cal C}}_{3}^{{\rm lead}} & = & \frac{FM_{{\rm pl}}}{c_{s}^{2}H}\left[\left(\frac{1}{c_{s}^{2}}-1-\frac{2\lambda}{\Sigma}\right)\epsilon_{s}+4\delta{\cal C}_{6}-\frac{4}{c_{s}^{2}}\delta{\cal C}_{7}\right]\,,\label{C3lead}
\end{eqnarray}
and $\delta{\cal C}_{i}$'s are the first-order slow-variation terms
given by 
\begin{eqnarray}
\delta{\cal C}_{1} & = & \epsilon_{s}-\eta_{s}-4\delta_{G3X}-12\delta_{G4X}-32\delta_{G4XX}
+12\delta_{G5\phi}-10\delta_{G5X}-8\delta_{G5XX}\,,\label{delC1} \\
\delta{\cal C}_{2} & = & \epsilon_{s}+\eta_{s}-2s+4\delta_{G4X}+2\delta_{G5X}-4\delta_{G5\phi}\,,
\label{delC2} \\
\delta{\cal C}_{3} & = & \frac{1}{c_{s}^{2}}(\delta_{G3X}+4\delta_{G4X}+3\delta_{G5X}
-\delta_{G4\phi}-4\delta_{G5\phi}+8\delta_{G4XX}+2\delta_{G5XX})
-\left(3+2\lambda_{3X}\right)\delta_{G3X}-8\left(5+2\lambda_{4X}\right)\delta_{G4XX}\nonumber \\
&  & {}-4(4+\lambda_{5X})\delta_{G5XX}+\delta_{G4\phi}+8\delta_{G5\phi}
-8\delta_{G4X}-9\delta_{G5X}-6\frac{c_{s}^{2}}{\epsilon_{s}}\left[(1+\lambda_{3X})
\delta_{G3X}^{2}+\xi(\delta^{2})\right]\,,\\
\delta{\cal C}_{6} & = & (1+\lambda_{3X})\delta_{G3X}+4(3+2\lambda_{4X})
\delta_{G4XX}+\delta_{G5X}+(5+2\lambda_{5X})\delta_{G5XX}\,,\\
\delta{\cal C}_{7} & = & \delta_{G3X}+6\delta_{G4XX}+\delta_{G5X}+\delta_{G5XX}\,,\\
\delta{\cal C}_{8} & = & \delta_{G3X}+4\delta_{G4XX}\,.
\end{eqnarray}
In order to derive the full expression of ${\cal A}_{{\cal R}}$ to
the order of $\epsilon$, we need to compute the corrections to the
first three integrations in Eq.~(\ref{L3d}). 
As studied in Ref.~\cite{Chen} in the context of k-inflation, 
there are several corrections to the bispectrum (\ref{ARl2}).

The first one comes from the variation of the coefficients $\tilde{{\cal C}}_{i}$
($i=1,2,3$), i.e., 
\begin{equation}
\tilde{{\cal C}}_{i}(\tau)=\tilde{{\cal C}}_{i}(\tau_{K})-\frac{d\tilde{{\cal C}}_{i}}{dt}
\frac{1}{H_{K}}\ln\frac{\tau}{\tau_{K}}+{\cal O}(\epsilon^{2}\tilde{{\cal C}}_{i})\,.
\label{Cvari}
\end{equation}
We evaluate all the physical variables at the time $\tau_{K}=-1/(Kc_{sK})$,
which corresponds to the moment when the wave number $K=k_1+k_2+k_3$ 
crosses the Hubble radius $Kc_{sK}=a_{K}H_{K}$.

The second one follows from the correction to the scale factor 
$a\simeq-1/(H\tau)$, i.e., 
\begin{equation}
a=-\frac{1}{H_{K}\tau}-\frac{\epsilon}{H_{K}\tau}+\frac{\epsilon}{H_{K}\tau}
\ln(\tau/\tau_{K})+{\cal O}(\epsilon^{2})\,.
\label{avari}
\end{equation}
Thirdly, the mode function (\ref{usol}) is subject to change by taking
into account the ${\cal O}(\epsilon)$ terms on the r.h.s. of Eq.~(\ref{zdd}) :
\begin{equation}
u(y)=-\frac{\sqrt{\pi}}{2\sqrt{2}}\frac{1}{\sqrt{\epsilon_{s}Fc_{s}}}\frac{H}
{M_{{\rm pl}}}\frac{y^{3/2}}{k^{3/2}}\left(1+\frac{1}{2}\epsilon
+\frac{1}{2}s\right)e^{i\frac{\pi}{2}(\epsilon+\frac{1}{2}\eta_{sF})}H_{\nu}^{(1)}[(1+\epsilon+s)y]\,,
\end{equation}
where $y= c_{s}k/(aH)$, $\nu=3/2+\epsilon+\eta_{sF}/2+s/2$,
and $H_{\nu}^{(1)}(x)$ is the Hankel function of the first kind.
In the large-scale limit ($y\to0$) the Hankel function behaves as
$H_{\nu}^{(1)}(x)\to-i/[\sin(\pi\nu)\Gamma(1-\nu)](x/2)^{-\nu}$ and
hence the mode function approaches 
\begin{equation}
u(0)=\frac{i}{2k^{3/2}}\frac{1}{\sqrt{(\epsilon_{s}F)_{k}\, c_{sk}}}\frac{H_{k}}{M_{{\rm pl}}}\frac{1}{k^{3/2}}\left[1-(\gamma_{2}+1)\epsilon-\frac{\gamma_{2}}{2}\eta_{sF}-\left(\frac{\gamma_{2}}{2}+1\right)s\right]e^{i\frac{\pi}{2}(\epsilon+\frac{1}{2}\eta_{sF})}\,,\label{u0}
\end{equation}
where $\gamma_{2}=\gamma_{1}-2+\ln2\simeq-0.7296...$ and 
$\gamma_{1}=0.5772...$
is the Euler-Mascheroni constant. Note that the quantities with the subscript
$k$ in Eq.~(\ref{u0}) are evaluated at $c_{sk}k/(a_{k}H_{k})=1$.
For the wave number $k_{i}$ there is the running from
$k_{i}$ to $K$, as 
\begin{equation}
\frac{1}{\sqrt{(\epsilon_{s}F)_{k_{i}}\, c_{sk_{i}}}}\frac{H_{k_{i}}}{M_{{\rm pl}}}=\frac{1}{\sqrt{(\epsilon_{s}F)_{K}\, c_{sK}}}\frac{H_{K}}{M_{{\rm pl}}}\left[1-\left(\epsilon+\frac{1}{2}\eta_{sF}+\frac{1}{2}s\right)\ln\frac{k_{i}}{K}\right]+{\cal O}(\epsilon^{2})\,.
\end{equation}
Writing the correction to the leading order solution (\ref{usol})
as $\Delta u^{*}(\tau,k)$, it follows that 
\begin{eqnarray}
\hspace{-0.5cm}\Delta u^{*}(\tau,k_{i}) & = & -\frac{1}{2k_{i}^{3/2}\sqrt{(\epsilon_{s}F)_{K}\, c_{sK}}}\frac{H_{K}}{M_{{\rm pl}}}e^{-i\frac{\pi}{2}(\epsilon+\frac{1}{2}\eta_{sF})}e^{-ix}\biggl[(\epsilon+s)(x-i)+isx^{2}\nonumber \\
\hspace{-0.5cm} &  & +\left\{ \left(\epsilon+\frac{1}{2}\eta_{sF}+\frac{1}{2}s\right)(i-x)-isx^{2}\right\} \ln\frac{\tau}{\tau_{K}}+\sqrt{\frac{\pi}{2}}e^{ix}\left(\epsilon+\frac{1}{2}\eta_{sF}+\frac{1}{2}s\right)x^{3/2}
 \left( \frac{dH_{\nu}^{(1)*}}{d\nu} \right)_{\nu=3/2}\biggr],\label{usvari}\\
\hspace{-0.5cm}\frac{d}{d\tau}\Delta u^{*}(\tau,k_{i}) & = & \frac{1}{2k_{i}^{3/2}\sqrt{(\epsilon_{s}F)_{K}\, c_{sK}}}\frac{H_{K}}{M_{{\rm pl}}}e^{-i\frac{\pi}{2}(\epsilon+\frac{1}{2}\eta_{sF})}k_{i}c_{sK}\, e^{-ix}\biggl[x(sx-i\epsilon)+\left(\epsilon+\frac{1}{2}\eta_{sF}+\frac{1}{2}s\right)\left(\frac{i}{x}-1\right)\nonumber \\
\hspace{-0.5cm} &  & +i\left(\epsilon+\frac{1}{2}\eta_{sF}-\frac{3}{2}s+isx\right)x\ln\frac{\tau}{\tau_{K}}+\sqrt{\frac{\pi}{2}}e^{ix}\left(\epsilon+\frac{1}{2}\eta_{sF}+\frac{1}{2}s\right)\frac{d}{dx}\left(x^{3/2}
 \biggl( \frac{dH_{\nu}^{(1)*}}{d\nu} \biggr)_{\nu=3/2} \right)\biggr]\,,
 \label{us2vari}
\end{eqnarray}
where $x\equiv-k_{i}c_{sK}\,\tau$.

In Appendix A we give the explicit forms of corrections to the first
three terms in Eq.~(\ref{ARl2}). 
Each correction can be expressed as 
\begin{equation}
\Delta{\cal A}_{{\cal R}}^{(1)}=\left(\frac{c_{s}^{2}}{4\epsilon_{s}F}\tilde{{\cal C}}_{1}^{{\rm lead}}\right)_{K}\delta Q_{1}\,,\qquad\Delta{\cal A}_{{\cal R}}^{(2)}=-\left(\frac{1}{4\epsilon_{s}F}\tilde{{\cal C}}_{2}^{{\rm lead}}\right)_{K}\delta Q_{2}\,,\qquad\Delta{\cal A}_{{\cal R}}^{(3)}=\biggl(\frac{3c_{s}^{2}H}{4\epsilon_{s}FM_{{\rm pl}}}\tilde{{\cal C}}_{3}^{{\rm lead}}\biggr)_{K}\delta Q_{3}\,,
\end{equation}
where $\delta Q_{1}$, $\delta Q_{2}$, and $\delta Q_{3}$ are the
${\cal O}(\epsilon)$ terms derived by summing up the contributions
(\ref{C1a})-(\ref{C1e}), (\ref{C2a})-(\ref{C2d}), and (\ref{C3a})-(\ref{C3d}),
respectively. On using Eqs.~(\ref{C1lead})-(\ref{C3lead}), it follows
that 
\begin{eqnarray}
\Delta{\cal A}_{{\cal R}} & = & \Delta{\cal A}_{{\cal R}}^{(1)}+\Delta{\cal A}_{{\cal R}}^{(2)}+\Delta{\cal A}_{{\cal R}}^{(3)}\nonumber \\
 & = & -\frac{1}{4}\left(\frac{1}{c_{s}^{2}}-1-\frac{2}{c_{s}^{2}}\frac{\delta{\cal C}_{7}}{\epsilon_{s}}\right)(3\delta Q_{1}+\delta Q_{2})+\frac{3}{4}\left( \frac{1}{c_{s}^{2}}-1-\frac{2\lambda}{\Sigma}
+4\frac{\delta{\cal C}_{6}}{\epsilon_{s}}-\frac{4}{c_{s}^{2}}\frac{\delta{\cal C}_{7}}{\epsilon_{s}}
\right) \delta Q_{3}\,.\label{ARco}
\end{eqnarray}
The explicit forms of $3\delta Q_{1}+\delta Q_{2}$ and $\delta Q_{3}$
are 
\begin{eqnarray}
 &  & 3\delta Q_{1}+\delta Q_{2}\nonumber \\
 &  & =\biggl[-2(7+2\gamma_{1}+6\gamma_{2})\epsilon+3(1-2\gamma_{1}-2\gamma_{2})\eta_{sF}-(21-14\gamma_{1}+6\gamma_{2})s-2(1-2\gamma_{1})\tilde{\eta}_{1}-2(2\epsilon+\eta_{sF}+s)\ln\frac{k_{1}k_{2}k_{3}}{K^{3}}\biggr]\nonumber \\
 &  & ~~~~\times\frac{1}{K}\sum_{i>j}k_{i}^{2}k_{j}^{2}+\biggl[2(3+\gamma_{1}+3\gamma_{2})\epsilon-3(1-\gamma_{1}-\gamma_{2})\eta_{sF}+\biggl(17-\frac{35}{2}\gamma_{1}+3\gamma_{2}\biggr)s+2(1-\gamma_{1})\tilde{\eta}_{1}\nonumber \\
 &  & ~~~~+(2\epsilon+\eta_{sF}+s)\ln\frac{k_{1}k_{2}k_{3}}{K^{3}}\biggr]\frac{1}{K^{2}}\sum_{i\neq j}k_{i}^{2}k_{j}^{3}+3(2\gamma_{1}-1)s\biggl[\frac{1}{K^{3}}\sum_{i\neq j}k_{i}^{2}k_{j}^{4}+\frac{2}{K^{3}}\sum_{i>j}k_{i}^{3}k_{j}^{3}-3\frac{(k_{1}k_{2}k_{3})^{2}}{K^{3}}\biggr]\nonumber \\
 &  & ~~~~ -\frac14 \biggl[ 22 \epsilon+9 \eta_{sF}+2\tilde{\eta}_1+(11+4\gamma_1)s \biggr]k_{1}k_{2}k_{3}
 -\biggl(\frac{1}{2}\epsilon+\frac{3}{4}\eta_{sF}-\frac{1}{2}\tilde{\eta}_{1}-\frac{1}{4}s\biggr)\sum_{i\neq j}k_{i}k_{j}^{2}\nonumber \\
 &  & ~~~~+\biggl[\frac{1}{2}(3+\gamma_{1}+3\gamma_{2})\epsilon-\frac{3}{4}(1-\gamma_{1}-\gamma_{2})\eta_{sF}+\frac{1}{2}(1-\gamma_{1})\tilde{\eta}_{1}+\frac{1}{4}(13-\gamma_{1}+3\gamma_{2})s+\frac{1}{4}(2\epsilon+\eta_{sF}+s)\ln\frac{k_{1}k_{2}k_{3}}{K^{3}}\biggr]\sum_{i}k_{i}^{3}\nonumber \\
 &  & ~~~~-\frac{1}{2}(1+\gamma_{1})s\frac{1}{K}\sum_{i}k_{i}^{4}-\frac{1}{2}\gamma_{1}s\frac{1}{K^{2}}\sum_{i\neq j}k_{i}k_{j}^{4}+\frac{3}{2}(2\epsilon+\eta_{sF}+s)\biggl(2\sum_{i}k_{i}^{3}+2\sum_{i\neq j}k_{i}k_{j}^{2}-{\cal V}\biggr)\,,\label{delQ12}\\
 &  & \delta Q_{3}\nonumber \\
 &  & =\biggl[-2(2+2\gamma_{1}+3\gamma_{2})\epsilon+3\left(\frac{3}{2}-\gamma_{1}-\gamma_{2}\right)\eta_{sF}+\left(3\gamma_{1}-3\gamma_{2}-\frac{29}{2}\right)s+(2\gamma_{1}-3)\tilde{\eta}_{3}-(2\epsilon+\eta_{sF}+s)\ln\frac{k_{1}k_{2}k_{3}}{K^{3}}\biggr]\nonumber \\
 &  & ~~~~\times\frac{(k_{1}k_{2}k_{3})^{2}}{K^{3}}+\left(\epsilon+\frac{1}{2}\eta_{sF}+\frac{1}{2}s\right)\biggl(\frac{1}{K^{2}}\sum_{i\neq j}k_{i}^{2}k_{j}^{3}-\frac{2}{K}\sum_{i>j}k_{i}^{2}k_{j}^{2}+{\cal U}\biggr)\,,\label{delQ3}
\end{eqnarray}
where $\tilde{\eta}_{i}=(d\tilde{{\cal C}}_{i}^{{\rm lead}}/dt)/(H{\cal \tilde{C}}_{i}^{{\rm lead}})$, 
and
\begin{eqnarray}
{\cal V} & \equiv & {\cal M}-(k_{2}^{3}+k_{3}^{3}){\rm Re}\biggl[\int_{0}^{\infty}dx_{1}\frac{e^{-iKx_{1}/k_1}}{x_{1}}\biggr]-\frac{k_{2}^{2}k_{3}^{2}}{k_{1}}{\cal G}-\frac{1}{3}{\cal N}+{\rm perm.}\nonumber \\
 & = & -k_{1}{\rm Re}\biggl[\int_{0}^{\infty}dx_{1}\,\frac{1}{x_{1}}\left(k_{2}^{2}+k_{3}^{2}+ik_{2}k_{3}\frac{k_{2}+k_{3}}{k_{1}}x_{1}\right)e^{-i\frac{k_{2}+k_{3}}{k_{1}}x_{1}}\frac{dh^{*}(x_{1})}{dx_{1}}\biggr]-(k_{2}^{3}+k_{3}^{3}){\rm Re}\biggl[\int_{0}^{\infty}dx_{1}\frac{e^{-iKx_{1}/k_1}}{x_{1}}\biggr]\nonumber \\
 &  & +\frac{k_{1}}{6}\sum_{i}k_{i}^{2}\,{\rm Re}\biggl[\int_{0}^{\infty}dx_{1}\,\frac{1}{x_{1}^{2}}e^{-ix_{1}\frac{k_{2}+k_{3}}{k_{1}}}\left(1+i\frac{k_{2}+k_{3}}{k_{1}}x_{1}-\frac{k_{2}k_{3}}{k_{1}^{2}}x_{1}^{2}\right)h^{*}(x_{1})\biggr]\nonumber \\
 &  & -\frac{k_{2}^{2}k_{3}^{2}}{k_{1}}{\rm Re}\left[\int_{0}^{\infty}dx_{1}\, h^{*}(x_{1})e^{-i\frac{k_{2}+k_{3}}{k_{1}}x_{1}}\right]+{\rm perm.}\,,\label{calV}\\
{\cal U} & \equiv & \frac{k_{2}^{2}k_{3}^{2}}{k_{1}}\,{\rm Re}\left[\int_{0}^{\infty}dx_{1}\, h^{*}(x_{1})\left(1-i\frac{k_{2}+k_{3}}{k_{1}}x_{1}\right)e^{-i\frac{k_{2}+k_{3}}{k_{1}}x_{1}}\right]+{\rm perm.}\,,\label{calU}\\
h(x) & \equiv & \sqrt{\frac{\pi}{2}}\, x^{3/2}\left[\frac{dH_{\nu}^{(1)}(x)}{d\nu}\right]_{\nu=3/2}=-2ie^{ix}+ie^{-ix}(1+ix)\left[{\rm Ci}(2x)+i\,{\rm Si}(2x)\right]-i\pi\sin x+i\pi x\cos x\,.
\end{eqnarray}
The definition of ${\cal M}$, ${\cal G}$, and ${\cal N}$ is given
in Appendix A.
In Eq.~(\ref{calV}), we have used the relation between the variables $x_{K}=-Kc_{sK}\tau$ and $x_{i}=-k_{i}c_{sK}\tau$
($i=1,2,3$), as $x_{K}=(K/k_{i})x_{i}$ (without summation over $i$).
 The variable $x_{K}=-Kc_{sK}\tau$ is related to $x_{i}=-k_{i}c_{sK}\tau$
($i=1,2,3$), as $x_{K}=(K/k_{i})x_{i}$. The symbol ``perm.'' stands
for cyclic permutations with respect to $k_{1}$, $k_{2}$, and $k_{3}$.
In Eq.~(\ref{delQ12}) we also used the relation $\tilde{\eta}_{2}=\tilde{\eta}_{1}+2s$
to eliminate $\tilde{\eta}_{2}$ (which follows from ${\cal \tilde{C}}_{2}^{{\rm lead}}
=-(c_{s}^{2}/3){\cal \tilde{C}}_{1}^{{\rm lead}}$).
In Appendix B we evaluate the values of ${\cal V}$ and ${\cal U}$
as functions of $r_2 \equiv k_2/k_1$ and $r_3 \equiv k_3/k_1$.

The total bispectrum ${\cal A}_{{\cal R}}$ is the sum of Eqs.~(\ref{ARl2})
and (\ref{ARco}), which can be written as 
\begin{equation}
{\cal A}_{{\cal R}}={\cal A}_{{\cal R}}^{{\rm lead}}+{\cal A}_{{\cal R}}^{{\rm corre}}\,,
\label{calAfi}
\end{equation}
where 
\begin{eqnarray}
{\cal A}_{{\cal R}}^{{\rm lead}} & = & \left[\frac{1}{4}\left(1-\frac{1}{c_{s}^{2}}\right)+\frac{1}{2c_{s}^{2}}\frac{\delta{\cal C}_{7}}{\epsilon_{s}}\right](3S_{1}-S_{2})+\left[\frac{3}{2}\left(\frac{1}{c_{s}^{2}}-1\right)-\frac{3\lambda}{\Sigma}+\frac{6\delta{\cal C}_{6}}{\epsilon_{s}}-\frac{6}{c_{s}^{2}}\frac{\delta{\cal C}_{7}}{\epsilon_{s}}\right]S_{3}\,,\label{Ale}\\
{\cal A}_{{\cal R}}^{{\rm corre}} & = & \frac{1}{4c_{s}^{2}}\left[\delta{\cal C}_{1}+(2\epsilon+7\eta_{sF}-5\eta_{7})\frac{\delta{\cal C}_{7}}{\epsilon_{s}}+3\delta{\cal C}_{8}\right]S_{1}+\frac{1}{4c_{s}^{2}}\left[\delta{\cal C}_{2}+(2\epsilon-\eta_{sF}-\eta_{7}+4s)\frac{\delta{\cal C}_{7}}{\epsilon_{s}}-\delta{\cal C}_{8}\right]S_{2}\nonumber \\
 &  & +\left[\frac{3}{2}\delta{\cal C}_{3}+(3\eta_{sF}-\eta_{6}-4s)\frac{\delta{\cal C}_{6}}{\epsilon_{s}}-\frac{1}{c_{s}^{2}}(3\eta_{sF}-2\eta_{7}-s)\frac{\delta{\cal C}_{7}}{\epsilon_{s}}\right]S_{3}-\frac{1}{4c_{s}^{2}}(\epsilon_{s}-3\delta{\cal C}_{8})S_{4}\nonumber \\
 &  & -\frac{1}{4}\left(\frac{1}{c_{s}^{2}}-1-\frac{2}{c_{s}^{2}}\frac{\delta{\cal C}_{7}}{\epsilon_{s}}\right)(3\delta Q_{1}+\delta Q_{2})
+\frac{3}{4}\left( \frac{1}{c_{s}^{2}}-1-\frac{2\lambda}{\Sigma}
+4\frac{\delta{\cal C}_{6}}{\epsilon_{s}}-\frac{4}{c_{s}^{2}}
 \frac{\delta{\cal C}_{7}}{\epsilon_{s}}\right) \delta Q_{3}\,.\label{Aco}
\end{eqnarray}
The leading-order bispectrum ${\cal A}_{{\cal R}}^{{\rm lead}}$
(given already in Refs.~\cite{Gao,ATnongau}) and the
correction ${\cal A}_{{\cal R}}^{{\rm corre}}$ are of the orders
of ${\cal O}(\epsilon^{0})$ and ${\cal O}(\epsilon)$, respectively.

\section{Local, equilateral, and enfolded non-Gaussianities}
\label{leesec} 

The non-linear parameter characterizing the strength of non-Gaussianities
is defined by 
\begin{equation}
f_{{\rm NL}}=\frac{10}{3}\frac{{\cal A}_{{\cal R}}}{\sum_{i=1}^{3}k_{i}^{3}}\,.
\end{equation}
In the following we estimate $f_{{\rm NL}}$ for three different shapes
of non-Gaussianities.

\subsection{Local non-Gaussianities}

The local shape corresponds to $k_{3} \to 0$ and $k_{2} \to k_{1} \equiv k$, 
in which case $f_{{\rm NL}}^{{\rm local}}=(5/3){\cal A}_{{\cal R}}/k^{3}$.
Since $S_{1}=k^{3}/2$, $S_{2}=3k^{3}/2=3S_{1}$, and $S_{3}=S_{4}=0$,
the leading-order bispectrum (\ref{Ale}) vanishes. In the limit that
$k_{3}\to0$ the function ${\cal U}$ given by Eq.~(\ref{calU})
approaches $k^{3}/2$ \cite{Chen} (see also Appendix B), so that the term $\delta Q_{3}$
in Eq.~(\ref{delQ3}) vanishes. Then the bispectrum (\ref{Aco})
reduces 
\begin{equation}
{\cal A}_{{\cal R}}^{{\rm corre}}=\frac{k^{3}}{8c_{s}^{2}}(\delta{\cal C}_{1}+3\delta{\cal C}_{2})-\frac{1}{4}\left(\frac{1}{c_{s}^{2}}-1\right)(3\delta Q_{1}+\delta Q_{2})+\frac{\delta{\cal C}_{7}}{8c_{s}^{2}\epsilon_{s}}\left[4(2\epsilon+\eta_{sF}-2\eta_{7}+3s)k^{3}+4(3\delta Q_{1}+\delta Q_{2})\right]\,.\label{ARlo}
\end{equation}
Using Eqs.~(\ref{delC1}) and (\ref{delC2}) together with the
relations (\ref{delF}), (\ref{eps}), and (\ref{etasF}), we have
$\delta{\cal C}_{1}+3\delta{\cal C}_{2}=4\epsilon+2\eta_{sF}-6s$.
In the limit $k_{3}\to0$ the function ${\cal V}$ behaves as ${\cal V}\to20k^{3}/3$ \cite{Chen} (see also Appendix B),
so that Eq.~(\ref{delQ12}) reduces to $3\delta Q_{1}+\delta Q_{2}\to(2\epsilon-\eta_{sF}+5s+2\tilde{\eta}_{1})k^{3}$.
Then Eq.~(\ref{ARlo}) reads 
\begin{equation}
{\cal A}_{{\cal R}}^{{\rm corre}}=\frac{k^{3}}{4}(2\epsilon-\eta_{sF}+5s+2\tilde{\eta}_{1})+\frac{k^{3}}{2c_{s}^{2}}(\eta_{sF}-4s-\tilde{\eta}_{1})+k^{3}\frac{\delta{\cal C}_{7}}{c_{s}^{2}\epsilon_{s}}(2\epsilon+4s+\tilde{\eta}_{1}-\eta_{7})\,.\label{ARlo2}
\end{equation}
Taking the time-derivatives of $\tilde{{\cal C}}_{1}^{{\rm lead}}$
and ${\cal C}_{7}$, we obtain the following relation 
\begin{equation}
\frac{\delta{\cal C}_{7}}{c_{s}^{2}\epsilon_{s}}(2\epsilon+4s+\tilde{\eta}_{1}-\eta_{7})=\frac{1}{2}(\eta_{sF}-\tilde{\eta}_{1}-2s)-\frac{1}{2c_{s}^{2}}(\eta_{sF}-4s-\tilde{\eta}_{1})\,.\label{delC71}
\end{equation}
Substituting Eq.~(\ref{delC71}) into Eq.~(\ref{ARlo2}), it follows
that 
\begin{equation}
{\cal A}_{{\cal R}}^{{\rm corre}}=\frac{k^{3}}{4}(2\epsilon+\eta_{sF}+s)\,.\label{ARlo3}
\end{equation}
Using the spectral index $n_{{\cal R}}$ given in Eq.~(\ref{nR}),
the non-linear parameter is expressed as 
\begin{equation}
f_{{\rm NL}}^{{\rm local}}=\frac{5}{12}(1-n_{{\cal R}})\,.\label{consistency}
\end{equation}
This matches with the Maldacena's result \cite{Maldacena} derived
for a canonical scalar field (see also Refs.~\cite{Chen,Cheung,Watanabe} for 
the derivation of the same relation in other single field models). 
Creminelli and Zaldarriaga \cite{Cremi}
pointed out that the consistency relation (\ref{consistency}) should
hold for any slow-variation single-field inflation.
In fact we have shown that this holds for most general single-field
scalar-tensor theories with second-order equations of motion by explicitly
computing the slow-variation corrections to the bispectrum (\ref{ARl2}).
Since $|f_{{\rm NL}}^{{\rm local}}|$ is much smaller than 1 in such 
models, the observational detection of local non-Gaussianities
with $|f_{{\rm NL}}^{{\rm local}}|\gtrsim1$ implies that we need
to go beyond the slow-variation single-field scenario.

In the limit that $k_{3} \to 0$ the shape functions $S_{6}$, $S_{7}$, and $S_{8}$
vanish. This means that the functions
${\cal C}_{6}$, ${\cal C}_{7}$, ${\cal C}_{8}$ on the r.h.s. of  
Eqs.~(\ref{C1trans})-(\ref{C4trans})
do not contribute to the local non-Gaussianities. In fact, we can
derive the consistency relation (\ref{consistency}) by setting $\delta{\cal C}_{7}=0$
in Eqs.~(\ref{ARlo}) and (\ref{delC71}). In this sense the situation
is analogous to that in k-inflation.

Let us consider the not-so squeezed case in which the ratio $r_{3}=k_{3}/k_{1}$
is non-vanishing, i.e., $0<r_{3}\ll1$ and $k_{1}=k_{2}$. The leading-order
non-linear parameter following from Eq.~(\ref{Ale}) is given by
\begin{equation}
f_{{\rm NL}}^{{\rm lead}}=\frac{5r_{3}^{2}\left[\alpha_{1}(22+4r_{3}
-4r_{3}^{2}-r_{3}^{3})(2+r_{3})+2\alpha_{2}\right]}{3(2+r_{3})^{3}(2+r_{3}^{3})}\,,\label{flead}
\end{equation}
where 
\begin{equation}
\alpha_{1}=\frac{1}{4}\left(1-\frac{1}{c_{s}^{2}}\right)+\frac{1}{2c_{s}^{2}}\frac{\delta{\cal C}_{7}}{\epsilon_{s}}\,,\qquad\alpha_{2}=\frac{3}{2}\left(\frac{1}{c_{s}^{2}}-1\right)-\frac{3\lambda}{\Sigma}+\frac{6\delta{\cal C}_{6}}{\epsilon_{s}}-\frac{6}{c_{s}^{2}}\frac{\delta{\cal C}_{7}}{\epsilon_{s}}\,.
\end{equation}
In the regime $r_{3}\ll1$ we have $f_{{\rm NL}}^{{\rm lead}}\simeq5(22\alpha_{1}+\alpha_{2})r_{3}^{2}/24$.
The values of $\alpha_{1}$ and $\alpha_{2}$ depend on the models,
but for $c_{s}^{2}\ll1$ they are at most of the order of $1/c_{s}^{2}$.
In this case the leading-order non-linear parameter can be estimated
as 
\begin{equation}
\left|f_{{\rm NL}}^{{\rm lead}}\right|\approx\frac{r_{3}^{2}}{c_{s}^{2}}\,.\label{flead2}
\end{equation}
Then the transition from the value (\ref{consistency}) to the value
(\ref{flead2}) occurs at 
\begin{equation}
r_{3}\approx c_{s}\,\sqrt{1-n_{{\cal R}}}\,.
\end{equation}
The effect of the term $1/c_{s}^{2}$ in Eq.~(\ref{flead2}) becomes
important for $r_{3}>c_{s}\,\sqrt{1-n_{{\cal R}}}$. When $n_{{\cal R}}=0.96$
this condition translates into $r_{3}>0.2\, c_{s}$. If $c_{s}=0.1$
and $r_{3}>0.1$, for example, the non-linear parameter (\ref{flead2})
can be larger than the order of 1.
However, for $c_{s}=0.1$, we also expect the dominant contribution to 
$f_{\rm NL}$ to come from other shapes (equilateral, orthogonal, etc.)

For the models in which $c_{s}^{2}$ is close to 1, $\alpha_{1}$
and $\alpha_{2}$ are of the order of unity. Hence the leading-order
non-linear parameter can be estimated as $\left|f_{{\rm NL}}^{{\rm lead}}\right|\approx r_{3}^{2}\ll1$
in the regime $r_{3}\ll1$. By increasing the value of $r_{3}$ from
0, we can observationally discriminate between the models with $c_{s}^{2}\ll1$
and $c_{s}^{2}\approx1$.

\subsection{Equilateral non-Gaussianities}

The equilateral shape is characterized by $k_{1}=k_{2}=k_{3}\equiv k$,
in which case the non-linear parameter is $f_{{\rm NL}}^{{\rm equil}}=(10/9)({\cal A}_{{\cal R}}/k^{3})$.
Since $S_{1}=4k^{3}/3$, $S_{2}=17k^{3}/6$, and $S_{3}=k^{3}/27$,
the bispectrum (\ref{Ale}) gives the leading-order non-linear parameter
\begin{eqnarray}
f_{{\rm NL}}^{{\rm equil,lead}} & = & \frac{85}{324}\left(1-\frac{1}{c_{s}^{2}}\right)-\frac{10}{81}\frac{\lambda}{\Sigma}+\frac{20}{81\epsilon_{s}}\left[(1+\lambda_{3X})\delta_{G3X}+4(3+2\lambda_{4X})\delta_{G4XX}+\delta_{G5X}+(5+2\lambda_{5X})\delta_{G5XX}\right]\nonumber \\
 &  & +\frac{65}{162c_{s}^{2}\epsilon_{s}}(\delta_{G3X}+6\delta_{G4XX}+\delta_{G5X}+\delta_{G5XX})\,.\label{equile}
\end{eqnarray}
If $c_{s}^{2}\ll1$, then we have $|f_{{\rm NL}}^{{\rm equil,lead}}|\gg1$.

In the equilateral limit the functions ${\cal V}$ and ${\cal U}$
are given by ${\cal V}=15[1+(1/2)\ln(2/3)]k^{3}$ and ${\cal U}=[6 \ln (3/2)-1]k^{3}$,
respectively (see Appendix B). Then the functions $\delta\tilde{Q}_{12}\equiv(3\delta Q_{1}+\delta Q_{2})/k^{3}$
and $\delta\tilde{Q}_{3}\equiv\delta Q_{3}/k^{3}$ 
reduce to
\begin{eqnarray}
\delta\tilde{Q}_{12} & = &
\epsilon [6-78\ln (2/3)-14\gamma_1]/3-\eta_{sF}[7\gamma_1+26\ln (2/3)]/2
\nonumber \\
&  & +s[16-39\ln (2/3)]/3+\tilde{\eta}_1(7\gamma_1+20)/6\,,  \\
\delta\tilde{Q}_{3} & = & -\epsilon[10\gamma_{1}+55+168\ln(2/3)]/27
-\eta_{sF}[2\gamma_{1}+7+28\ln(2/3)]/9 \nonumber \\
 &  & -4s[10+21\ln(2/3)]/27+\tilde{\eta}_{3}(2\gamma_{1}-3)/27.
\end{eqnarray}
The correction to $f_{{\rm NL}}^{{\rm equil,lead}}$ coming from Eq.~(\ref{ARlo2})
is given by 
\begin{eqnarray}
f_{{\rm NL}}^{{\rm equil,corre}} & = & \frac{5}{972c_{s}^{2}\epsilon_{s}}\biggl[3\epsilon_{s}(24\delta{\cal C}_{1}+51\delta{\cal C}_{2}+4c_{s}^{2}\delta{\cal C}_{3})+8c_{s}^{2}(3\eta_{sF}-4s-\eta_{6})\delta{\cal C}_{6}+(450\epsilon+327\eta_{sF}+620s-497\eta_{7})\delta{\cal C}_{7}\nonumber \\
 &  & ~~~~~~~~~~-153\epsilon_{s}\delta{\cal C}_{8}+72\epsilon_{s}^{2}\biggr]-\frac{5}{18}\left(\frac{1}{c_{s}^{2}}-1-\frac{2\delta{\cal C}_{7}}{c_{s}^{2}\epsilon_{s}}\right)\delta\tilde{Q}_{12}+\frac{5}{6}\left(\frac{1}{c_{s}^{2}}-1-\frac{2\lambda}{\Sigma}+\frac{4\delta{\cal C}_{6}}{\epsilon_{s}}-\frac{4\delta{\cal C}_{7}}{c_{s}^{2}\epsilon_{s}}\right)\delta\tilde{Q}_{3}.
 \label{equicorrection}
\end{eqnarray}
For the theories in which $f_{{\rm NL}}^{{\rm equil,lead}}$ vanishes,
the next-order correction $f_{{\rm NL}}^{{\rm equil,corre}}$ is the
dominant contribution. In the case of a canonical scalar field with
the Lagrangian $P=X-V(\phi)$, $G_{3}=0$, $G_{4}=0$, $G_{5}=0$,
for example, it follows that $f_{{\rm NL}}^{{\rm equil,lead}}=0$
and $f_{{\rm NL}}^{{\rm equil,corre}}=55\epsilon_{s}/36+5\eta_{s}/12$.

For the theories with $c_s^2 \neq 1$ the non-linear parameters
(\ref{equile}) and (\ref{equicorrection}) reproduce the results known in literature for specific 
models of inflation.
For example, this is the case for k-inflation \cite{Seery,Chen}, 
k-inflation with the Galileon 
terms \cite{Mizuno,ATnongau}, potential-driven 
Galileon inflation \cite{Tavakol}, and inflation with a field derivative
coupling to the Einstein tensor \cite{Watanabe}.
Generally we require that $c_s^2 \ll 1$ to realize 
the large equilateral non-linear parameter.

\subsection{Enfolded non-Gaussianities}

The enfolded shape is characterized by $k_{2}+k_{3}=k_{1}$. Taking
the momenta $k_{1}=k$ and $k_{2} \to k_{3}=k/2$, 
the non-linear parameter\footnote{At the point $k_{1}=k$ and $k_{2} \to k_{3}=k/2$, the equilateral shape gives no contribution. However, the contribution from the orthogonal shape will be of the same order of the enfolded one, since, by definition, in this case we have $f_{\rm NL}^{\rm ortho}\to-2 f_{\rm NL}^{\rm enfold}$.}
is $f_{{\rm NL}}^{{\rm enfold}}=8{\cal A}_{{\cal R}}/(3k^{3})$. Since
$S_{1}=23k^{3}/64$, $S_{2}=63k^{3}/64$, and $S_{3}=k^{3}/128$ in
this case, the leading-order non-linear parameter is given by 
\begin{equation}
f_{{\rm NL}}^{{\rm enfold,lead}}=\frac{1}{32}\left(1-\frac{1}{c_{s}^{2}}\right)-\frac{1}{16}\frac{\lambda}{\Sigma}+\frac{1}{8\epsilon_{s}}\left[(1+\lambda_{3X})\delta_{G3X}+4(3+2\lambda_{4X})\delta_{G4XX}+\delta_{G5X}+(5+2\lambda_{5X})\delta_{G5XX}\right]\,,\label{enle}
\end{equation}
where, unlike the equilateral case, the $\delta{\cal C}_{7}$-dependent
term in Eq.~(\ref{Ale}) disappears.

In the enfolded limit one has ${\cal V}=[315/64-\ln (2)/2]k^{3}$
and ${\cal U}=[5/24+\ln (2)/4]k^{3}$ (see Appendix B). The functions $\delta\tilde{Q}_{12}$
and $\delta\tilde{Q}_{3}$ are
\begin{eqnarray}
\delta\tilde{Q}_{12} & = & \epsilon [45 -24\gamma_1+108 \ln (2)]/64
-\eta_{sF} [87+36\gamma_1-108 \ln (2)]/128
\nonumber \\
& &+s[333+108 \ln (2)]/128+\tilde{\eta}_1 (39+3\gamma_1)/32 \,,\\
\delta\tilde{Q}_{3} & = & -\epsilon[15\gamma_{1}+17-54\ln(2)]/192
-\eta_{sF}[36\gamma_{1}-5-108\ln(2)]/768\nonumber \\
 &  & -s[109-108\ln(2)]/768+\tilde{\eta}_{3}(2\gamma_{1}-3)/128\,.
\end{eqnarray}
The correction to $f_{{\rm NL}}^{{\rm enfold,lead}}$ is
\begin{eqnarray}
f_{{\rm NL}}^{{\rm enfold,corre}} & = & \frac{1}{96c_{s}^{2}\epsilon_{s}}[\epsilon_{s}(23\delta{\cal C}_{1}+63\delta{\cal C}_{2}+3c_{s}^{2}\delta{\cal C}_{3})-2c_{s}^{2}(\eta_{6}+4s-3\eta_{sF})\delta{\cal C}_{6}+(172\epsilon+254s+92\eta_{sF}-174\eta_{7})\delta{\cal C}_{7}\nonumber \\
 &  & +2\epsilon_{s}^{2}]-\frac{2}{3}\left(\frac{1}{c_{s}^{2}}-1-\frac{2\delta{\cal C}_{7}}{c_{s}^{2}\epsilon_{s}}\right)\delta\tilde{Q}_{12}+2\left(\frac{1}{c_{s}^{2}}-1-\frac{2\lambda}{\Sigma}+\frac{4\delta{\cal C}_{6}}{\epsilon_{s}}-\frac{4\delta{\cal C}_{7}}{c_{s}^{2}\epsilon_{s}}\right)\delta\tilde{Q}_{3}\,.
\end{eqnarray}
For a canonical scalar field we have that $f_{{\rm NL}}^{{\rm enfold,lead}}=0$
and $f_{{\rm NL}}^{{\rm enfold,corre}}=7\epsilon_{s}/8+5\eta_{s}/12$.

\section{Shapes of non-Gaussianities}
\label{shapesec} 

The leading-order bispectrum (\ref{ARl}) can be written in terms
of the sum of each component, as ${\cal A}_{{\cal R}}=\sum_{i=1}^{8}{\cal A}_{{\cal R}}^{(i)}$.
Then the function ${\cal F}_{{\cal R}}(k_{1},k_{2},k_{3})$ in Eq.~(\ref{calF})
is decomposed into eight components 
\begin{equation}
{\cal F}_{{\cal R}}^{(i)}=(2\pi)^{4}\frac{{\cal A}_{{\cal R}}^{(i)}}
{\prod_{i=1}^{3}k_{i}^{3}}=\frac{(2\pi)^{4}}{(k_{1}k_{2}k_{3})^{3}}{\cal B}_{i}S_{i}\,,
\end{equation}
where ${\cal B}_{i}$'s are the coefficients appearing in front of
each shape function $S_{i}$ in Eq.~(\ref{ARl}), 
say, ${\cal B}_{1}=c_{s}^{2}{\cal C}_{1}/(4\epsilon_{s}F)$.

In order to estimate the correlation between two different shapes, 
we define the following quantity \cite{Fergu}
\begin{equation}
C({\cal F}_{{\cal R}}^{(i)},{\cal F}_{{\cal R}}^{(j)})=\frac{{\cal I}({\cal F}_{{\cal R}}^{(i)},
{\cal F}_{{\cal R}}^{(j)})}{\sqrt{{\cal I}({\cal F}_{{\cal R}}^{(i)},
{\cal F}_{{\cal R}}^{(i)})\,{\cal I}({\cal F}_{{\cal R}}^{(j)},{\cal F}_{{\cal R}}^{(j)})}}\,,\label{corre}
\end{equation}
where 
\begin{equation}
{\cal I}({\cal F}_{{\cal R}}^{(i)},{\cal F}_{{\cal R}}^{(j)})=\int d{\cal V}_{k}\,
{\cal F}_{{\cal R}}^{(i)}(k_{1},k_{2},k_{3}){\cal F}_{{\cal R}}^{(j)}(k_{1},k_{2},k_{3})
\frac{(k_{1}k_{2}k_{3})^{4}}{(k_{1}+k_{2}+k_{3})^{3}}\,.
\end{equation}
The integration should be done in the region $0\le k_{1}<\infty$,
$0<k_{2}/k_{1}<1$, and $1-k_{2}/k_{1}\le k_{3}/k_{1}\le1$. Note
that the above integral can be expressed in terms of $r_{2}=k_{2}/k_{1}$
and $r_{3}=k_{3}/k_{1}$ with the integral of $k_{1}$ factorized
out. For $|C({\cal F}_{{\cal R}}^{(i)},{\cal F}_{{\cal R}}^{(j)})|$
close to 1 the correlation is large, whereas for $|C({\cal F}_{{\cal R}}^{(i)},{\cal F}_{{\cal R}}^{(j)})|$
close to 0 the two shapes are almost orthogonal with a small correlation.

The CMB data analysis of non-Gaussianities has been carried out by using
the factorizable shape functions which are written as the sums of monomials
of $k_{1}$, $k_{2}$, and $k_{3}$. There are a number of templates
${\cal F}_{{\cal R}}$ which resemble model predictions of the bispectrum. 
The templates corresponding to local and equilateral
non-Gaussianities are given, respectively, by \cite{Cremishape1,Cremishape2}
\begin{equation}
{\cal F}_{{\cal R}}^{{\rm local}}(k_{1},k_{2},k_{3})=(2\pi)^{4}
\left(\frac{3}{10}f_{{\rm NL}}^{{\rm local}}\right)
\left(\frac{1}{k_{1}^{3}k_{2}^{3}}+\frac{1}{k_{2}^{3}k_{3}^{3}}+\frac{1}{k_{3}^{3}k_{1}^{3}}\right)\,,\label{slocal}
\end{equation}
and 
\begin{equation}
{\cal F}_{{\cal R}}^{{\rm equil}}(k_{1},k_{2},k_{3})=
(2\pi)^{4}\left(\frac{9}{10}f_{{\rm NL}}^{{\rm equil}}\right)
\left[-\frac{1}{k_{1}^{3}k_{2}^{3}}-\frac{1}{k_{2}^{3}k_{3}^{3}}
-\frac{1}{k_{3}^{3}k_{1}^{3}}-\frac{2}{k_{1}^{2}k_{2}^{2}k_{3}^{2}}
+\left(\frac{1}{k_{1}k_{2}^{2}k_{3}^{3}}+5~{\rm perm.}\right)\right]\,.
\label{sequilateral}
\end{equation}
Since the local non-Gaussianities are small in the Horndeski's 
theories, we do not consider the correlation
with the local template.

The orthogonal template, which has a small correlation with the
equilateral one, is given by \cite{SSZ} 
\begin{equation}
{\cal F}_{{\cal R}}^{{\rm ortho}}(k_{1},k_{2},k_{3})=
(2\pi)^{4}\left(\frac{9}{10}f_{{\rm NL}}^{{\rm ortho}}\right)
\left[-\frac{3}{k_{1}^{3}k_{2}^{3}}-\frac{3}{k_{2}^{3}k_{3}^{3}}
-\frac{3}{k_{3}^{3}k_{1}^{3}}-\frac{8}{k_{1}^{2}k_{2}^{2}k_{3}^{2}}
+\left(\frac{3}{k_{1}k_{2}^{2}k_{3}^{3}}+5~{\rm perm.}\right)\right]\,.\label{sortho}
\end{equation}
The enfolded template, which is a linear combination of the
orthogonal and equilateral templates, is defined by \cite{Meerburg}
\begin{equation}
{\cal F}_{{\cal R}}^{{\rm enfold}}(k_{1},k_{2},k_{3})=
(2\pi)^{4}\left(\frac{9}{10}f_{{\rm NL}}^{{\rm enf}}\right)
\left[\frac{1}{k_{1}^{3}k_{2}^{3}}+\frac{1}{k_{2}^{3}k_{3}^{3}}
+\frac{1}{k_{3}^{3}k_{1}^{3}}+\frac{3}{k_{1}^{2}k_{2}^{2}k_{3}^{2}}
-\left(\frac{1}{k_{1}k_{2}^{2}k_{3}^{3}}+5~{\rm perm.}\right)\right]\,.\label{senfolded}
\end{equation}

In Table \ref{scorre} we show the correlation between ${\cal F}_{{\cal R}}^{(i)}$
($i=3,4,5,7,8$) and the three templates given above. Since the correlations
between ${\cal F}_{{\cal R}}^{(i)}$ ($i=4,5,7,8$) and ${\cal F}_{{\cal R}}^{{\rm equil}}$
are close to 1, ${\cal F}_{{\cal R}}^{(i)}$ ($i=4,5,7,8$) are well 
approximated by the equilateral shape.
In particular, both $C({\cal F}_{{\cal R}}^{(7)},{\cal F}_{{\cal R}}^{{\rm ortho}})$
and $C({\cal F}_{{\cal R}}^{(8)},{\cal F}_{{\cal R}}^{{\rm ortho}})$
are close to 0 with the same level of correlation between ${\cal F}_{{\cal R}}^{{\rm equil}}$
and ${\cal F}_{{\cal R}}^{{\rm ortho}}$. Hence the shape functions
$S_{7}$ and $S_{8}$ highly mimic the equilateral template.
They also vanish in the local limit
($k_{3}\to0$) and in the enfolded limit ($k_{2}+k_{3}\to k_{1}$).
Note that both $|C({\cal F}_{{\cal R}}^{(7)},{\cal F}_{{\cal R}}^{{\rm enfold}})|$
and $|C({\cal F}_{{\cal R}}^{(8)},{\cal F}_{{\cal R}}^{{\rm enfold}})|$
are close to 0.5 (which is similar to the value 
$|C({\cal F}_{{\cal R}}^{{\rm equil}},{\cal F}_{{\cal R}}^{{\rm enfold}})|=0.511911$).
Since there is the relation ${\cal F}_{{\cal R}}^{{\rm enfold}}
=({\cal F}_{{\cal R}}^{{\rm equil}}-{\cal F}_{{\cal R}}^{{\rm ortho}})/2$,
we have $|C({\cal F}_{{\cal R}}^{(i)},{\cal F}_{{\cal R}}^{{\rm enfold}})|\simeq0.5$
for the shape function ${\cal F}_{{\cal R}}^{(i)}$ very close to
the equilateral one.

\begin{table}[ht]
\centering %
\begin{tabular}{||c|c|c|c|c|c|c|c|c|c|c||}
\hline 
 & ${\cal F}_{{\cal R}}^{{\rm equil}}$  & ${\cal F}_{{\cal R}}^{{\rm ortho}}$  & ${\cal F}_{{\cal R}}^{{\rm enfold}}$  
 & ${\cal F}_{{\cal R}}^{(3)}$  & ${\cal F}_{{\cal R}}^{(4)}$  & ${\cal F}_{{\cal R}}^{(5)}$  & ${\cal F}_{{\cal R}}^{(7{\rm equil})}$  & ${\cal F}_{{\cal R}}^{(8)}$ & ${\cal F}_{{\cal R}}^{(7{\rm ortho})}$  & $3{\cal F}_{{\cal R}}^{(1)}-{\cal F}_{{\cal R}}^{(2)}$\tabularnewline
\hline 
${\cal F}_{{\cal R}}^{{\rm equil}}$ & 1  & 0.0254062  & 0.511911  & 0.936177  & -0.998757  & -0.994234  & 0.999892  & -0.999994 & -0.00693357 & 0.986954\tabularnewline
\hline 
${\cal F}_{{\cal R}}^{{\rm ortho}}$  &  & 1  & -0.845755  & -0.290742  & 0.0177139  & -0.117961  & 0.0353534  & -0.0277283 & 0.904843 & -0.116557\tabularnewline
\hline 
${\cal F}_{{\cal R}}^{{\rm enfold}}$  &  &  & 1  & 0.749518 & -0.548302 & -0.4293  & 0.503306  & -0.509913 & -0.781246 & 0.626939\tabularnewline
\hline 
${\cal F}_{{\cal R}}^{(3)}$  &  &  &  & 1 & -0.952469 & -0.893224 & 0.933797 & -0.935615 & -0.357802  & 0.980504\tabularnewline
\hline 
${\cal F}_{{\cal R}}^{(4)}$  &  &  &  &  & 1 & 0.987653 & -0.998384  & 0.998686 & 0.056402 & -0.993745\tabularnewline
\hline 
${\cal F}_{{\cal R}}^{(5)}$  &  &  &  &  &  & 1 & -0.994696 & 0.99437 & -0.0995524 & -0.964012\tabularnewline
\hline 
${\cal F}_{{\cal R}}^{(7{\rm equil})}$ &  &  &  &  &  &  & 1 & -0.999936 & 0 & 0.9859\tabularnewline
\hline 
${\cal F}_{{\cal R}}^{(8)}$ &  &  &  &  &  &  &  & 1 & 0.00524865 & -0.986715\tabularnewline
\hline 
${\cal F}_{{\cal R}}^{(7{\rm ortho})}$ &  &  &  &  &  &  &  &  & 1 & -0.167335\tabularnewline
\hline 
\end{tabular}\caption[scorre]{
The correlation (\ref{corre}) between two different shape functions. 
${\cal F}_{{\cal R}}^{(7{\rm equil})}$ is the normalized shape of ${\cal F}_{{\cal R}}^{(7)}$, whereas
${\cal F}_{{\cal R}}^{(7{\rm ortho})}$ is the shape function
orthogonal to ${\cal F}_{{\cal R}}^{(7{\rm equil})}$ (or, equivalently, to ${\cal F}_{{\cal R}}^{(7)}$).}
\label{scorre} 
\end{table}

We recall that the functions $3S_{1}-S_{2}$ and $S_{3}$ are related
with $S_{7}$ {[}see Eq.~(\ref{S78}){]}.
We can use this property in order to rewrite the leading-order part of the bispectrum in a convenient basis.
We introduce the following shape
\begin{equation}
S_7^{\rm equil}=-\frac{12}{13}\,S_7\,,
\end{equation}
whose minus sign has been chosen so that ${\cal F}_{{\cal R}}^{(7{\rm equil})}$ has 
a positive high correlation with the equilateral profile. Furthermore, we can analytically show
that the following shape is exactly orthogonal to $S_{7}^{\rm equil}$: 
\begin{equation}
S_{7}^{{\rm ortho}}=\frac{12}{14-13\beta}\,(\beta S_{7}+3S_{1}-S_{2})\,,
\label{S7orthdef}
\end{equation}
where 
\begin{equation}
\beta=\frac{16}{3}\frac{248041-25200\pi^{2}}{1986713-201600\pi^{2}}=1.1967996\dots
\end{equation}
The normalizations of $S_7^{\rm equil}$ and $S_7^{\rm ortho}$ have been 
done such that, at the equilateral configuration ($k_1=k_2=k_3=k$), 
we have $S_7^{\rm equil}=k^3=S_7^{\rm ortho}$. 
This normalization follows from the standard definition of the previous 
templates introduced in the literature.

We note here that the leading-order bispectrum (\ref{Ale}) includes the term
$3S_{1}-S_{2}$, so that we also consider the correlation between the combination
$3{\cal F}_{{\cal R}}^{(1)}-{\cal F}_{{\cal R}}^{(2)}$ and other
shapes in Table I. The shape $3{\cal F}_{{\cal R}}^{(1)}-{\cal F}_{{\cal R}}^{(2)}$
has a high correlation with ${\cal F}_{{\cal R}}^{{\rm equil}}$.
Compared to ${\cal F}_{{\cal R}}^{(7)}$ and ${\cal F}_{{\cal R}}^{(8)}$,
however, it is not very close to the equilateral shape. 
Moreover ${\cal F}_{{\cal R}}^{(3)}$ has some correlation 
with the orthogonal shape, i.e., 
$C({\cal F}_{{\cal R}}^{(3)},{\cal F}_{{\cal R}}^{{\rm ortho}})=-0.290742$.

As we see in Table~\ref{scorre}, the correlation between ${\cal F}_{{\cal R}}^{(7{\rm ortho})}$
and ${\cal F}_{{\cal R}}^{{\rm equil}}$ is very small. 
In terms of $S_{7}^{\rm equil}$ and $S_{7}^{{\rm ortho}}$, 
the functions $3S_{1}-S_{2}$ and $S_{3}$
are expressed as 
\begin{equation}
3S_{1}-S_{2}=\frac{13}{12}\,\beta\, S_{7}^{\rm equil}+\frac{14-13\beta}{12}\,S_{7}^{{\rm ortho}}\,,
\qquad S_{3}=\frac{13}{432}\,(3\beta-2)\,S_7^{\rm equil}+\frac{14-13\beta}{144}\,S_{7}^{{\rm ortho}}\,.
\end{equation}
Using these relations, the leading-order bispectrum (\ref{Ale})
can be written in terms of the equilateral basis $S_{7}^{\rm equil}$
and the orthogonal basis $S_{7}^{{\rm ortho}}$, as
\begin{equation}
{\cal A}_{\cal R}^{\rm lead}=c_1 S_{7}^{\rm equil} +
c_2 S_{7}^{{\rm ortho}}\,,
\label{ARleading}
\end{equation}
where 
\begin{eqnarray}
c_1 &=& \frac{13}{12}\left[\frac{1}{24}\left(1-\frac{1}{c_{s}^{2}}\right)
\left(2+3\beta\right)+\frac{\lambda}{12\Sigma}\left(2-3\beta\right)
-\frac{\delta{\cal C}_{6}}{6\epsilon_{s}}\left(2-3\beta\right)
+\frac{\delta{\cal C}_{7}}{3\epsilon_{s}c_{s}^{2}}\right]\,,\\
c_2 &=& \frac{14-13\beta}{12}\left[\frac{1}{8}\left(1-\frac{1}{c_{s}^{2}}\right)
-\frac{\lambda}{4\Sigma}
+\frac{\delta{\cal C}_{6}}{2\epsilon_{s}}\right]\,.
\end{eqnarray}
The coefficients $c_1$ and $c_2$ characterize the magnitudes 
of the three-point correlation function coming from equilateral and
orthogonal contributions, respectively.
Finally, we also introduce the enfolded shape function
\begin{equation}
S_7^{\rm enfold}=(S_7^{\rm equil}-S_7^{\rm ortho})/2\,,
\end{equation}
which has a maximum at $k_1 \to 2k_2$, $k_2\to k_3$.
Note that $S_7^{\rm enfold}$ vanishes at the equilateral configuration.

\begin{figure}
\includegraphics[width=3.2in]{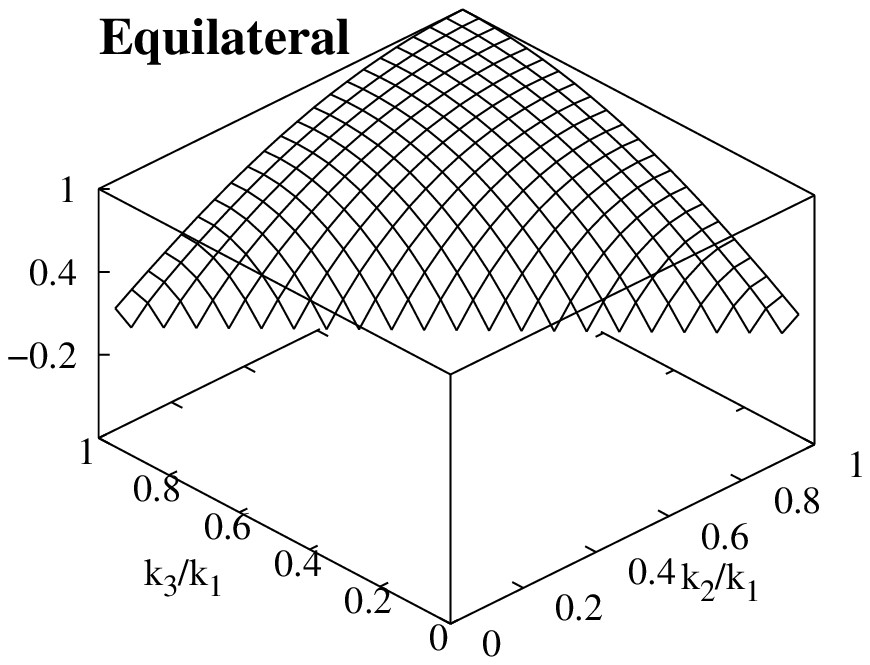}
\includegraphics[width=3.2in]{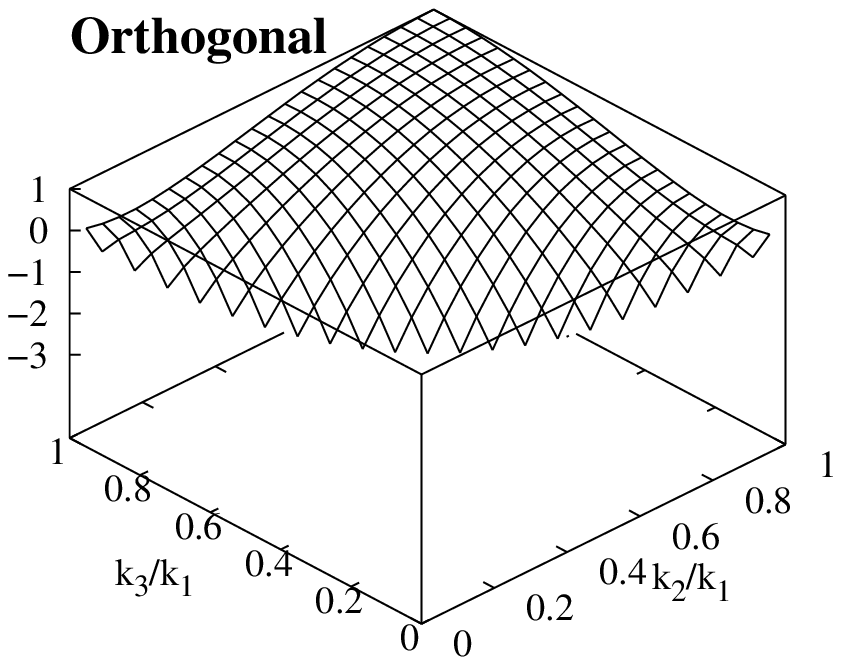}
\includegraphics[width=3.2in]{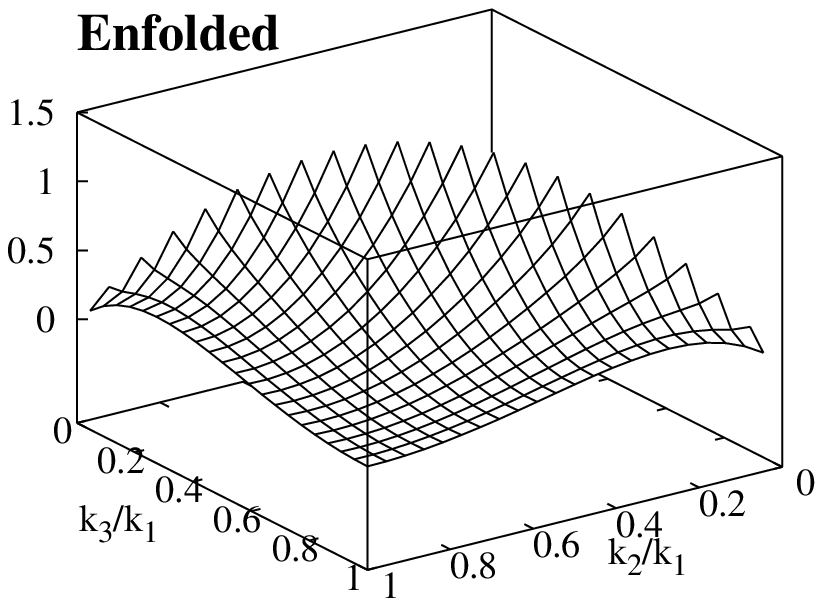}
\caption{\label{fig12}
The shape functions 
${\cal F}_{{\cal R}}^{(7{\rm equil})}(1,k_{2}/k_{1},k_{3}/k_{1})
(k_{2}/k_{1})^{2}(k_{3}/k_{1})^{2}$ (top left),
${\cal F}_{{\cal R}}^{(7{\rm ortho})}(1,k_{2}/k_{1},k_{3}/k_{1})
(k_{2}/k_{1})^{2}(k_{3}/k_{1})^{2}$ (top right), and 
${\cal F}_{{\cal R}}^{(7{\rm enfold})}(1,k_{2}/k_{1},k_{3}/k_{1})
(k_{2}/k_{1})^{2}(k_{3}/k_{1})^{2}$ (bottom).}
\end{figure}

In Fig.~\ref{fig12} we plot the three shape functions 
${\cal F}_{{\cal R}}^{(7{\rm equil})}$, ${\cal F}_{{\cal R}}^{(7{\rm ortho})}$
and ${\cal F}_{{\cal R}}^{(7{\rm enfold})}$  multiplied by the functions
$(k_{2}/k_{1})^{2}(k_{3}/k_{1})^{2}$.
The correlations of ${\cal F}_{\cal R}^{\rm lead}=(2\pi)^4{\cal A}_{\cal R}^{\rm lead}/\prod_{i=1}^3k_i^3$ 
with the shape functions ${\cal F}_{{\cal R}}^{{\rm equil}}$, ${\cal F}_{{\cal R}}^{{\rm ortho}}$, and 
${\cal F}_{{\cal R}}^{{\rm enfold}}$ can be evaluated as 
\begin{eqnarray}
C^{\rm equil} &\equiv& C({\cal F}_{\cal R}^{\rm lead}, {\cal F}_{{\cal R}}^{{\rm equil}})
=\frac{8.25104 \left( 1.44717 \times 10^{-2} c_1-1.70045 \times 10^{-4} c_2\right)}
{\sqrt{1.42610 \times 10^{-2} c_1^2+4.09480 \times 10^{-2} c_2^2}}\,,\label{Ceq} \\
C^{\rm ortho} &\equiv& C({\cal F}_{\cal R}^{\rm lead}, {\cal F}_{{\cal R}}^{{\rm ortho}})
=\frac{5.12494 \left( 8.23792 \times 10^{-4} c_1+3.57273 \times 10^{-2} c_2\right)}
{\sqrt{1.42610 \times 10^{-2} c_1^2+4.09480 \times 10^{-2} c_2^2}}\,,\\
C^{\rm enfold} &\equiv& C({\cal F}_{\cal R}^{\rm lead}, {\cal F}_{{\cal R}}^{{\rm enfold}})
=\frac{8.80788 \left( 6.82395 \times 10^{-3} c_1-1.79487 \times 10^{-2} c_2\right)}
{\sqrt{1.42610 \times 10^{-2} c_1^2+4.09480 \times 10^{-2} c_2^2}}\,,\label{Cen}
\end{eqnarray}
which depend on the coefficients $c_1$ and $c_2$. 
In particular, we find $C({\cal F}_{{\cal R}}^{(7{\rm enfold})},{\cal F}_{{\cal R}}^{{\rm enfold}})=0.928621$.

\section{Shapes of non-Gaussianities in concrete models}
\label{concretesec} 

Let us study the non-Gaussianities of concrete models of inflation
in which the bispectrum (\ref{ARleading}) can be large due to the small
scalar propagation speed $c_{s}$. As we will see below, there are
some models where the orthogonal shape provides an important 
contribution to the bispectrum.

\subsection{Power-law k-inflation}

We first consider k-inflation characterized by 
\begin{equation}
P(\phi,X)=K(\phi)(-X+X^{2})\,,\quad G_{3}=0\,,\quad G_{4}=0\,,\quad G_{5}=0\,,
\end{equation}
where $K(\phi)$ is a function in terms of $\phi$. {}From
the background equations (\ref{E1d}) and (\ref{E1}) it follows that
\begin{equation}
K(\phi)=\frac{3M_{{\rm pl}}^{2}H^{2}}{X(3X-1)}\,,\qquad
\frac{\dot{H}}{H^{2}}=-\frac{3(2X-1)}{3X-1}\,.\label{KH}
\end{equation}
As an example, we study power-law inflation characterized by $a\propto t^{1/\gamma}$
and $H=1/(\gamma t)$, where $\gamma~(\ll1)$ is constant. 
Substituting the Hubble parameter into the second of Eq.~(\ref{KH}), we obtain
$X=(3-\gamma)/[3(2-\gamma)]$ and $\phi=\phi_{0}+\sqrt{(3-\gamma)/[3(2-\gamma)]}\, t$ 
($\phi_{0}$ is the initial value of the field). {}From
the first of Eq.~(\ref{KH}) we find that power-law inflation is
realized for the choice \cite{Mukhanov} 
\begin{equation}
K(\phi)=\frac{6(2-\gamma)M_{{\rm pl}}^{2}}{\gamma^{2}(\phi-\phi_{0})^{2}}\,.
\end{equation}
In this case $c_{s}^{2}$ and $\lambda/\Sigma$ in Eq.~(\ref{ARleading})
are given by 
\begin{equation}
c_{s}^{2}=\frac{\gamma}{3(4-\gamma)}\,,\qquad\frac{\lambda}{\Sigma}=\frac{1}{2}(1-c_{s}^{2})\,.
\end{equation}
For $\gamma\ll1$ we have that $1/c_{s}^{2}\simeq12/\gamma\gg1$.

In this model (dubbed Model A) the leading-order bispectrum (\ref{ARleading})
reduces to
\begin{equation}
{\cal A}_{{\cal R}}^{{\rm lead}}=\frac{13}{12}\left[-\frac{2+3\beta}{24c_{s}^{2}}+\frac{1}{6}
-\frac{1}{24}(2-3\beta)c_{s}^{2}\right]S_{7}^{\rm equil}
-\frac{14-13\beta}{96}\left(\frac{1}{c_{s}^{2}}
-c_{s}^{2}\right)S_{7}^{{\rm ortho}}\,.\label{Amodel1}
\end{equation}
In the limit $c_{s}^{2}\ll1$ it follows that  
${\cal A}_{{\cal R}}^{{\rm lead}}\simeq(-0.252/c_{s}^{2})S_{7}^{\rm equil}
+(0.016/c_{s}^{2})S_{7}^{{\rm ortho}}$.
Since in this limit the non-linear parameters (\ref{equile}) and (\ref{enle}) are given 
by $f_{\rm NL}^{\rm equil,lead} \simeq -85/(324c_s^2)$
and $f_{\rm NL}^{\rm enfold,lead} \simeq -1/(32c_s^2)$ respectively, 
the WMAP9 year constraint (\ref{equilcon}) of the equilateral shape 
gives the bound $c_s^2>1.2 \times 10^{-3}$ (95\,\%\,CL).  

The scalar spectral index (\ref{nR}) and the tensor-to-scalar
ratio (\ref{tsratio}) can be expressed in terms of $c_s$, as
\begin{equation}
n_{\cal R}=1-\frac{24c_s^2}{1+3c_s^2}\,,\qquad
r=\frac{192c_s^3}{1+3c_s^2}\,.
\label{moan}
\end{equation}
For the $\Lambda$CDM model without the running scalar spectral index, 
the bounds on $n_s$ and $r$ from the WMAP9 data alone are 
$n_{\cal R}=0.992 \pm 0.019$ (68\,\%\,CL)
and $r<0.38$ (95\,\%\,CL), respectively.
If we combine the WMAP9 data with the measurements of high-$l$ CMB 
anisotropies, baryon acoustic oscillations, and the Hubble constant, 
the constraints are $0.9636 \pm 0.0084$ (68\,\%\,CL)
and $r<0.13$ (95\,\%\,CL).
If we employ the bound $0.95<n_{\cal R}<1$ then the scalar propagation 
is constrained to be $c_s^2<2.1 \times 10^{-3}$.
Since in this case $r<0.018$ from Eq.~(\ref{moan}),
the observational constraint on $r$ is satisfied.
Combining the bound of $c_s^2$ with that of the scalar non-Gaussianity, 
it follows that $1.2 \times 10^{-3}<c_s^2<2.1 \times 10^{-3}$.

In Fig.~\ref{fig34} we plot the shape function 
${\cal F}_{{\cal R}}^{{\rm lead}}(1,k_{2}/k_{1},k_{3}/k_{1})(k_{2}/k_{1})^{2}(k_{3}/k_{1})^{2}$
for $c_{s}^2=2.0 \times 10^{-3}$ (labelled as ``A'' in the figure), 
where ${\cal F}_{{\cal R}}^{{\rm lead}}=(2\pi)^{4}{\cal A}_{{\cal R}}^{{\rm lead}}/\prod_{i=1}^{3}k_{i}^{3}$.
When $c_s^2=2.0 \times 10^{-3}$, the correlations (\ref{Ceq})-(\ref{Cen}) are 
$C^{\rm equil}=-0.99474$, $C^{\rm ortho}=0.06305$, and $C^{\rm enfold}=-0.58511$, respectively.
Hence the shape of non-Gaussianities is close to the equilateral one illustrated
in Fig.~\ref{fig12}, whose property is 
is independent of the choice of $c_s^2$.

\subsection{k-inflation with the term $G_{3}(X)$}

We study k-inflation in the presence of the covariant Galileon term
$G_{3}(X)$ characterized by \cite{KYY,Mizuno} 
\begin{equation}
P(X)=-X+\frac{X^{2}}{2M^{4}}\,,\qquad G_{3}(X)=\frac{\mu X}{M^{4}}\,,
\label{model2}
\end{equation}
where $M$ and $\mu$ are constants having a dimension of mass. 
In this model (dubbed Model B) the de Sitter solution is present 
when $\epsilon=\delta_{PX}+3\delta_{G3X}=0$.
Using Eq.~(\ref{E1d}) as well, we obtain 
\begin{equation}
H^{2}=\frac{M^{4}}{18\mu^{2}}\frac{(1-x)^{2}}{x}\,,\qquad
\frac{\mu}{M_{{\rm pl}}}=\frac{1-x}{x\sqrt{3(2-x)}}\,,
\end{equation}
where $x=X/M^{4}$. As long as inflation is realized in the regime
$\mu/M_{{\rm pl}}\ll1$, $x$ is close to 1. In what follows we replace
$x$ for 1 except for the terms including $1-x$.

\begin{figure}
\includegraphics[width=3.2in]{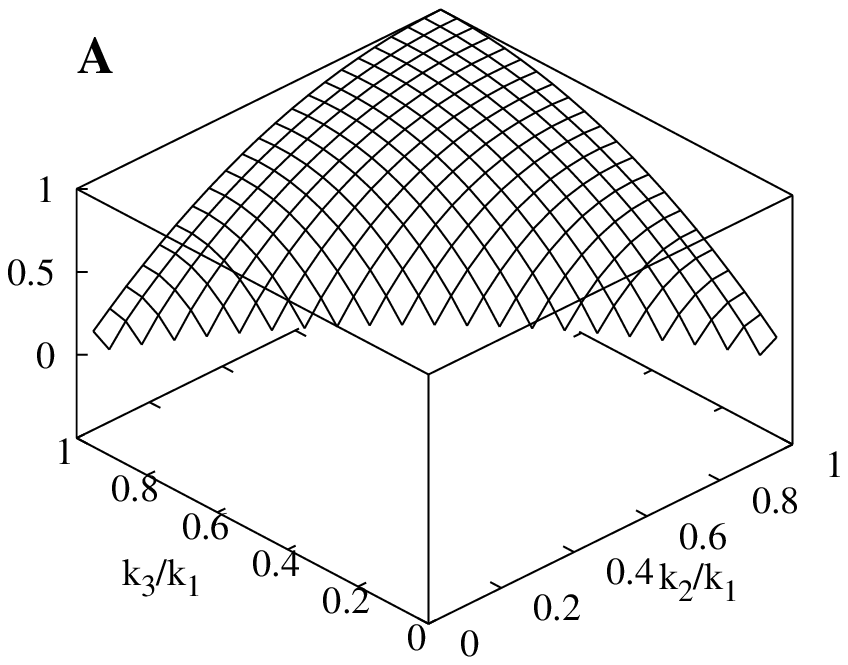}
\includegraphics[width=3.2in]{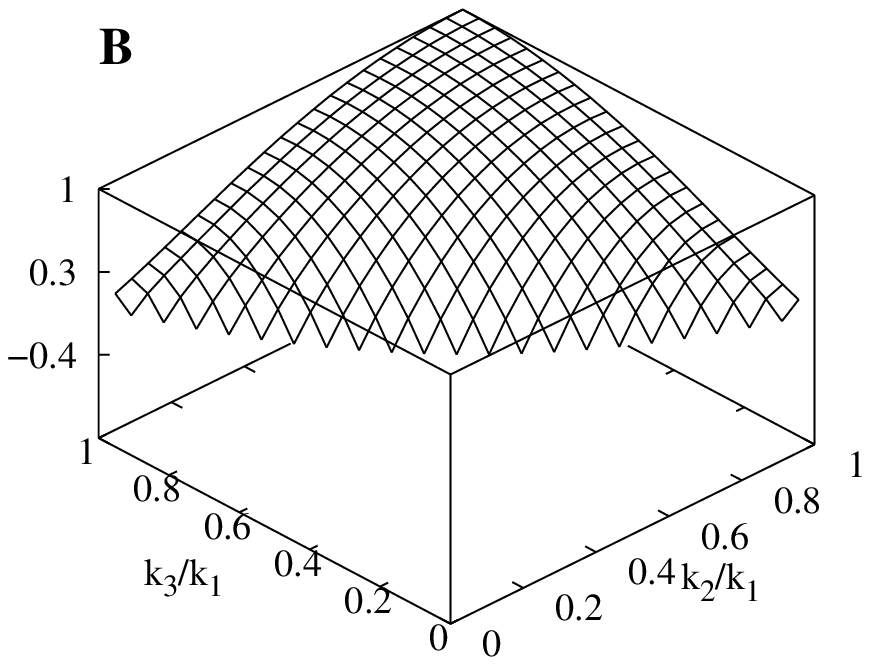}
\includegraphics[width=3.2in]{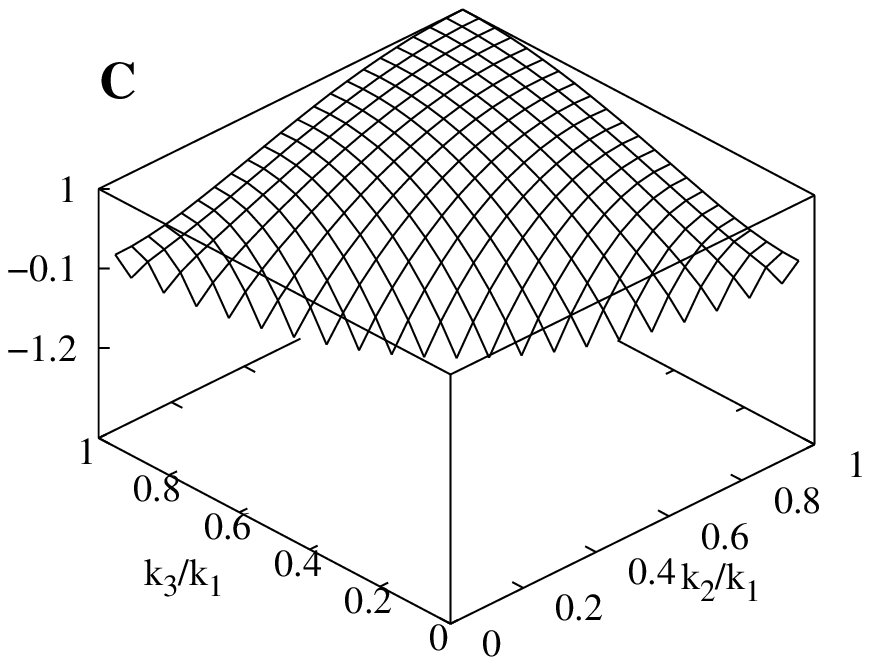}
\includegraphics[width=3.2in]{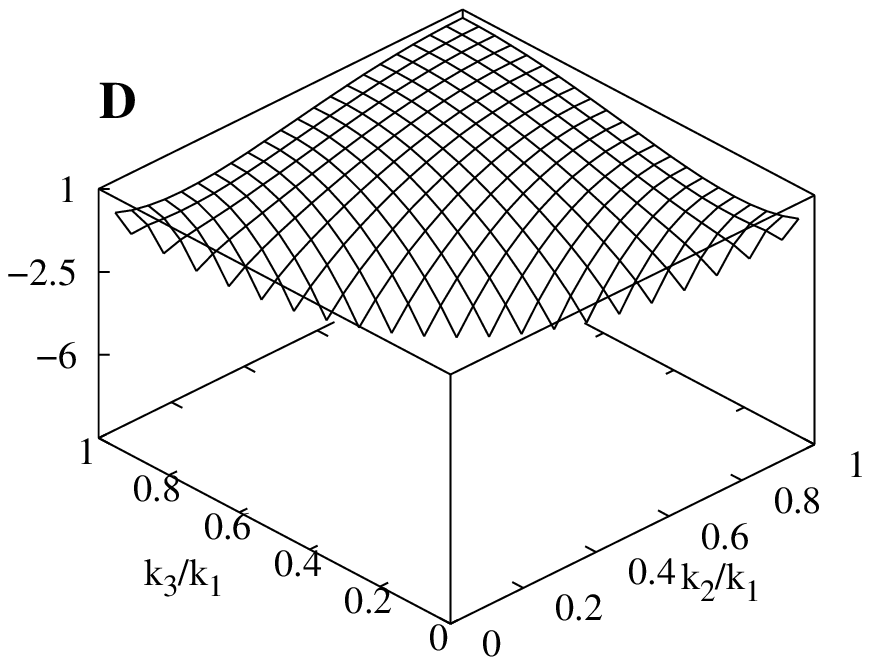}
\caption{\label{fig34}
The leading-order shape function 
${\cal F}^{\rm lead}_{{\cal R}}(1,k_{2}/k_{1},k_{3}/k_{1})
(k_{2}/k_{1})^{2}(k_{3}/k_{1})^{2}$ for $c_s^2=2.0 \times 10^{-3}$ 
normalized to 1 at the equilateral configuration.
Each panel corresponds to 
the models A, B, C, D.
}
\end{figure}

Along the de Sitter solution we have that $\lambda/\Sigma=6c_{s}^{2}/\epsilon_{s}$,
$\delta{\cal C}_{6}=\delta{\cal C}_{7}=\delta_{G3X}=2(1-x)$. Since
$\epsilon_{s}=2(1-x)$ and $c_{s}^{2}=(1-x)/6$, the
bispectrum (\ref{ARleading}) reads 
\begin{equation}
{\cal A}_{{\cal R}}^{{\rm lead}}=\frac{13}{12}\left[\frac{1}{4c_{s}^{2}}
\left(1-\frac{\beta}{2}\right)-\frac{1}{6}(1-3\beta)\right]S_{7}^{\rm equil}
-\frac{14-13\beta}{12}\left(\frac{1}{8c_{s}^{2}}-\frac{1}{2}\right)S_{7}^{{\rm ortho}}\,.\label{Amodel2}
\end{equation}
If $c_{s}^{2}\ll1$, we obtain ${\cal A}_{{\cal R}}^{{\rm lead}}\simeq
(0.109/c_{s}^{2})S_{7}^{\rm equil}+(0.016/c_{s}^{2})S_{7}^{{\rm ortho}}$ together with
the non-linear parameters $f_{\rm NL}^{\rm equil,lead} \simeq 5/(36c_s^2)$
and $f_{\rm NL}^{\rm enfold,lead} \simeq -1/(32c_s^2)$. 
Notice that the sign of $f_{\rm NL}^{\rm equil,lead}$ is opposite compared to the 
k-inflation model A. {}From the WMAP 9 year bound (\ref{equilcon})
it follows that $c_s^2>4.3 \times 10^{-4}$ (95\,\%\,CL).  
In Model B the scalar spectral index is $n_{\cal R}=1$, whereas
$r$ is related to $c_s^2$ via
\begin{equation}
c_s^2=\frac{3^{1/3}}{48}r^{2/3}\,.
\label{csr}
\end{equation}
The WMAP9 bound $r<0.38$ gives the constraint
$c_s^2<1.6 \times 10^{-2}$. 
Using the severer bound 
$r<0.13$, it follows that $c_s^2<7.7 \times 10^{-3}$.
In both cases there are viable parameter spaces compatible
with the constraint from the scalar non-Gaussainity.
When $c_s^2=2.0 \times 10^{-3}$, for example, 
we have $C^{\rm equil}=0.96865$, $C^{\rm ortho}=0.25272$, and $C^{\rm enfold}=0.29984$.
As we see in Fig.~\ref{fig34}, the shape of non-Gaussianities 
for $c_s^2=2.0 \times 10^{-3}$ is approximately close to the equilateral one.

For potential-driven inflation ($P=X-V(\phi)$) with the term
$G_3(X)=\mu X/M^4$ \cite{Kamada}, the non-Gaussianities are 
small because $c_s^2$ is not much smaller than 1 \cite{Tavakol}. 
In such models, if the Galileon 
self-interaction dominates over the standard kinetic term even after 
inflation, there is an instability associated with the appearance of 
the negative $c_s^2$ after the field velocity $\dot{\phi}$ changes 
its sign \cite{Junko}\footnote{In the framework of the effective field theory 
the Lagrangian is valid at the energy scale of inflation, but some higher dimensional 
operators can appear during the reheating stage.
There may be a possibility that the Laplacian instability can be
avoided by such operators.}. 
In the model B discussed above, reheating needs to occur 
gravitationally \cite{KYY}, so the situation should be different from that 
in Galileon inflation driven by a potential with a minimum.

\subsection{k-inflation with the term $G_{4}(X)$}

The next model (dubbed Model C) is k-inflation with the covariant Galileon term $G_{4}(X)$
\cite{ATnongau}, i.e., 
\begin{equation}
P(X)=-X+\frac{X^{2}}{2M^{4}}\,,\qquad G_{4}(X)=\frac{\mu X^{2}}{M^{7}}\,.
\end{equation}
Similar to the case (\ref{model2}), there is a de Sitter solution
satisfying the conditions 
\begin{equation}
H^{2}=\frac{M^{3}}{36\mu}\frac{1-x}{x}\,,\qquad\frac{\mu M}{M_{{\rm pl}}^{2}}
=\frac{1-x}{6x^{2}(3-2x)}\,.
\end{equation}
Inflation occurs in the regime where $x=X/M^{4}$ is close to 1. 
Employing the similar approximation to that used previously, 
we have $\delta{\cal C}_{6}=2\delta{\cal C}_{7}=12\delta_{G4XX}$,
$\epsilon_{s}=8\delta_{G4XX}$, $\delta_{G4XX}=(1-x)/3$, $c_{s}^{2}=2(1-x)/9$,
and $\lambda/\Sigma=1/2$. Hence the bispectrum (\ref{ARleading})
reduces to
\begin{equation}
{\cal A}_{{\cal R}}^{{\rm lead}}=\frac{13}{12}\left[\frac{1}{24c_{s}^{2}}
\left(4-3\beta\right)-\frac{1}{12}(4-9\beta)\right]S_{7}^{\rm equil}
-\frac{14-13\beta}{12}\left(\frac{1}{8c_{s}^{2}}-\frac{3}{4}\right)S_{7}^{{\rm ortho}}\,.\label{Amodel3}
\end{equation}

In the limit $c_{s}^{2}\ll1$, we have ${\cal A}_{{\cal R}}^{{\rm lead}}
\simeq (0.018/c_{s}^{2})S_{7}^{\rm equil}+(0.016/c_{s}^{2})S_{7}^{{\rm ortho}}$.
The equilateral and orthogonal non-linear parameters in this limit 
are given by $f_{\rm NL}^{\rm equil,lead} \simeq 25/(648c_s^2)$
and $f_{\rm NL}^{\rm enfold,lead} \simeq -1/(32c_s^2)$, respectively.
The tighter constraint on $c_s^2$ comes from the enfolded
bound (\ref{enfoldcon}) rather than the equilateral bound (\ref{equilcon}), 
i.e., $c_s^2>3.6 \times 10^{-4}$.
Note that the scalar spectral index is $n_{\cal R}=1$ and that 
the same relation as Eq.~(\ref{csr}) holds in Model C.
Hence the upper bound of $c_s^2$ coming from the 
observational constraint of $r$ is the same as that of Model B, 
i.e., $c_s^2<1.6 \times 10^{-2}$ for $r<0.38$ and  
$c_s^2<7.7 \times 10^{-3}$ for $r=0.13$.

In Fig.~\ref{fig34} we plot the shape of non-Gaussianities for $c_s^2=2.0 \times 10^{-3}$, 
in which case the shape is between the equilateral and orthogonal ones
shown in Fig.~\ref{fig12}.
In fact, we have $C^{\rm equil}=0.58145$, $C^{\rm ortho}=0.75322$, 
and $C^{\rm enfold}=-0.33691$ for $c_s^2=2.0 \times 10^{-3}$.
The orthogonal contribution tends to be less important 
for the values of $c_s^2$ larger than the order of $10^{-2}$.

\subsection{k-inflation with the term $G_{5}(X)$}

Finally we study the following model (dubbed Model D) \cite{ATnongau} 
\begin{equation}
P(X)=-X+\frac{X^{2}}{2M^{4}}\,,\qquad G_{5}(X)=\frac{\mu X^{2}}{M^{10}}\,.
\end{equation}
In this case, for $x=X/M^{4}$ close to 1, there is a de Sitter solution
satisfying the conditions $H^{2}=M^{4}/(6M_{{\rm pl}}^{2})$ and $\mu^{2}M^{4}/M_{{\rm pl}}^{6}=27(1-x)^{2}/25$.
Since $\delta{\cal C}_{6}=3\delta{\cal C}_{7}=36(1-x)/5$, $\epsilon_{s}=18(1-x)/5$,
$c_{s}^{2}=3(1-x)/10$, and $\lambda/\Sigma=1/2$, the bispectrum
is given by 
\begin{equation}
{\cal A}_{{\cal R}}^{{\rm lead}}=\frac{13}{12}\left[-\frac{1}{c_{s}^{2}}\left(\frac{\beta}{8}-\frac{5}{36}\right)
-\frac{1}{2}+\beta\right]S_{7}^{\rm equil}-\frac{14-13\beta}{12}\left(\frac{1}{8c_{s}^{2}}-1\right)S_{7}^{{\rm ortho}}\,.\label{Amodel4}
\end{equation}

In the limit $c_{s}^{2}\ll1$ we have 
${\cal A}_{{\cal R}}^{{\rm lead}}\simeq(-0.012/c_{s}^{2})S_{7}^{\rm equil}+(0.016/c_{s}^{2})S_{7}^{{\rm ortho}}=
-(0.023/c_s^2) S_7^{\rm enfold}+(0.005/c_s^2)\,S_7^{\rm ortho}$, 
in which case the sign in front of the shape $S_7^{\rm equil}$ in Eq.~(\ref{Amodel4}) is opposite
to those in the models B and C.
In the same limit the equilateral and enfolded non-linear parameters 
are $f_{{\rm NL}}^{{\rm equil,lead}}\simeq 5/(972c_{s}^{2})$ and 
$f_{\rm NL}^{\rm enfold,lead} \simeq -1/(32c_s^2)$, respectively.
In this case $f_{{\rm NL}}^{{\rm equil,lead}}$ is smaller than 
$|f_{\rm NL}^{\rm enfold,lead}|$ by one order of magnitude.
The WMAP 9 year enfolded bound (\ref{enfoldcon}) gives the 
constraint $c_s^2>3.6 \times 10^{-4}$.
In Model D we have that $n_{\cal R}=1$ and that the relation between $r$ and 
$c_s^2$ is the same as Eq.~(\ref{csr}).
Hence the upper bound on $c_s^2$ is the same as 
that of Models B and C.

\begin{figure}
\includegraphics[width=3.2in]{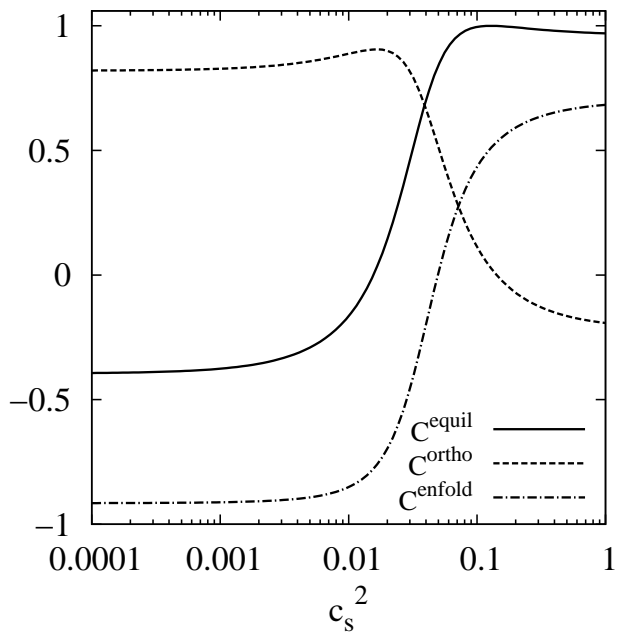}
\caption{\label{modeld}
The correlations (\ref{Ceq})-(\ref{Cen}) versus
$c_s^2$ for Model D. 
In the limit that $c_s^2 \ll 1$, $|C^{\rm enfold}|$ and $|C^{\rm ortho}|$ 
are larger than $|C^{\rm equil}|$.
On the other hand, for $c_s^2 \gtrsim 0.04$, $|C^{\rm equil}|$ is largest.
}
\end{figure}

The shape of non-Gaussianities for $c_s^2=2.0 \times 10^{-3}$
is plotted in Fig.~\ref{fig34}. 
When $c_s^2=2.0 \times 10^{-3}$ the correlations  (\ref{Ceq})-(\ref{Cen}) 
are $C^{\rm equil}=-0.35587$, 
$C^{\rm ortho}=0.83545$, and $C^{\rm enfold}=-0.90787$, respectively, 
in which case the shape has quite high (anti)-correlations with 
both the orthogonal and enfolded templates.
In Fig.~\ref{modeld} we show the correlations  $C^{\rm equil}$, 
$C^{\rm ortho}$, and $C^{\rm enfold}$ versus $c_s^2$. 
For $c_s^2 \gtrsim 0.04$ the correlation 
with the equilateral template is larger than 
those with other templates.
For $c_s^2 \lesssim 0.04$ the contributions of the orthogonal 
and enfolded shapes tend to be important.
In the limit that $c_s^2 \ll 1$, $|C^{\rm enfold}|$ is largest 
among other correlations.
Model D is an explicit example where the orthogonal (or enfolded) 
shape provides a significant contribution
to the bispectrum.

\section{Conclusions}
\label{concludesec} 

In the Horndeski's most general scalar-tensor theories we derived the three-point 
correlation function of primordial curvature perturbations generated during inflation 
in the presence of slow-variation corrections to the leading-order bispectrum.
Unlike previous works \cite{Gao,ATnongau}, the bispectrum (\ref{calAfi}) is 
valid for any shape of non-Gaussianities at first order of $\epsilon$.

In the squeezed limit ($k_3 \to 0$, $k_1 \to k_2$) the leading-order bispectrum 
(\ref{Ale}) vanishes, so that the correction (\ref{Aco}) is the dominant contribution 
to ${\cal A}_{\cal R}$. By using Eq.~(\ref{Aco}), we showed that the non-linear 
parameter in this limit is given by $f_{\rm NL}^{\rm local}=(5/12)(1-n_{\cal R})$.
This agrees with the result of Refs.~\cite{Maldacena,Cremi} in which the 
three-point correlation function was derived by dealing with the long-wavelength 
curvature perturbation (mode ${\bm k}_3$) as a classical background.
As demonstrated in Ref.~\cite{Cremi}, this result should be valid for any single-field
inflation in which the decaying mode of ${\cal R}$ is neglected relative 
to the growing mode. Our direct computation of the three-point 
correlation function in the presence of all possible slow-variation corrections
is another independent proof that the non-Gaussianity consistency 
relation holds for most general single-field theories with second-order
equations of motion.

The result of the local non-Gaussianities shows that $|f_{\rm NL}^{\rm local}|$
is much smaller than 1, e.g., $f_{\rm NL}^{\rm local}=0.0125$ for $n_{\cal R}=0.97$.
In the case where the shape of non-Gaussianities is not exactly the squeezed 
one ($0<r_3= k_3/k_1 \ll 1$), 
the leading-order bispectrum gives the non-linear parameter 
$|f_{\rm NL}^{\rm lead}| \approx r_3^2/c_s^2$.
Hence the leading-order term dominates over the correction 
for $r_3>c_s \sqrt{1-n_{\cal R}}$.
For the models with $c_s^2 \ll 1$ the non-linear 
parameter can be as large as $|f_{\rm NL}|>1$
with the growth of $r_3$.
By measuring the shape which is not so squeezed, it should be 
possible to discriminate the models with different values of $c_s^2$.
The leading-order non-linear parameters in the equilateral and 
enfolded limits are given by Eqs.~(\ref{equile}) and 
(\ref{enle}), respectively, whose magnitudes can be 
larger than 1 for the models with $c_s^2 \ll 1$.
These results will be useful to constrain concrete models 
of inflation in future high-precision observations.

We also showed that the leading-order bispectrum can be expressed 
in terms of the sum of the two bases $S_7^{\rm equil}$ and $S_7^{\rm ortho}$.
The shape $S_7^{\rm equil}$ is very highly correlated with the equilateral 
template (\ref{sequilateral}). It also vanishes in both local and 
enfolded limits. The shape $S_7^{\rm ortho}$, which is defined 
by (\ref{S7orthdef}), is exactly orthogonal to $S_7^{\rm equil}$.
The coefficients $c_1$ and $c_2$ in front of $S_7^{\rm equil}$ and $S_7^{\rm ortho}$ in 
Eq.~(\ref{ARleading}) characterize the equilateral and orthogonal 
contributions, respectively. 

In Sec.~\ref{concretesec} we presented concrete models in which 
the orthogonal shape can provide important contributions to the bispectrum.
In power-law k-inflation the shape of non-Gaussianities 
is well approximated by the equilateral type.
However, in k-inflation described by the Lagrangian $P(X)=-X+X^2/(2M^4)$
with a number of different Galileon terms like 
$G_4(X)=\mu X^2/M^7$ and $G_5(X)=\mu X^2/M^{10}$, we found 
that the orthogonal contribution is crucially important for $c_s^2 \ll 1$.
In the presence of the term $G_5(X)=\mu X^2/M^{10}$, 
the correlations with the orthogonal and enfolded templates 
in the regime $c_s^2 \ll 1$
are higher than that with the equilateral template.

It will be interesting to see how the observations such as Planck \cite{PLANCK}
provide the constraints on the scalar non-Gaussianities as well as the 
scalar spectral index and the tensor-to-scalar ratio.
In particular, if future observations confirm the value
$|f_{\rm NL}^{\rm local}|>1$ at more than 95\,\%\,CL, 
this implies that we need to go beyond the slow-variation 
single-field inflationary scenario (including 
the Horndeski's theories).
The information of other shapes of non-Gaussianities 
(including the not so squeezed one) will be useful to 
discriminate between many different models.

\section*{ACKNOWLEDGEMENTS}
We are grateful to Eiichiro Komatsu for motivating us to derive the
full bispectrum with slow-variation corrections in the Horndeski's theories.
We also thank Xingang Chen, Shinji Mukohyama, and Misao Sasaki for
useful correspondence and discussions. This work was supported in
part by the Grant-in-Aid for Scientific Research Fund of No.~24540286
and Scientific Research on Innovative Areas (No.~21111006).

\appendix

\section{Corrections to the bispectrum}

Following the calculations of Ref.~\cite{Chen} we present the explicit
forms of the corrections to the bispectrum (\ref{ARl2}). They come
from the first three integrals in the action (\ref{L3d}).
We write each ${\cal O}(\epsilon)$ contribution to the bispectrum as 
$\Delta{\cal A}_{{\cal R}}^{(i)}$ ($i=1,2,3$). 
Note that similar calculations were also carried out 
in Ref.~\cite{Burrage}.

\vspace{0.5cm}
 (i) $\int dt\, d^{3}x\, a^{3}\tilde{{\cal C}}_{1}\Mpl^{2}{\cal R}\dot{{\cal R}}^{2}$
\begin{itemize}
\item (a) The correction from the variation of $\tilde{{\cal C}}_{1}$ 
in Eq.~(\ref{Cvari})
\begin{equation}
\Delta{\cal A}_{{\cal R}}^{(1)}\supset-\biggl(\frac{c_{s}^{2}}{4\epsilon_{s}F}\frac{1}{H}\frac{d{\tilde{{\cal C}}}_{1}^{{\rm lead}}}{dt}\biggr)_{K}\biggl[(1-2\gamma_{1})\frac{1}{K}\sum_{i>j}k_{i}^{2}k_{j}^{2}-(1-\gamma_{1})\frac{1}{K^{2}}\sum_{i\neq j}k_{i}^{2}k_{j}^{3}\biggr]\,.\label{C1a}
\end{equation}

\item (b) The correction from the scale factor $a$ in Eq.~(\ref{avari})
\begin{equation}
\Delta{\cal A}_{{\cal R}}^{(1)}\supset\biggl(\frac{c_{s}^{2}\,\epsilon}{2\epsilon_{s}F}\tilde{{\cal C}}_{1}^{{\rm lead}}\biggr)_{K}\biggl[(1+2\gamma_{1})\frac{1}{K}\sum_{i>j}k_{i}^{2}k_{j}^{2}-\gamma_{1}\frac{1}{K^{2}}\sum_{i\neq j}k_{i}^{2}k_{j}^{3}\biggr]\,.
\end{equation}

\item (c) The contribution from the correction to $u (0,k_{i})$ 
in Eq.~(\ref{u0})
\begin{eqnarray}
\Delta{\cal A}_{{\cal R}}^{(1)} & \supset & -\biggl(\frac{c_{s}^{2}}{4\epsilon_{s}F}\tilde{{\cal C}}_{1}^{{\rm lead}}\biggr)_{K}\left[3(1+\gamma_{2})\epsilon+\frac{3\gamma_{2}}{2}\eta_{sF}+3\left(1+\frac{\gamma_{2}}{2}\right)s+\left(\epsilon+\frac{\eta_{sF}}{2}+\frac{s}{2}\right)\ln\frac{k_{1}k_{2}k_{3}}{K^{3}}\right]\nonumber \\
 &  & \times\biggl(\frac{2}{K}\sum_{i>j}k_{i}^{2}k_{j}^{2}-\frac{1}{K^{2}}\sum_{i\neq j}k_{i}^{2}k_{j}^{3}\biggr)\,.
\end{eqnarray}

\item (d) The contribution from the correction to $u^{*}(\tau,k_{i})$ 
in Eq.~(\ref{usvari})
\begin{eqnarray}
\Delta{\cal A}_{{\cal R}}^{(1)} & \supset & \biggl(\frac{c_{s}^{2}}{4\epsilon_{s}F}\tilde{{\cal C}}_{1}^{{\rm lead}}\biggr)_{K}\biggl[3(1-2\gamma_{1})s\frac{(k_{1}k_{2}k_{3})^{2}}{K^{3}}-\left\{ (1+2\gamma_{1})\epsilon+\left(\gamma_{1}-\frac{1}{2}\right)\eta_{sF}+\left(\frac{3}{2}+\gamma_{1}\right)s\right\} \frac{1}{K}\sum_{i>j}k_{i}^{2}k_{j}^{2}\nonumber \\
 &  & +\left\{ \gamma_{1}\epsilon-\frac{1}{2}(1-\gamma_{1})\eta_{sF}+\frac{1}{2}(1+\gamma_{1})s\right\} \frac{1}{K^{2}}\sum_{i\neq j}k_{i}^{2}k_{j}^{3}+\left(\epsilon+\frac{\eta_{sF}}{2}+\frac{s}{2}\right)\left(\frac{k_{2}^{2}k_{3}^{2}}{k_{1}}{\cal G}+{\rm perm.}\right)\biggr]\,,\label{C1d}
\end{eqnarray}
where 
\begin{equation}
{\cal G}(k_{1},k_{2},k_{3})\equiv{\rm Re}\left[\int_{0}^{\infty}dx_{1}\, h^{*}(x_{1})e^{-\frac{k_{2}+k_{3}}{k_{1}}x_{1}}\right]\,,
\nonumber 
\end{equation}
and $x_{1}=-k_{1}c_{sK}\tau$. 
\item (e) The contribution from the correction to $\frac{d}{d\tau}u^{*}(\tau,k_{i})$
in Eq.~(\ref{us2vari})
\begin{eqnarray}
\Delta{\cal A}_{{\cal R}}^{(1)} & \supset & -\biggl(\frac{c_{s}^{2}}{4\epsilon_{s}F}\tilde{{\cal C}}_{1}^{{\rm lead}}\biggr)_{K}\biggl[\left\{ 2(1+2\gamma_{1})\epsilon+(2\gamma_{1}-1)(\eta_{sF}-3s)\right\} \frac{1}{K}\sum_{i>j}k_{i}^{2}k_{j}^{2}+(2\epsilon+\eta_{sF}+s)k_{1}k_{2}k_{3}\nonumber \\
 &  & -\left\{ 2\gamma_{1}\epsilon+(\gamma_{1}-1)\eta_{sF}+(4-6\gamma_{1})s\right\} \frac{1}{K^{2}}\sum_{i\neq j}k_{i}^{2}k_{j}^{3}+(1-2\gamma_{1})s\biggl(\frac{1}{K^{3}}\sum_{i\neq j}k_{i}^{2}k_{j}^{4}+\frac{2}{K^{3}}\sum_{i>j}k_{i}^{3}k_{j}^{3}\biggr)\nonumber \\
 &  & -(2\epsilon+\eta_{sF}+s)\biggl\{\sum_{i}k_{i}^{3}+\sum_{i\neq j}k_{i}k_{j}^{2}+\sum_{i}k_{i}^{3}\,{\rm Re}\int_{0}^{\infty}dx_{K}\frac{e^{-ix_{K}}}{x_{K}}-\frac{1}{2}({\cal M}+{\rm perm.})\biggr\}\biggr]\,,\label{C1e}
\end{eqnarray}
where $x_{K}=-Kc_{sK}\tau$, and 
\[
{\cal M}(k_{1},k_{2},k_{3})\equiv-k_{1}{\rm Re}\biggl[\int_{0}^{\infty}dx_{1}\,\frac{1}{x_{1}}\left(k_{2}^{2}+k_{3}^{2}+ik_{2}k_{3}\frac{k_{2}+k_{3}}{k_{1}}x_{1}\right)e^{-i\frac{k_{2}+k_{3}}{k_{1}}x_{1}}\frac{dh^{*}(x_{1})}{dx_{1}}\biggr]\,.
\]

\end{itemize}
\vspace{0.3cm}
 (ii) $\int dt\, d^{3}x\, a\,\tilde{{\cal C}}_{2}\Mpl^{2}{\cal R}(\partial{\cal R})^{2}$
\begin{itemize}
\item (a) The correction from the variation of $\tilde{{\cal C}}_{2}$ 
\begin{equation}
\Delta{\cal A}_{{\cal R}}^{(2)}\supset-\biggl(\frac{1}{4\epsilon_{s}F}\frac{1}{H}\frac{d{\tilde{{\cal C}}}_{2}^{{\rm lead}}}{dt}\biggr)_{K}\biggl[(1-\gamma_{1})\biggl(\frac{1}{2}\sum_{i}k_{i}^{3}-\frac{1}{K^{2}}\sum_{i\neq j}k_{i}^{2}k_{j}^{3}\biggr)-\frac{1}{2}k_{1}k_{2}k_{3}+\frac{1-2\gamma_{1}}{K}\sum_{i>j}k_{i}^{2}k_{j}^{2}+\frac{1}{2}\sum_{i\neq j}k_{i}k_{j}^{2}\biggr]\,.\label{C2a}
\end{equation}

\item (b) The correction from the scale factor $a$ 
\begin{equation}
\Delta{\cal A}_{{\cal R}}^{(2)}\supset\biggl(\frac{\epsilon}{2\epsilon_{s}F}\tilde{{\cal C}}_{2}^{{\rm lead}}\biggr)_{K}\biggl[\gamma_{1}\biggl(\frac{1}{2}\sum_{i}k_{i}^{3}-\frac{1}{K^{2}}\sum_{i\neq j}k_{i}^{2}k_{j}^{3}\biggr)+\frac{1}{2}k_{1}k_{2}k_{3}+\frac{1+2\gamma_{1}}{K}\sum_{i>j}k_{i}^{2}k_{j}^{2}-\frac{1}{2}\sum_{i\neq j}k_{i}k_{j}^{2}\biggr]\,.
\end{equation}

\item (c) The contribution from the correction to $u(0,k_{i})$ 
\begin{eqnarray}
\Delta{\cal A}_{{\cal R}}^{(2)} & \supset & -\biggl(\frac{1}{4\epsilon_{s}F}\tilde{{\cal C}}_{2}^{{\rm lead}}\biggr)_{K}\left[3(1+\gamma_{2})\epsilon+\frac{3\gamma_{2}}{2}\eta_{sF}+3\left(1+\frac{\gamma_{2}}{2}\right)s+\left(\epsilon+\frac{\eta_{sF}}{2}+\frac{s}{2}\right)\ln\frac{k_{1}k_{2}k_{3}}{K^{3}}\right]\nonumber \\
 &  & \times\biggl(\frac{1}{2}\sum_{i}k_{i}^{3}+\frac{2}{K}\sum_{i>j}k_{i}^{2}k_{j}^{2}-\frac{1}{K^{2}}\sum_{i\neq j}k_{i}^{2}k_{j}^{3}\biggr)\,.
\end{eqnarray}

\item (d) The contribution from the correction to $u^{*}(\tau,k_{i})$ 
\begin{eqnarray}
\Delta{\cal A}_{{\cal R}}^{(2)} & \supset & -\biggl(\frac{1}{8\epsilon_{s}F}\tilde{{\cal C}}_{2}^{{\rm lead}}\biggr)_{K}\biggl[\biggl\{3\gamma_{1}\epsilon+\frac{3}{2}(\gamma_{1}-1)\eta_{sF}+\frac{3}{2}(1+\gamma_{1})s\biggr\}\sum_{i}k_{i}^{3}-3\left(\epsilon+\frac{\eta_{sF}}{2}+\frac{s}{2}\right)\sum_{i\neq j}k_{i}k_{j}^{2}\nonumber \\
 &  & +\left\{ 3\epsilon+\frac{3}{2}\eta_{sF}+\left(\frac{5}{2}-2\gamma_{1}\right)s\right\} k_{1}k_{2}k_{3}+\left\{ 6(1+2\gamma_{1})\epsilon-3(1-2\gamma_{1})\eta_{sF}+(5+6\gamma_{1})s\right\} \frac{1}{K}\sum_{i>j}k_{i}^{2}k_{j}^{2}\nonumber \\
 &  & -\left\{ 6\gamma_{1}\epsilon+3(\gamma_{1}-1)\eta_{sF}+(1+6\gamma_{1})s\right\} \frac{1}{K^{2}}\sum_{i\neq j}k_{i}^{2}k_{j}^{3}-(1+\gamma_{1})s\frac{1}{K}\sum_{i}k_{i}^{4}-\gamma_{1}s\frac{1}{K^{2}}\sum_{i\neq j}k_{i}k_{j}^{4}\nonumber \\
 &  & +(2\epsilon+\eta_{sF}+s)({\cal N}+{\rm perm.})\biggr]\,,\label{C2d}
\end{eqnarray}
where 
\[
{\cal N}(k_{1},k_{2},k_{3})\equiv\frac{k_{1}}{2}\sum_{i}k_{i}^{2}\,{\rm Re}\biggl[\int_{0}^{\infty}dx_{1}\,\frac{1}{x_{1}^{2}}e^{-ix_{1}\frac{k_{2}+k_{3}}{k_{1}}}\left(-1-i\frac{k_{2}+k_{3}}{k_{1}}x_{1}+\frac{k_{2}k_{3}}{k_{1}^{2}}x_{1}^{2}\right)h^{*}(x_{1})\biggr]\,.
\]
Note that the coefficient $\tilde{{\cal C}}_{2}^{{\rm lead}}$ is
related to $\tilde{{\cal C}}_{1}^{{\rm lead}}$ via $\tilde{{\cal C}}_{2}^{{\rm lead}}=-(c_{s}^{2}/3)\tilde{{\cal C}}_{1}^{{\rm lead}}$. 
\end{itemize}
\vspace{0.3cm}
 (iii) $\int dt\, d^{3}x\, a^{3}\tilde{{\cal C}}_{3}\Mpl\dot{{\cal R}}^{3}$
\begin{itemize}
\item (a) The correction from the variation of $\tilde{{\cal C}}_{3}$ 
\begin{equation}
\Delta{\cal A}_{{\cal R}}^{(3)}\supset\biggl(\frac{3(2\gamma_{1}-3)c_{s}^{2}}{4\epsilon_{s}FM_{{\rm pl}}}\frac{d{\tilde{{\cal C}}}_{3}^{{\rm lead}}}{dt}\biggr)_{K}\frac{(k_{1}k_{2}k_{3})^{2}}{K^{3}}\,.\label{C3a}
\end{equation}

\item (b) The correction from the scale factor $a$ 
\begin{equation}
\Delta{\cal A}_{{\cal R}}^{(3)}\supset\biggl(\frac{3(2\gamma_{1}-1)c_{s}^{2}\,\epsilon\, H}{4\epsilon_{s}FM_{{\rm pl}}}\tilde{{\cal C}}_{3}^{{\rm lead}}\biggr)_{K}\frac{(k_{1}k_{2}k_{3})^{2}}{K^{3}}\,.
\end{equation}

\item (c) The contribution from the correction to $u (0,k_{i})$ 
\begin{equation}
\Delta{\cal A}_{{\cal R}}^{(3)}\supset-\biggl(\frac{3Hc_{s}^{2}}{2\epsilon_{s}FM_{{\rm pl}}}\tilde{{\cal C}}_{3}^{{\rm lead}}\biggr)_{K}\left[3(1+\gamma_{2})\epsilon+\frac{3}{2}\gamma_{2}\eta_{sF}+3\left(1+\frac{\gamma_{2}}{2}\right)s+\left(\epsilon+\frac{\eta_{sF}}{2}+\frac{s}{2}\right)\ln\frac{k_{1}k_{2}k_{3}}{K^{3}}\right]\frac{(k_{1}k_{2}k_{3})^{2}}{K^{3}}.
\end{equation}

\item (d) The contribution from the correction to $\frac{d}{d\tau}u^{*}(\tau,k_{i})$
\begin{eqnarray}
\Delta{\cal A}_{{\cal R}}^{(3)} & \supset & \biggl(\frac{3Hc_{s}^{2}}{4\epsilon_{s}FM_{{\rm pl}}}\tilde{{\cal C}}_{3}^{{\rm lead}}\biggr)_{K}\biggl[\left\{ 3(1-2\gamma_{1})\epsilon+3\left(\frac{3}{2}-\gamma_{1}\right)\eta_{sF}+\left(3\gamma_{1}-\frac{17}{2}\right)s\right\} \frac{(k_{1}k_{2}k_{3})^{2}}{K^{3}}\nonumber \\
 &  & +\left(\epsilon+\frac{\eta_{sF}}{2}+\frac{s}{2}\right)\biggl\{\biggl(\frac{1}{K^{2}}\sum_{i\neq j}k_{i}^{2}k_{j}^{3}-\frac{2}{K}\sum_{i>j}k_{i}^{2}k_{j}^{2}\biggr)+{\cal U}\biggr\}\biggr]\,,\label{C3d}
\end{eqnarray}
where ${\cal U}$ is defined in Eq.~(\ref{calU}). 
\end{itemize}

\section{Expressions for ${\cal U}$ and ${\cal V}$}

We evaluate the integrals ${\cal U}$ and ${\cal V}$ defined in Eqs.~(\ref{calU})
and (\ref{calV}) for general values of $r_{2}=k_{2}/k_{1},r_{3}=k_{3}/k_{1}$, 
and in three different limits: local shapes ($r_{3}\to0$, $r_{2}=1$), equilateral
($r_{3}=1,r_{2}=1$), and enfolded ($r_{3}\to1/2$,
$r_{2}\to1/2$) shapes. In doing so, we need to use 
the following relations 
\begin{eqnarray}
h^{*}(x) & = & 2\sin(x)+[\sin(x)-x\cos(x)]{\rm Ci}(2x)-[\cos(x)+x\sin(x)]{\rm Si}(2x)\nonumber \\
 &  & +i\,\{2\cos(x)+\pi\sin(x)-\pi x\cos(x)-[\cos(x)+x\sin(x)]{\rm Ci}(2x)-[\sin(x)-x\cos(x)]{\rm Si}(2x)\}\,,\label{hstar}\\
\frac{dh^{*}(x)}{dx} & = & \cos(x)-\sin(x)/x+x\sin(x){\rm Ci}(2x)-x\cos(x){\rm Si}(2x)\nonumber \\
 &  & +i\,[\pi x\sin(x)-\sin(x)-\cos(x)/x-x\cos(x){\rm Ci}(2x)-x\sin(x){\rm Si}(2x)]\,.
\end{eqnarray}

\subsection*{The ${\cal U}$ integral}

Let us first evaluate the integral ${\cal U}$.
We employ a few different procedures of regularizations, 
but they lead to the same final results. 
One possibility is to solve directly the integral, as
\begin{equation}
{\cal U}=\int_{0}^{\infty}f_{{\cal U}}(x_{1})\, dx_{1}+{\rm perm.}
=\lim_{y\to\infty}\int_{0}^{y}f_{{\cal U}}(x_{1})\, dx_{1}
+{\rm perm.}\,,
\end{equation}
where $f_{{\cal U}}$ is the integrand in Eq.~(\ref{calU}).
We can either neglect the terms which rapidly oscillate around
0, or we can, equivalently, shift $y$ in the complex domain for
the oscillating terms like $e^{\pm iky}\to e^{\pm ik\rho^{2}(1\pm i/\rho)}\to0$
(Method I). Another possible method (Method II) consists of solving
the following limit 
\begin{equation}
{\cal U}=\int_{0}^{\infty}f_{{\cal U}}(x_{1})\, dx_{1}+{\rm perm.}
=\lim_{c\to0}\int_{0}^{\infty}e^{-c x_{1}}\, f_{{\cal U}}(x_{1})\, dx_{1}+{\rm perm.}\,.
\end{equation}
We need to check whether both methods lead to the same results,
and we can -- at least for some values of $r_{2},r_{3}$ this proves
to be possible -- check the result against numerical integrations.

By employing the Method I mentioned above, the result of the integration
is 
\begin{eqnarray}
{\cal U}(r_{2},r_{3}) & = & k_{1}^{3}\left[\frac{4r_{2}^{2}r_{3}^{3}\left(3r_{2}^{2}-6r_{2}+r_{3}^{2}+3\right)}{\left(r_{2}-r_{3}-1\right){}^{3}\left(r_{2}+r_{3}-1\right){}^{3}}-\frac{2r_{2}^{2}r_{3}^{2}}{\left(r_{2}-r_{3}+1\right){}^{3}}\right]\ln2\nonumber \\
 &  & {}+\frac{k_{1}^{3}}{\left(r_{2}+r_{3}+1\right){}^{2}}\left\{ \frac{[(r_{2}+1){}^{3}-6r_{3}^{2}\left(r_{2}+1\right)+r_{3}^{3}]r_{2}^{2}}{(r_{2}-r_{3}+1){}^{2}}+\frac{r_{3}^{2}[r_{2}^{3}+3r_{3}r_{2}^{2}+3(r_{3}^{2}-2)r_{2}+r_{3}^{3}-6r_{3}+1]r_{2}^{2}}{(r_{2}+r_{3}-1){}^{2}}\right.\nonumber \\
 &  & {}+\frac{r_{3}^{2}[r_{2}^{3}-6\left(r_{3}+1\right)r_{2}^{2}+(r_{3}+1){}^{3}]}{\left(1-r_{2}+r_{3}\right){}^{2}}+\frac{2r_{3}^{2}\left(r_{2}+r_{3}+1\right){}^{2}r_{2}^{2}}{\left(r_{2}-r_{3}+1\right){}^{3}}\,\ln\!\left(\frac{r_{2}+r_{3}+1}{r_{3}}\right)\nonumber \\
 &  & {}+\left.\frac{2r_{3}^{2}\left(r_{2}+r_{3}+1\right){}^{2}r_{2}^{2}}{\left(1-r_{2}+r_{3}\right){}^{3}}\,\ln\!\left(\frac{r_{2}+r_{3}+1}{r_{2}}\right)+\frac{2r_{3}^{2}\left(r_{2}+r_{3}+1\right){}^{2}r_{2}^{2}}{\left(r_{2}+r_{3}-1\right){}^{3}}\,\ln\!\left(r_{2}+r_{3}+1\right)\right\} ,
\end{eqnarray}
where ${\cal U}(r_1,r_2)={\cal U}(r_2,r_1)$, that is ${\cal U}$ is symmetric under the exchange $r_1\leftrightarrow r_2$.

Let us consider this general expression in several different cases. Some of
these cases look -- only apparently -- singular: this behavior takes
place as the triangle of the momenta ${\bm k}_{i}$ degenerates into
a line. 
\begin{itemize}
\item \emph{Equilateral case}, $r_{2}=r_{3}=1$. In this case we find 
\begin{equation}
{\cal U}(1,1)={\cal U}_{{\rm equil}}=[6\ln(3/2)-1]\, k_{1}^{3}\approx1.43279\, k_{1}^{3}\,.
\end{equation}
Since there is no apparent singular behavior for ${\cal U}$, this
result can also be confirmed numerically%
\footnote{Mathematica, working in high precision, returns this same numerical
value. The same value can also be found by applying the Method II
mentioned above.%
}. 
\item \emph{Local case}, $r_{2}=1$, $r_{3}\to0$ (or $r_{2}\to0$, $r_{3}=1$).
In this case, the limit exists and gives 
\begin{equation}
\lim_{r_{3}\to0}{\cal U}(1,r_{3})={\cal U}_{{\rm local}}
=\frac{1}{2}k_{1}^{3}\,.
\end{equation}
The result in the local case matches with that derived by Chen \textit{et
al.} \cite{Chen}. 
\item Singular line, $r_{3}=\lim_{\epsilon\to0}(1-r_{2}+\epsilon)$, with
$0<r_{2}<1$. In this case we find 
\begin{equation}
\lim_{\epsilon\to0}{\cal U}=\frac{r_{2}\{(r_{2}-1)r_{2}
[20\left(r_{2}-1\right)r_{2}+9]+6\}(r_{2}-1)-6(r_{2}-1){}^{3}\ln(1-r_{2})+6r_{2}^{3}\ln r_{2}}{24\left(r_{2}-1\right)r_{2}}\, k_{1}^{3}\,.
\end{equation}
The limits $r_{2}\to0,1$ give again the local result, as expected. 
\item \emph{Enfolded case}, $r_{2}=1/2$, $r_{3}=\lim_{\epsilon\to0}(1/2+\epsilon)$.
This is a particular case of the previous one. Then we find 
\begin{equation}
\lim_{\epsilon\to0}{\cal U}={\cal U}_{{\rm enfold}}
=\frac{1}{24}\,(5+6\ln2)k_{1}^{3} \approx0.38162\, k_{1}^{3}\,.
\end{equation}

\item Singular line, $r_{3}=\lim_{\epsilon\to0}(r_{2}-1+\epsilon)$, with
$r_{2}>1$. In this case we obtain
\begin{equation}
\lim_{\epsilon\to0}{\cal U}=\frac{\left(r_{2}-1\right)\{r_{2}[r_{2}\left(6r_{2}^{2}-9r_{2}
+29\right)-40]+20\}+6r_{2}^{3}\Bigl[(r_{2}-1){}^{3}\ln\!\left(\frac{r_{2}}{r_{2}-1}\right)+\ln r_{2}\Bigr]}{24\left(r_{2}-1\right)r_{2}}\, k_{1}^{3}.
\end{equation}
In the limit $r_{2}\to1^{+}$, this reproduces the local
limit result, as expected. 
\item Singular line, $r_{3}=\lim_{\epsilon\to0}(1+r_{2}-\epsilon)$. 
In this case we find 
\begin{equation}
\lim_{\epsilon\to0}{\cal U}=\frac{r_{2}\bigl\{ r_{2}\{r_{2}[3r_{2}(2r_{2}+5)+38]+15\}+6\bigr\}+6r_{2}^{3}(r_{2}+1){}^{3}\ln\!\left(\frac{1}{r_{2}}+1\right)+6(r_{2}+1){}^{3}\ln(r_{2}+1)}{24r_{2}\left(r_{2}+1\right)}\, k_{1}^{3}\,,
\end{equation}
which recovers the local limit as $r_{2}=\epsilon\to0$. 
\end{itemize}
Therefore, we have shown that the physical limits are all finite.

\subsection*{The ${\cal V}$ integral}

The next step is to compute the integral (\ref{calV}). 
It is convenient to solve the integral 
by studying the limit 
\begin{equation}
{\cal V}=\lim_{\varepsilon\to0}\lim_{y\to\infty}\int_{\varepsilon}^{y}
f_{{\cal V}}(x_{1})dx_{1}+{\rm perm.}\,,
\end{equation}
where $f_{{\cal V}}$ is the integrand in Eq.~(\ref{calV}).
As for the $y \to \infty$ limit, we set the rapidly oscillating functions to
vanish by regularizing $e^{\pm iky}$ as in the Method I of 
the integral ${\cal U}$.
After taking the limit $\varepsilon\to0$, we finally obtain
\begin{eqnarray}
{\cal V} & = & \frac{k_{1}^{3}}{3(r_{2}-r_{3}-1)(r_{2}+r_{3}-1)(r_{2}+r_{3}+1)}[3r_{2}^{6}+6r_{3}r_{2}^{5}+r_{2}^{4}(r_{3}^{2}-9r_{3}-3)+r_{3}r_{2}^{3}(6r_{3}^{2}-13r_{3}+3)+3r_{3}^{4}r_{2}^{2}+3\nonumber \\
 &  & {}-15r_{3}^{3}r_{2}^{2}-4r_{3}^{2}r_{2}^{2}+3r_{2}^{2}(r_{3}-1)-12r_{3}^{5}(r_{2}+1)-r_{2}r_{3}(19r_{3}^{3}+15r_{3}^{2}+13r_{3}+9)-7r_{3}^{6}+3r_{3}^{4}+6r_{3}^{3}+r_{3}^{2}+6r_{3}]\nonumber \\
 &  & {}+\frac{k_{1}^{3}}{6(r_{2}-r_{3}+1)(r_{2}+r_{3}+1)}\,(7r_{2}^{5}+5r_{3}r_{2}^{4}+19r_{2}^{4}-6r_{3}^{2}r_{2}^{3}+5r_{3}r_{2}^{3}+14r_{2}^{3}-6r_{3}^{3}r_{2}^{2}+6r_{3}^{2}r_{2}^{2}+4r_{3}r_{2}^{2}+14r_{2}^{2}\nonumber \\
 &  & {}-r_{3}^{3}r_{2}+6r_{3}^{2}r_{2}+5r_{3}r_{2}+19r_{2}+r_{3}^{5}-r_{3}^{4}-6r_{3}^{3}-6r_{3}^{2}+5r_{3}+7-r_{3}^{4}r_{2})\nonumber \\
 &  & {}-\frac{k_{1}^{3}\ln2}{3(r_{2}-r_{3}-1){}^{2}(r_{2}+r_{3}-1){}^{2}}\, r_{3}^{3}\,(5r_{2}^{4}-20r_{2}^{3}-6r_{3}^{2}r_{2}^{2}+22r_{2}^{2}+4r_{3}^{2}r_{2}-20r_{2}+r_{3}^{4}-6r_{3}^{2}+5)\nonumber \\
 &  & {}+\frac{k_{1}^{3}\ln(r_{2}+r_{3}+1)}{6\left(r_{2}+r_{3}-1\right){}^{2}}(r_{2}^{5}+2r_{3}r_{2}^{4}-2r_{2}^{4}-3r_{3}^{2}r_{2}^{3}-2r_{3}r_{2}^{3}-3r_{2}^{3}-3r_{3}^{3}r_{2}^{2}+8r_{3}^{2}r_{2}^{2}-8r_{3}r_{2}^{2}+3r_{2}^{2}+2r_{3}^{4}r_{2}-2r_{3}^{3}r_{2}\nonumber \\
 &  & {}-8r_{3}^{2}r_{2}-2r_{3}r_{2}+2r_{2}+r_{3}^{5}-2r_{3}^{4}-3r_{3}^{3}+3r_{3}^{2}+2r_{3}-1)\nonumber \\
 &  & {}-\frac{k_{1}^{3}\ln\!\left(\frac{r_{2}+r_{3}+1}{r_{2}}\right)}{6\left(r_{2}-r_{3}-1\right){}^{2}}(r_{2}^{5}-2r_{3}r_{2}^{4}-2r_{2}^{4}-3r_{3}^{2}r_{2}^{3}+2r_{3}r_{2}^{3}-3r_{2}^{3}+3r_{3}^{3}r_{2}^{2}+8r_{3}^{2}r_{2}^{2}+8r_{3}r_{2}^{2}+3r_{2}^{2}+2r_{3}^{4}r_{2}+2r_{3}^{3}r_{2}\nonumber \\
 &  & {}-8r_{3}^{2}r_{2}+2r_{3}r_{2}+2r_{2}-r_{3}^{5}-2r_{3}^{4}+3r_{3}^{3}+3r_{3}^{2}-2r_{3}-1)\nonumber \\
 &  & {}+\frac{k_{1}^{3}\ln\!\left(\frac{r_{2}+r_{3}+1}{2r_{3}}\right)}{6\left(r_{2}-r_{3}+1\right){}^{2}}(r_{2}^{5}-2r_{3}r_{2}^{4}+2r_{2}^{4}-3r_{3}^{2}r_{2}^{3}-2r_{3}r_{2}^{3}-3r_{2}^{3}+3r_{3}^{3}r_{2}^{2}-8r_{3}^{2}r_{2}^{2}+8r_{3}r_{2}^{2}-3r_{2}^{2}+2r_{3}^{4}r_{2}-2r_{3}^{3}r_{2}\nonumber \\
 &  & {}-8r_{3}^{2}r_{2}-2r_{3}r_{2}+2r_{2}-r_{3}^{5}+2r_{3}^{4}+3r_{3}^{3}-3r_{3}^{2}-2r_{3}+1)\,,
\end{eqnarray}
where ${\cal V}(r_1,r_2)={\cal V}(r_2,r_1)$. Let us now analyze this expression, on the lines/points of physical
interest. 
\begin{itemize}
\item \emph{Equilateral} case, $r_{2}=r_{3}=1$. In this case we find 
\begin{equation}
{\cal V}(1,1)={\cal V}_{{\rm equil}}=\frac{15}{2}\,[2+\ln(2/3)]\, k_{1}^{3}\approx11.959\, k_{1}^{3}\,.
\end{equation}

\item \emph{Local} case, $r_{2}=1$, $r_{3}\to0$ (or $r_{2}\to0$, $r_{3}=1$).
In this case, the limit exists and gives 
\begin{equation}
\lim_{r_{3}\to0^{+}}{\cal V}(1,r_{3})={\cal V}_{{\rm local}}=\frac{20}{3}k_{1}^{3}\,,
\label{lolimi}
\end{equation}
where this value of ${\cal V}_{{\rm local}}$ matches with the one
derived by Chen \textit{et al.} \cite{Chen}. 
\item Singular line, $r_{3}=\lim_{\epsilon\to0}(1-r_{2}+\epsilon)$, with
$0<r_{2}<1$. In this case we obtain 
\begin{equation}
\lim_{\epsilon\to0}{\cal V}=\frac{\{(r_{2}-1)r_{2}[11(r_{2}-1)r_{2}-105]-86\}
r_{2}(1-r_{2})+6(r_{2}-1){}^{3}\ln(1-r_{2})-6r_{2}^{3}\ln r_{2}}{12\left(r_{2}-1\right)r_{2}}\, k_{1}^{3}\,.
\end{equation}
The limits $r_{2}\to0,1$ again give the result (\ref{lolimi}).
\item \emph{Enfolded} case, $r_{2}=1/2$, $r_{3}=\lim_{\epsilon\to0}(1/2+\epsilon)$.
This is a particular case of the previous one. Then we find 
\begin{equation}
\lim_{\epsilon\to0}{\cal V}={\cal V}_{{\rm enfold}}
=\left( \frac{315}{64}-\frac{1}{2}\ln2 \right)
k_{1}^{3}\approx4.5753\, k_{1}^{3}\,.
\end{equation}

\item Singular line, $r_{3}=\lim_{\epsilon\to0}(r_{2}-1+\epsilon)$, with
$r_{2}>1$. In this case we find 
\begin{equation}
\lim_{\epsilon\to0}{\cal V}=\frac{(r_{2}-1)\Bigl[\{r_{2}[r_{2}(86r_{2}-105)+94]+22\}r_{2}-6(r_{2}-1){}^{2}r_{2}^{3}\ln\!\left(\frac{r_{2}}{r_{2}-1}\right)-11\Bigr]-6r_{2}^{3}\ln r_{2}}{12\left(r_{2}-1\right)r_{2}}\, k_{1}^{3}\,.
\end{equation}
Taking the limit $r_{2}\to1^{+}$, we recover the result (\ref{lolimi}). 
\item Singular line, $r_{3}=\lim_{\epsilon\to0}(1+r_{2}-\epsilon)$. 
In this case we obtain 
\begin{equation}
\lim_{\epsilon\to0}{\cal V}=\frac{r_{2}\bigl(r_{2}\{r_{2}[r_{2}(86r_{2}+239)+295]+239\}+86\bigr)-6r_{2}^{3}(r_{2}+1){}^{3}\ln\!\left(\frac{1}{r_{2}}+1\right)-6(r_{2}+1){}^{3}\ln(r_{2}+1)}{12r_{2}\left(r_{2}+1\right)}\, k_{1}^{3}\,,
\end{equation}
which again reproduces the value (\ref{lolimi}) 
as $r_{2}=\epsilon\to0$. 
\end{itemize}
Therefore, the integral ${\cal V}$ remains finite in the physical
parameter space of $r_{2}$ and $r_{3}$.




\begin{thebibliography}{10}
\bibitem{inflation} A.~A.~Starobinsky, 
Phys.\ Lett.\ B \textbf{91}, 99 (1980); D.~Kazanas, 
Astrophys.\ J.\ \textbf{241} L59 (1980); K.~Sato, Mon.\ Not.\ R.\ Astron.\ Soc.
\textbf{195}, 467 (1981); Phys.\ Lett.\ \textbf{99B}, 66 (1981);
A.~H.~Guth, 
Phys.\ Rev.\ D \textbf{23}, 347 (1981).

\bibitem{oldper} V.~F.~Mukhanov and G.~V.~Chibisov, 
JETP Lett.\ \textbf{33}, 532 (1981); A.~H.~Guth and S.~Y.~Pi,
Phys.\ Rev.\ Lett.\ \textbf{49} (1982) 1110; S.~W.~Hawking, Phys.\ Lett.\ B
\textbf{115}, 295 (1982); A.~A.~Starobinsky, 
Phys.\ Lett.\ B \textbf{117} (1982) 175; J.~M.~Bardeen, P.~J.~Steinhardt
and M.~S.~Turner, Phys.\ Rev.\ D \textbf{28}, 679 (1983).

\bibitem{COBE} G.~F.~Smoot \textit{et al.}, 
Astrophys.\ J.\ \textbf{396}, L1 (1992).

\bibitem{WMAP1} D.~N.~Spergel \textit{et al.} {[}WMAP Collaboration{]},
Astrophys.\ J.\ Suppl.\ \textbf{148}, 175 (2003) {[}astro-ph/0302209{]}.

\bibitem{WMAP9} G.~Hinshaw \textit{et al.}, 
arXiv:1212.5226 {[}astro-ph.CO{]}.

\bibitem{curvaton} K.~Enqvist and M.~S.~Sloth, 
Nucl.\ Phys.\ B \textbf{626}, 395 (2002) {[}hep-ph/0109214{]}; D.~H.~Lyth
and D.~Wands, 
Phys.\ Lett.\ B \textbf{524}, 5 (2002) {[}hep-ph/0110002{]}; T.~Moroi
and T.~Takahashi, 
Phys.\ Lett.\ B \textbf{522}, 215 (2001) {[}Erratum-ibid.\ B \textbf{539},
303 (2002){]} {[}hep-ph/0110096{]}.

\bibitem{Salopek} D.~S.~Salopek and J.~R.~Bond, 
Phys.\ Rev.\ D \textbf{42}, 3936 (1990).

\bibitem{Gangui} A.~Gangui, F.~Lucchin, S.~Matarrese and S.~Mollerach,
Astrophys.\ J.\ \textbf{430}, 447 (1994).

\bibitem{Verde} L.~Verde, L.~M.~Wang, A.~Heavens and M.~Kamionkowski,
Mon.\ Not.\ Roy.\ Astron.\ Soc.\ \textbf{313}, L141 (2000) {[}astro-ph/9906301{]}.

\bibitem{KSpergel} E.~Komatsu and D.~N.~Spergel, 
Phys.\ Rev.\ \textbf{D63}, 063002 (2001).

\bibitem{Bartolo} N.~Bartolo, S.~Matarrese, A.~Riotto, 
Phys.\ Rev.\ \textbf{D65}, 103505 (2002) {[}hep-ph/0112261{]}; N.~Bartolo,
E.~Komatsu, S.~Matarrese and A.~Riotto, 
Phys.\ Rept.\ \textbf{402}, 103 (2004) {[}astro-ph/0406398{]}.

\bibitem{Maldacena} J.~M.~Maldacena, 
JHEP \textbf{0305}, 013 (2003) {[}astro-ph/0210603{]}.

\bibitem{Cremi2003} P.~Creminelli, 
JCAP \textbf{0310}, 003 (2003) {[}astro-ph/0306122{]}.

\bibitem{Rigo} G.~I.~Rigopoulos and E.~P.~S.~Shellard, 
Phys.\ Rev.\ D \textbf{68}, 123518 (2003) {[}astro-ph/0306620{]};
G.~I.~Rigopoulos and E.~P.~S.~Shellard, 
JCAP \textbf{0510}, 006 (2005) {[}astro-ph/0405185{]}.

\bibitem{Arkani} N.~Arkani-Hamed, P.~Creminelli, S.~Mukohyama
and M.~Zaldarriaga, 
JCAP \textbf{0404}, 001 (2004) {[}hep-th/0312100{]}.

\bibitem{Tong} M.~Alishahiha, E.~Silverstein and D.~Tong, 
Phys.\ Rev.\ D \textbf{70}, 123505 (2004) {[}hep-th/0404084{]}.

\bibitem{Bartolo2} N.~Bartolo, S.~Matarrese and A.~Riotto, 
Phys.\ Rev.\ D \textbf{69}, 043503 (2004) {[}hep-ph/0309033{]};
M.~Sasaki, J.~Valiviita and D.~Wands, 
Phys.\ Rev.\ D \textbf{74}, 103003 (2006) {[}astro-ph/0607627{]};
K.~A.~Malik and D.~H.~Lyth, 
JCAP \textbf{0609}, 008 (2006) {[}astro-ph/0604387{]}.

\bibitem{Seery} D.~Seery and J.~E.~Lidsey, 
JCAP \textbf{0506}, 003 (2005) {[}astro-ph/0503692{]}.

\bibitem{Seery2} D.~Seery and J.~E.~Lidsey, 
JCAP \textbf{0509}, 011 (2005) {[}astro-ph/0506056{]}.

\bibitem{Lyth} D.~H.~Lyth and Y.~Rodriguez, 
Phys.\ Rev.\ Lett.\ \textbf{95}, 121302 (2005) {[}astro-ph/0504045{]};
D.~H.~Lyth and Y.~Rodriguez, 
Phys.\ Rev.\ D \textbf{71}, 123508 (2005) {[}astro-ph/0502578{]};
D.~H.~Lyth, K.~A.~Malik and M.~Sasaki, 
JCAP \textbf{0505}, 004 (2005) {[}astro-ph/0411220{]}.

\bibitem{multifield} 
K.~Enqvist and A.~Vaihkonen, 
JCAP \textbf{0409}, 006 (2004) {[}hep-ph/0405103{]}; 
B.~A.~Bassett, S.~Tsujikawa and D.~Wands,
Rev.\ Mod.\ Phys.\  {\bf 78}, 537 (2006) [astro-ph/0507632];
F.~Vernizzi and D.~Wands, 
JCAP \textbf{0605}, 019 (2006); T.~Battefeld and R.~Easther, 
JCAP \textbf{0703}, 020 (2007) {[}astro-ph/0610296{]}; S.~Yokoyama,
T.~Suyama and T.~Tanaka, 
JCAP \textbf{0707}, 013 (2007) {[}arXiv:0705.3178 {[}astro-ph{]}{]};
F.~Arroja, S.~Mizuno and K.~Koyama, 
JCAP \textbf{0808}, 015 (2008) {[}arXiv:0806.0619 {[}astro-ph{]}{]};
C.~T.~Byrnes, K.~-Y.~Choi and L.~M.~H.~Hall, 
JCAP \textbf{0810}, 008 (2008) {[}arXiv:0807.1101 {[}astro-ph{]}{]};
C.~T.~Byrnes and G.~Tasinato, 
JCAP \textbf{0908}, 016 (2009) {[}arXiv:0906.0767 {[}astro-ph.CO{]}{]};
H.~R.~S.~Cogollo, Y.~Rodriguez and C.~A.~Valenzuela-Toledo,
JCAP {\bf 0808} (2008) 029 [arXiv:0806.1546 [astro-ph]];
C.~T.~Byrnes and K.~-Y.~Choi, 
Adv.\ Astron.\ \textbf{2010}, 724525 (2010) {[}arXiv:1002.3110 {[}astro-ph.CO{]}{]};
Y.~Rodriguez and C.~A.~Valenzuela-Toledo,
Phys.\ Rev.\ D {\bf 81} (2010) 023531
[arXiv:0811.4092 [astro-ph]];
S.~Renaux-Petel, S.~Mizuno and K.~Koyama, 
JCAP \textbf{1111}, 042 (2011) {[}arXiv:1108.0305 {[}astro-ph.CO{]}{]};
J.~Elliston, D.~J.~Mulryne, D.~Seery and R.~Tavakol, 
JCAP \textbf{1111}, 005 (2011) {[}arXiv:1106.2153 {[}astro-ph.CO{]}{]}.

\bibitem{Easther} X.~Chen, R.~Easther and E.~A.~Lim, 
JCAP \textbf{0706}, 023 (2007) {[}astro-ph/0611645{]}; X.~Chen, R.~Easther
and E.~A.~Lim, 
JCAP \textbf{0804}, 010 (2008) {[}arXiv:0801.3295 {[}astro-ph{]}{]}.

\bibitem{Barnaby} N.~Barnaby and J.~M.~Cline, 
JCAP \textbf{0707}, 017 (2007) {[}arXiv:0704.3426 {[}hep-th{]}{]};
N.~Barnaby and J.~M.~Cline, 
JCAP \textbf{0806}, 030 (2008) {[}arXiv:0802.3218 {[}hep-th{]}{]}.

\bibitem{ekprotic} K.~Koyama, S.~Mizuno, F.~Vernizzi and D.~Wands,
JCAP \textbf{0711}, 024 (2007) {[}arXiv:0708.4321 {[}hep-th{]}{]};
E.~I.~Buchbinder, J.~Khoury and B.~A.~Ovrut, 
Phys.\ Rev.\ Lett.\ \textbf{100}, 171302 (2008) {[}arXiv:0710.5172
{[}hep-th{]}{]}.

\bibitem{Suyama} T.~Suyama and M.~Yamaguchi, 
Phys.\ Rev.\ D \textbf{77}, 023505 (2008) {[}arXiv:0709.2545 {[}astro-ph{]}{]};
K.~Ichikawa, T.~Suyama, T.~Takahashi and M.~Yamaguchi, 
Phys.\ Rev.\ D \textbf{78}, 023513 (2008) {[}arXiv:0802.4138 {[}astro-ph{]}{]};
T.~Suyama, T.~Takahashi, M.~Yamaguchi and S.~Yokoyama, 
JCAP \textbf{1012}, 030 (2010) {[}arXiv:1009.1979 {[}astro-ph.CO{]}{]}.

\bibitem{Langlois} D.~Langlois, S.~Renaux-Petel, D.~A.~Steer
and T.~Tanaka, 
Phys.\ Rev.\ Lett.\ \textbf{101}, 061301 (2008) {[}arXiv:0804.3139
{[}hep-th{]}{]}; D.~Langlois, S.~Renaux-Petel, D.~A.~Steer and
T.~Tanaka, 
Phys.\ Rev.\ D \textbf{78}, 063523 (2008) {[}arXiv:0806.0336 {[}hep-th{]}{]}.

\bibitem{multibrid} M.~Sasaki, 
Prog.\ Theor.\ Phys.\ \textbf{120}, 159 (2008) {[}arXiv:0805.0974
{[}astro-ph{]}{]}; A.~Naruko and M.~Sasaki, 
Prog.\ Theor.\ Phys.\ \textbf{121}, 193 (2009) {[}arXiv:0807.0180
{[}astro-ph{]}{]}.

\bibitem{Piazza} J.~Khoury and F.~Piazza, 
JCAP \textbf{0907}, 026 (2009) {[}arXiv:0811.3633 {[}hep-th{]}{]}.

\bibitem{Takamizu} 
Y.~-i.~Takamizu, S.~Mukohyama, M.~Sasaki and Y.~Tanaka, 
JCAP \textbf{1006}, 019 (2010) {[}arXiv:1004.1870 {[}astro-ph.CO{]}{]}.

\bibitem{Matteo} 
N.~Bartolo, M.~Fasiello, S.~Matarrese and A.~Riotto,
JCAP {\bf 1008}, 008 (2010)
[arXiv:1004.0893 [astro-ph.CO]];
JCAP {\bf 1009}, 035 (2010)
[arXiv:1006.5411 [astro-ph.CO]];
JCAP {\bf 1012}, 026 (2010)
[arXiv:1010.3993 [astro-ph.CO]];
M.~Fasiello,
arXiv:1106.2189 [astro-ph.CO].

\bibitem{Burrage} 
C.~Burrage, R.~H.~Ribeiro and D.~Seery,
JCAP {\bf 1107}, 032 (2011)
[arXiv:1103.4126 [astro-ph.CO]];
R.~H.~Ribeiro and D.~Seery,
JCAP {\bf 1110}, 027 (2011)
[arXiv:1108.3839 [astro-ph.CO]].

\bibitem{Libanov} 
M.~Libanov, S.~Mironov and V.~Rubakov, 
Phys.\ Rev.\ D \textbf{84}, 083502 (2011) {[}arXiv:1105.6230 {[}astro-ph.CO{]}{]}.

\bibitem{Sugiyama} 
N.~S.~Sugiyama, E.~Komatsu and T.~Futamase,
Phys.\ Rev.\ Lett.\ \textbf{106}, 251301 (2011).

\bibitem{Ribeiro} 
R.~H.~Ribeiro,
JCAP {\bf 1205}, 037 (2012)
[arXiv:1202.4453 [astro-ph.CO]].

\bibitem{Wandelt} E.~Komatsu, D.~N.~Spergel and B.~D.~Wandelt,
Astrophys.\ J.\ \textbf{634}, 14 (2005) {[}astro-ph/0305189{]}.

\bibitem{Babich} D.~Babich, P.~Creminelli and M.~Zaldarriaga,
JCAP \textbf{0408}, 009 (2004) {[}astro-ph/0405356{]}.

\bibitem{Cremishape1} 
P.~Creminelli, A.~Nicolis, L.~Senatore,
M.~Tegmark and M.~Zaldarriaga, 
JCAP \textbf{0605}, 004 (2006) {[}astro-ph/0509029{]}.

\bibitem{Liguori} M.~Liguori, F.~K.~Hansen, E.~Komatsu, S.~Matarrese
and A.~Riotto, 
Phys.\ Rev.\ D \textbf{73}, 043505 (2006) {[}astro-ph/0509098{]}.

\bibitem{Smith} K.~M.~Smith and M.~Zaldarriaga, 
Mon.\ Not.\ Roy.\ Astron.\ Soc.\ \textbf{417}, 2 (2011) {[}astro-ph/0612571{]}.

\bibitem{Yadav} A.~P.~S.~Yadav, E.~Komatsu and B.~D.~Wandelt,
Astrophys.\ J.\ \textbf{664}, 680 (2007) {[}astro-ph/0701921{]}.

\bibitem{Cremishape2} 
P.~Creminelli, L.~Senatore and M.~Zaldarriaga,
JCAP \textbf{0703}, 019 (2007) {[}astro-ph/0606001{]}.

\bibitem{Fergu} 
J.~R.~Fergusson and E.~P.~S.~Shellard, 
Phys.\ Rev.\ D \textbf{80}, 043510 (2009) {[}arXiv:0812.3413 {[}astro-ph{]}{]}.

\bibitem{SSZ} L.~Senatore, K.~M.~Smith and M.~Zaldarriaga, 
JCAP \textbf{1001}, 028 (2010) {[}arXiv:0905.3746 {[}astro-ph.CO{]}{]}.

\bibitem{Meerburg} P.~D.~Meerburg, J.~P.~van der Schaar and P.~S.~Corasaniti,
JCAP \textbf{0905}, 018 (2009) {[}arXiv:0901.4044 {[}hep-th{]}{]}.

\bibitem{CremiGa} P.~Creminelli, G.~D'Amico, M.~Musso, J.~Norena
and E.~Trincherini, 
JCAP \textbf{1102}, 006 (2011) {[}arXiv:1011.3004 {[}hep-th{]}{]}.

\bibitem{WMAP9fnl} C.~L.~Bennett \textit{et al.}, 
arXiv:1212.5225 {[}astro-ph.CO{]}.

\bibitem{Cremi} P.~Creminelli and M.~Zaldarriaga, 
JCAP \textbf{0410}, 006 (2004) {[}astro-ph/0407059{]}.

\bibitem{Arroja} F.~Arroja, A.~E.~Romano and M.~Sasaki, 
Phys.\ Rev.\ D \textbf{84}, 123503 (2011) {[}arXiv:1106.5384 {[}astro-ph.CO{]}{]};
P.~Adshead, C.~Dvorkin, W.~Hu and E.~A.~Lim, 
Phys.\ Rev.\ D \textbf{85}, 023531 (2012) {[}arXiv:1110.3050 {[}astro-ph.CO{]}{]}.

\bibitem{Namjoo} M.~H.~Namjoo, H.~Firouzjahi and M.~Sasaki, 
arXiv:1210.3692 {[}astro-ph.CO{]}.

\bibitem{Cheung} 
C.~Cheung, A.~L.~Fitzpatrick, J.~Kaplan and L.~Senatore, 
JCAP \textbf{0802}, 021 (2008) {[}arXiv:0709.0295 {[}hep-th{]}{]}.

\bibitem{Ganc} 
J.~Ganc and E.~Komatsu, 
JCAP \textbf{1012}, 009 (2010) {[}arXiv:1006.5457 {[}astro-ph.CO{]}{]}.

\bibitem{Petelsqueezed} 
S.~Renaux-Petel, 
JCAP \textbf{1010}, 020 (2010) {[}arXiv:1008.0260 {[}astro-ph.CO{]}{]}.

\bibitem{Watanabe} 
C.~Germani and Y.~Watanabe,
JCAP {\bf 1107}, 031 (2011)
[Addendum-ibid.\  {\bf 1107}, A01 (2011)]
[arXiv:1106.0502 [astro-ph.CO]].

\bibitem{Mukhanov} C.~Armendariz-Picon, T.~Damour and V.~F.~Mukhanov,
Phys.\ Lett.\ B \textbf{458}, 209 (1999) {[}hep-th/9904075{]}.

\bibitem{Garriga} J.~Garriga and V.~F.~Mukhanov, 
Phys.\ Lett.\ B \textbf{458}, 219 (1999) {[}hep-th/9904176{]}.

\bibitem{Chen} X.~Chen, M.~-x.~Huang, S.~Kachru and G.~Shiu,
JCAP \textbf{0701}, 002 (2007) {[}hep-th/0605045{]}.

\bibitem{Horndeski} G.~W.~Horndeski, Int.\ J.\ Theor.\ Phys.\ 10,
363-384 (1974).

\bibitem{DGSZ} C.~Deffayet, X.~Gao, D.~A.~Steer and G.~Zahariade,
Phys.\ Rev.\ D \textbf{84}, 064039 (2011) {[}arXiv:1103.3260 {[}hep-th{]}{]}.

\bibitem{Char} C.~Charmousis, E.~J.~Copeland, A.~Padilla and
P.~M.~Saffin, 
Phys.\ Rev.\ Lett.\ \textbf{108}, 051101 (2012) {[}arXiv:1106.2000
{[}hep-th{]}{]}.

\bibitem{Koba11} T.~Kobayashi, M.~Yamaguchi and J.~'i.~Yokoyama,
Prog.\ Theor.\ Phys.\ \textbf{126}, 511 (2011) {[}arXiv:1105.5723
{[}hep-th{]}{]}.

\bibitem{Gao} X.~Gao and D.~A.~Steer, 
JCAP \textbf{1112}, 019 (2011) {[}arXiv:1107.2642 {[}astro-ph.CO{]}{]}.

\bibitem{ATnongau} 
A.~De Felice and S.~Tsujikawa, 
Phys.\ Rev.\ D \textbf{84}, 083504 (2011) {[}arXiv:1107.3917 {[}gr-qc{]}{]}.

\bibitem{Mizuno} 
S.~Mizuno and K.~Koyama, 
Phys.\ Rev.\ D \textbf{82}, 103518 (2010) {[}arXiv:1009.0677 {[}hep-th{]}{]}.

\bibitem{ATnongau0} 
A.~De Felice and S.~Tsujikawa, 
JCAP \textbf{1104}, 029 (2011) {[}arXiv:1103.1172 {[}astro-ph.CO{]}{]}.

\bibitem{KobaGa} T.~Kobayashi, M.~Yamaguchi and J.~'i.~Yokoyama,
Phys.\ Rev.\ D \textbf{83}, 103524 (2011) {[}arXiv:1103.1740 {[}hep-th{]}{]}.

\bibitem{Gaotensor} 
X.~Gao, T.~Kobayashi, M.~Yamaguchi and J.~'i.~Yokoyama,
Phys.\ Rev.\ Lett.\ \textbf{107}, 211301 (2011) {[}arXiv:1108.3513
{[}astro-ph.CO{]}{]}; 
X.~Gao \textit{et al.}, 
arXiv:1207.0588 {[}astro-ph.CO{]}.

\bibitem{ADM} 
R.~L.~Arnowitt, S.~Deser and C.~W.~Misner,
Phys.\ Rev.\  {\bf 117}, 1595 (1960).

\bibitem{DeFe2012} 
A.~De Felice and S.~Tsujikawa,
JCAP {\bf 1202}, 007 (2012)
[arXiv:1110.3878 [gr-qc]].

\bibitem{Petel} S.~Renaux-Petel, 
JCAP \textbf{1202}, 020 (2012) {[}arXiv:1107.5020 {[}astro-ph.CO{]}{]}.

\bibitem{Senatore10}
L.~Senatore, K.~M.~Smith and M.~Zaldarriaga,
JCAP {\bf 1001} (2010) 028
[arXiv:0905.3746 [astro-ph.CO]].

\bibitem{KYY} 
T.~Kobayashi, M.~Yamaguchi and J.~'i.~Yokoyama,
Phys.\ Rev.\ Lett.\ \textbf{105}, 231302 (2010) {[}arXiv:1008.0603
{[}hep-th{]}{]}.

\bibitem{Kamada} 
K.~Kamada, T.~Kobayashi, M.~Yamaguchi and J.~'i.~Yokoyama,
Phys.\ Rev.\ D {\bf 83}, 083515 (2011)
[arXiv:1012.4238 [astro-ph.CO]].

\bibitem{Tavakol} 
A.~De Felice, S.~Tsujikawa, J.~Elliston and R.~Tavakol,
JCAP {\bf 1108}, 021 (2011)
[arXiv:1105.4685 [astro-ph.CO]].

\bibitem{Junko} 
J.~Ohashi and S.~Tsujikawa,
JCAP {\bf 1210}, 035 (2012)
[arXiv:1207.4879 [gr-qc]].

\bibitem{PLANCK} {[}PLANCK Collaboration{]}, 
arXiv:astro-ph/0604069.\end{thebibliography}
\end{document}